\documentclass[prd,twocolumn,tightenlines,preprintnumbers,showpacs,superscriptaddress,notitlepage,nofootinbib,eqsecnum,floatfix,longbibliography]{revtex4-1}
\usepackage[export]{adjustbox} 
\usepackage{dcolumn}   
\usepackage{bm}        
\usepackage{amssymb}   
\usepackage{standalone}
\usepackage{enumitem}
\usepackage[pdftex]{color}
\usepackage{xcolor}
\usepackage{slashed}
\usepackage{booktabs}
\usepackage{multirow}
\usepackage{amsmath}
\usepackage{stackrel}
\usepackage{rotating}
\usepackage{CJKutf8}
\usepackage{pifont}
\usepackage{mathtools}
\usepackage{hyperref}
\hypersetup{
    colorlinks=true,       
    linkcolor=blue,          
    citecolor=blue,        
    filecolor=blue,      
    urlcolor=blue           
}
\usepackage{simplewick}

\newcommand{\eqnref}[1]{Eq.~(\ref{#1})}
\newcommand{\secref}[1]{Sec.~\ref{#1}}
\newcommand{\tabref}[1]{Tab.~\ref{#1}}
\newcommand{\figref}[1]{Fig.~\ref{#1}}

\newcommand{\nxlo}[1]{N${}^{#1}$LO}

\definecolor{kngrey}{HTML}{A6AAA9}
\definecolor{knred}{HTML}{EC5D57}
\definecolor{knorange}{HTML}{F39019}
\definecolor{knyellow}{HTML}{F5D328}
\definecolor{kngreen}{HTML}{70BF41}
\definecolor{knblue}{HTML}{51A7F9}
\definecolor{knpurple}{HTML}{B36AE2}

\def\mc#1{{\mathcal #1}}

\def\a{{\alpha}}

\def\d{{\delta}}
\def\D{{\Delta}}
\def\t{\tau}
\def\e{{\varepsilon}}
\def\g{{\gamma}}
\def\G{{\Gamma}}

\def\l{{\lambda}}
\def\L{{\Lambda}}

\def\s{{\sigma}}

\definecolor{gray}{rgb}{0.6,0.6,0.6}

\def\a{{\alpha}}

\def\d{{\delta}}
\def\D{{\Delta}}
\def\e{{\epsilon}}
\def\g{{\gamma}}
\def\G{{\Gamma}}

\def\l{{\lambda}}
\def\L{{\Lambda}}

\def\s{{\sigma}}
\def\t{{\tau}}

\def\mc#1{{\mathcal #1}}

\def\Lbar{{\bar{L}}}

\newcommand{\ithems}{
    Interdisciplinary Theoretical and Mathematical Sciences Program (iTHEMS),
    RIKEN, 2-1 Hirosawa,
    Wako, Saitama 351-0198, Japan
}
\newcommand{\arithmer}{
    Arithmer Inc., R\&D Headquarters,
    Minato, Tokyo 106-6040, Japan
}

\newcommand{\jlabt}{
	Theory Center,
	Thomas Jefferson National Accelerator Facility,
	Newport News, VA 23606, USA
	}
\newcommand{\jlabc}{
	Scientific Computing Group,
	Thomas Jefferson National Accelerator Facility,
	Newport News, VA 23606, USA
	}
\newcommand{\julich}{
 Institut f\"{u}r Kernphysik and Institute for Advanced Simulation,
 Forschungszentrum J\"{u}lich, 54245 J\"{u}lich Germany
 }

\newcommand{\lblnsd}{
    Nuclear Science Division,
    Lawrence Berkeley National Laboratory,
	Berkeley, CA 94720, USA
	}

\newcommand{\llnl}{
	Physics Division,
	Lawrence Livermore National Laboratory,
	Livermore, CA 94550, USA
	}
\newcommand{\llnldesign}{
	Design Physics Division,
	Lawrence Livermore National Laboratory,
	Livermore, CA 94550, USA
	}
\newcommand{\nvidia}{
    NVIDIA Corporation,
    2701 San Tomas Expressway, Santa Clara, CA 95050, USA
    }

\newcommand{\ucb}{
	Department of Physics,
	University of California,
	Berkeley, CA 94720, USA
	}
\newcommand{\umd}{
	Department of Physics,
	University of Maryland,
	College Park, MD 20742, USA
}
\newcommand{\unc}{
	Department of Physics and Astronomy,
	University of North Carolina,
	Chapel Hill, NC 27516-3255, USA
	}

\newcommand{\wm}{
	Department of Physics,
	The College of William \& Mary,
	Williamsburg, VA 23187, USA
	}

\newcount\hour \newcount\hourminute \newcount\minute
\hour=\time \divide \hour by 60
\hourminute=\hour \multiply \hourminute by 60
\minute=\time \advance \minute by -\hourminute
\newcommand{\mydate}{\ \today \ - \number\hour :\number\minute}

\begin{document}

\title{$F_K / F_\pi$ from M\"{o}bius domain-wall fermions solved on gradient-flowed HISQ ensembles}

\author{Nolan~Miller}
\affiliation{\unc}

\author{Henry~Monge-Camacho}
\affiliation{\unc}

\author{Chia~Cheng~Chang (\begin{CJK*}{UTF8}{bsmi}張家丞\end{CJK*})}
\affiliation{\ithems}
\affiliation{\lblnsd}
\affiliation{\ucb}

\author{Ben~H\"orz}
\affiliation{\lblnsd}

\author{Enrico~Rinaldi}
\affiliation{\arithmer}
\affiliation{\ithems}

\author{Dean Howarth}
\affiliation{\llnl}
\affiliation{\lblnsd}

\author{Evan~Berkowitz}
\affiliation{\umd}
\affiliation{\julich}

\author{David~A.~Brantley}
\affiliation{\llnl}

\author{Arjun~Singh~Gambhir}
\affiliation{\llnldesign}
\affiliation{\lblnsd}

\author{Christopher~K\"orber}
\affiliation{\ucb}
\affiliation{\lblnsd}

\author{Christopher~J.~Monahan}
\affiliation{\wm}
\affiliation{\jlabt}

\author{M.A.~Clark}
\affiliation{\nvidia}

\author{B\'{a}lint~Jo\'{o}}
\affiliation{\jlabc}

\author{Thorsten~Kurth}
\affiliation{\nvidia}

\author{Amy~Nicholson}
\affiliation{\unc}

\author{Kostas Orginos}
\affiliation{\wm}
\affiliation{\jlabt}

\author{Pavlos~Vranas}
\affiliation{\llnl}
\affiliation{\lblnsd}

\author{Andr\'{e}~Walker-Loud}
\affiliation{\lblnsd}
\affiliation{\llnl}
\affiliation{\ucb}

\date{\mydate}

\begin{abstract}
We report the results of a lattice quantum chromodynamics calculation of $F_K/F_\pi$ using M\"{o}bius domain-wall fermions computed on gradient-flowed $N_f=2+1+1$ highly-improved staggered quark ensembles.
The calculation is performed with five values of the pion mass ranging from $130 \lesssim m_\pi \lesssim 400$~MeV, four lattice spacings of $a\sim 0.15, 0.12, 0.09$ and $0.06$~fm and multiple values of the lattice volume.
The interpolation/extrapolation to the physical pion and kaon mass point, the continuum, and infinite volume limits are performed with a variety of different extrapolation functions utilizing both the relevant mixed-action effective field theory expressions as well as discretization-enhanced continuum chiral perturbation theory formulas.
We find that the $a\sim0.06$~fm ensemble is helpful, but not necessary to achieve a subpercent determination of $F_K/F_\pi$.
We also include an estimate of the strong isospin breaking corrections and arrive at a final result of $F_{\hat{K}^+}/F_{\hat{\pi}^+} = 1.1942(45)$ with all sources of statistical and systematic uncertainty included.
This is consistent with the Flavour Lattice Averaging Group average value, providing an important benchmark for our lattice action.
Combining our result with experimental measurements of the pion and kaon leptonic decays leads to a determination of $|V_{us}|/|V_{ud}| = 0.2311(10)$.
\end{abstract}

\preprint{LLNL-JRNL-809712, RIKEN-iTHEMS-Report-20, JLAB-THY-20-3192}

\maketitle
\tableofcontents
\section{Introduction \label{sec:intro}}

Leptonic decays of the charged pions and kaons provide a means for probing flavor-changing interactions of the Standard Model (SM). In particular, the SM predicts that the Cabibbo-Kobayashi-Maskawa (CKM) matrix is unitary, providing strict constraints on various sums of the matrix elements. Thus, a violation of these constraints is indicative of new, beyond the SM physics. There is a substantial flavor physics program dedicated to searching indirectly for potential violations.

CKM matrix elements may be determined through a combination of experimental leptonic decay widths and theoretical determinations of the meson decay constants. For example, the ratio of the kaon and pion decay constants, $F_K, F_\pi$, respectively, may be related to the ratio of light and strange CKM matrix elements $|V_{us}|, |V_{ud}|$ via~\cite{Marciano:2004uf,Aubin:2004fs},
\begin{align}
\label{eq:vus_vud}
\frac{\Gamma(K\to l\bar{\nu}_l)}{\Gamma(\pi \to l\bar{\nu}_l)}
    &= \frac{|V_{us}|^2}{|V_{ud}|^2} \frac{F_K^2}{F_{\pi}^2}\frac{m_K}{m_{\pi}}
        \frac{\left(1-\frac{m_l^2}{m_K^2}\right)^2}{\left(1-\frac{m_l^2}{m_\pi^2}\right)^2}
\nonumber\\&\phantom{=}
    \times\left[1+\d_{\mathrm{EM}} + \d_{SU(2)}\right] \, .
\end{align}
In this expression, $l=e,\mu$,  the one-loop radiative quantum electrodynamics (QED) correction is $\delta_{\mathrm{EM}}$~\cite{Decker:1994ea,Finkemeier:1995gi} and $\d_{SU(2)}$ is the strong isospin breaking correction that relates $F_K^2/F_\pi^2$ in the isospin limit to $F_{K^+}^2/F_{\pi^+}^2$ that includes $m_d-m_u$ corrections~\cite{Cirigliano:2011tm}
\begin{equation*}
    \frac{F_{\hat{K}^+}^2}{F_{\hat{\pi}^+}^2} =
        \frac{F_K^2}{F_\pi^2} \left[1 + \d_{SU(2)} \right]\, .
\end{equation*}

Using lattice quantum chromodynamics (QCD) calculations of the ratio of decay constants in the above expression yields one of the most precise determinations of $|V_{us}|/|V_{ud}|$~\cite{Tanabashi:2018oca}. Combining the results obtained through lattice QCD with independent determinations of the CKM matrix elements, such as semileptonic meson decays,  provides a means for testing the unitarity of the CKM matrix and obtaining signals of new physics.

$F_K/F_{\pi}$ is a so-called \textit{gold-plated} quantity~\cite{Davies:2003ik}  for calculating within lattice QCD. This dimensionless ratio skirts the issue of determining a physical scale for the lattices, and gives precise results due to the correlated statistical fluctuations between numerator and denominator, as well as the lack of signal-to-noise issues associated with calculations involving, for instance, nucleons.
Lattice QCD calculations of $F_K/F_{\pi}$ are now a mature endeavor, with state-of-the-art calculations determining this quantity consistently with subpercent precision. The most recent review by the Flavour Lattice Averaging Group (FLAG), which performs global averages of quantities that have been calculated and extrapolated to the physical point by multiple groups, quotes a value of
\begin{equation}
    \frac{F_{\hat{K}^+}}{F_{\pi^+}} = 1.1932(19)
\end{equation}
for $N_f = 2+1+1$ dynamical quark flavors, including strong-isospin breaking corrections~\cite{Aoki:2019cca}.

This average includes calculations derived from two different lattice actions, one~\cite{Carrasco:2014poa} with twisted-mass fermions~\cite{Frezzotti:2003xj,Frezzotti:2003ni} and the other two~\cite{Dowdall:2013rya,Bazavov:2017lyh} with the highly improved staggered quark (HISQ) action~\cite{Follana:2006rc}.  The results obtained using the HISQ action are approximately seven times more precise than those from twisted mass and so the universality of the continuum limit for $F_K/F_\pi$ from $N_f=2+1+1$ results has not been tested with precision yet:
in the continuum limit, all lattice actions should reduce to a single universal limit, that of SM QCD, provided all systematics are properly accounted for. Thus, in addition to lending more confidence to its global average, the calculation of a gold-plated quantity also allows for precise testing of new lattice actions, and the demonstration of control over systematic uncertainties for a given action.
FLAG also reports averages for $N_F=2+1$, $F_{K^\pm}/F_{\pi^\pm}=1.1917(37)$ from Refs.~\cite{Follana:2007uv,Bazavov:2010hj,Durr:2010hr,Blum:2014tka,Durr:2016ulb,Bornyakov:2016dzn} and for $N_f=2$, $F_{K^\pm}/F_{\pi^\pm}=1.1205(18)$ from Refs.~\cite{Blossier:2009bx}, though we restrict our direct comparisons to the $N_f=2+1+1$ results just for simplicity.

In this work, we report a new determination of $F_K/F_\pi$ calculated with M\"{o}bius domain-wall fermions computed on gradient-flowed $N_f=2+1+1$ HISQ ensembles~\cite{Berkowitz:2017opd}.
Our final result in the isospin symmetric limit, \secref{sec:full_analysis}, including a breakdown in terms of statistical ($s$), pion and kaon mass extrapolation ($\chi$), continuum limit ($a$), infinite volume limit ($V$), physical point (phys) and model selection ($M$) uncertainties, is
\begin{align}\label{eq:final_isospin}
\frac{F_K}{F_\pi} &= 1.1964(32)^s(12)^\chi(20)^a(01)^V(15)^{\rm phys}(12)^M
\nonumber\\ &= 1.1964(44)\, .
\end{align}
With our estimated strong isospin breaking corrections, \secref{sec:isospin}, our result including $m_d-m_u$ effects is
\begin{align}\label{eq:final_FKFpi_plus}
\frac{F_{\hat{K}^+}}{F_{\hat{\pi}^+}} &= 1.1942(44)(07)^{\rm iso}
\nonumber\\
    &= 1.1942(45)\, ,
\end{align}
where the first uncertainty in the first line is the combination of those in \eqnref{eq:final_isospin}.

In the following sections we will discuss details of our lattice calculation, including a brief synopsis of the action and ensembles used, as well as our strategy for extracting the relevant quantities from correlation functions. We will then detail our procedure for extrapolating to the physical point via combined continuum, infinite volume, and physical pion and kaon mass limits and the resulting uncertainty breakdown.
We discuss the impact of the $a\sim0.06$~fm ensemble on our analysis, the convergence of the $SU(3)$-flavor chiral expansion, and the estimate of the strong isospin breaking corrections.
We conclude with an estimate of the impact our result has on improving the extraction of $|V_{us}|/|V_{ud}|$ and an outlook.

\section{Details of the lattice calculation \label{sec:LQCD}}

\subsection{MDWF on gradient-flowed HISQ \label{sec:action}}

\begingroup \squeezetable
\begin{table*}
\caption{\label{tab:lattice_params}
Input parameters for our lattice action.  The abbreviated ensemble name~\cite{Bhattacharya:2015wna} indicates the approximate lattice spacing in fm and pion mass in MeV.
The S, L, XL which come after an ensemble name denote a relatively small, large and extra-large volume with respect to $m_\pi L=4$.
}
\begin{ruledtabular}
\begin{tabular}{lccclll|cccllllcrr}
Ensemble& $\beta$& $N_{\rm cfg}$& volume& $am_l$& $am_s$& $am_c$&
    $L_5/a$& $aM_5$& $b_5, c_5$& $am_l^{\rm val}$& $am^{\rm res}_l\hspace{-0.2em}\times\hspace{-0.2em}10^{4}$&
    $am_s^{\rm val}$& $am^{\rm res}_s\hspace{-0.2em}\times\hspace{-0.2em}10^{4}$& $\s$ &$N$& $N_{\rm src}$\\
\hline
a15m400\footnote{Additional ensembles generated by CalLat using the MILC code.  The m350 and m400 ensembles were made on the Vulcan supercomputer at LLNL while the a15m135XL, a09m135, and a06m310L ensembles were made on the Sierra and Lassen supercomputers at LLNL and the Summit supercomputer at OLCF using \texttt{QUDA}~\cite{Clark:2009wm,Babich:2011np}. These configurations are available to any interested party upon request, and will be available for easy anonymous downloading---hopefully soon.}&
    5.80& 1000& $16^3\times48$& 0.0217& 0.065& 0.838&
    12& 1.3& 1.50, 0.50& 0.0278& 9.365(87)& 0.0902& 6.937(63)& 3.0& 30& 8\\
a15m350\footnotemark[1]&
    5.80& 1000& $16^3\times48$& 0.0166& 0.065& 0.838&
    12& 1.3& 1.50, 0.50& 0.0206& 9.416(90)& 0.0902& 6.688(62)& 3.0& 30& 16\\
a15m310&
    5.80& 1000& $16^3\times48$& 0.013& 0.065& 0.838&
    12& 1.3& 1.50, 0.50& 0.0158& 9.563(67)& 0.0902& 6.640(44)& 4.2& 45& 24\\
a15m220&
    5.80& 1000& $24^3\times48$& 0.0064& 0.064& 0.828&
    16& 1.3& 1.75, 0.75& 0.00712& 5.736(38)& 0.0902& 3.890(25)& 4.5& 60& 16\\
a15m135XL\footnotemark[1]&
    5.80& 1000& $48^3\times64$& 0.002426& 0.06730& 0.8447&
    24& 1.3& 2.25, 1.25& 0.00237& 2.706(08)& 0.0945& 1.860(09)& 3.0& 30& 32\\
\hline
a12m400\footnotemark[1]&
    6.00& 1000& $24^3\times64$& 0.0170& 0.0509& 0.635&
    8&  1.2& 1.25, 0.25& 0.0219& 7.337(50)& 0.0693& 5.129(35)& 3.0& 30& 8\\
a12m350\footnotemark[1]&
    6.00& 1000& $24^3\times64$& 0.0130& 0.0509& 0.635&
    8&  1.2& 1.25, 0.25& 0.0166& 7.579(52)& 0.0693& 5.062(34)& 3.0& 30& 8\\
a12m310&
    6.00& 1053& $24^3\times64$& 0.0102& 0.0509& 0.635&
    8&   1.2& 1.25, 0.25& 0.0126& 7.702(52)& 0.0693& 4.950(35)& 3.0& 30& 8\\
a12m220S&
    6.00& 1000& $24^4\times64$& 0.00507& 0.0507& 0.628&
    12&  1.2& 1.50, 0.50& 0.00600& 3.990(42)& 0.0693& 2.390(24)& 6.0& 90& 4\\
a12m220&
    6.00& 1000& $32^3\times64$& 0.00507& 0.0507& 0.628&
    12&  1.2& 1.50, 0.50& 0.00600& 4.050(20)& 0.0693& 2.364(15)& 6.0& 90& 4\\
a12m220L&
    6.00& 1000& $40^3\times64$& 0.00507& 0.0507& 0.628&
    12&  1.2& 1.50, 0.50& 0.00600& 4.040(26)& 0.0693& 2.361(19)& 6.0& 90& 4\\
a12m130&
    6.00& 1000& $48^3\times64$& 0.00184& 0.0507& 0.628&
    20& 1.2& 2.00, 1.00& 0.00195& 1.642(09)& 0.0693& 0.945(08)& 3.0& 30& 32\\
\hline
a09m400\footnotemark[1]&
    6.30& 1201& $32^3\times64$& 0.0124& 0.037& 0.44&
    6&  1.1& 1.25, 0.25& 0.0160& 2.532(23)& 0.0491& 1.957(17)& 3.5& 45& 8\\
a09m350\footnotemark[1]&
    6.30& 1201& $32^3\times64$& 0.00945& 0.037& 0.44&
    6&  1.1& 1.25, 0.25& 0.0121& 2.560(24)& 0.0491& 1.899(16)& 3.5& 45& 8\\
a09m310&
    6.30& 780& $32^3\times96$& 0.0074& 0.037& 0.44&
    6&  1.1& 1.25, 0.25& 0.00951& 2.694(26)& 0.0491& 1.912(15)& 6.7& 167& 8\\
a09m220&
    6.30& 1001& $48^3\times96$& 0.00363& 0.0363& 0.43&
    8&  1.1& 1.25, 0.25& 0.00449& 1.659(13)& 0.0491& 0.834(07)& 8.0& 150& 6\\
a09m135\footnotemark[1]&
    6.30& 1010& $64^3\times96$& 0.001326& 0.03636& 0.4313&
    12& 1.1& 1.50, 0.50& 0.00152& 0.938(06)& 0.04735& 0.418(04)& 3.5& 45& 16\\
\hline
a06m310L\footnotemark[1]&
    6.72& 1000& $72^3\times96$& 0.0048& 0.024& 0.286&
    6&  1.0& 1.25, 0.25& 0.00617& 0.225(03)& 0.0309& 0.165(02)& 3.5& 45& 8
\end{tabular}
\end{ruledtabular}
\end{table*}
\endgroup

There are many choices for discretizing QCD, with each choice being commonly referred to as a lattice action.
These actions correspond to different UV theories that share a common low-energy theory, QCD. Sufficiently close to the continuum limit, the discrete lattice actions can be expanded as a series of local operators known as the Symanzik expansion~\cite{Symanzik:1983dc,Symanzik:1983gh}, the low-energy effective field theory (EFT) for the discrete lattice action. The Symanzik EFT contains a series of operators having higher dimension than those in QCD, multiplied by appropriate powers of the lattice spacing, $a$.
For all lattice actions, the only operators of mass-dimension $\leq4$ are those of QCD, such that the explicit effects from the various discretizations are encoded only in higher-dimensional operators which are all irrelevant in the renormalization sense.
There is a universality of the continuum limit, $a\rightarrow 0$, in that all lattice actions, if calculated using sufficiently small lattice spacing, will recover the target theory of QCD, provided there are no surprises from nonperturbative effects.

Performing lattice QCD calculations with different actions is therefore valuable to test this universality, to help ensure a given action is not accidentally in a different phase of QCD, and to protect against unknown systematic uncertainties arising from a particular calculation with a particular action.
In this work, we use a mixed-action~\cite{Renner:2004ck} in which the discretization scheme for the valence quarks is the M\"{o}bius domain-wall fermion (MDWF) action~\cite{Brower:2004xi,Brower:2005qw,Brower:2012vk} while the discretization scheme for the sea-quarks is the HISQ action~\cite{Follana:2006rc}. Before solving the MDWF propagators, we apply a gradient-flow~\cite{Narayanan:2006rf,Luscher:2011bx,Luscher:2013cpa} smoothing algorithm~\cite{Luscher:2010iy,Lohmayer:2011si} to the gluons to dampen UV fluctuations, which also significantly improves the chiral symmetry properties of the MDWF action~\cite{Berkowitz:2017opd} (for example, the residual chiral symmetry breaking scale of domain-wall fermions $m^{\rm res}$ is held to less than 10\% of $m_l$ for reasonable values of $L_5$ and $M_5$, see \tabref{tab:lattice_params}).
Our motivation to perform this calculation is to improve our understanding of $F_K/F_\pi$ and to test the MDWF on gradient-flowed HISQ action we have used to compute the $\pi^-\rightarrow\pi^+$ neutrinoless double beta decay matrix elements arising from prospective higher-dimension lepton-number-violating physics~\cite{Nicholson:2018mwc}, and the axial coupling of the nucleon $g_A$~\cite{Berkowitz:2017gql,Chang:2018uxx}.
As there is an otherwise straightforward path to determining $g_A$ to subpercent precision with pre-exascale computing such as Summit at Oak Ridge Leadership Computing Facility (OLCF) and Lassen at Lawrence Livermore National Laboratory (LLNL)~\cite{Berkowitz:2018gqe}, it is important to ensure this action is consistent with known results at this level of precision.

There are several motivations for choosing this mixed-action (MA) scheme~\cite{Renner:2004ck,Bar:2002nr}.
The MILC Collaboration provides their gauge configurations to any interested party and we have made heavy use of them.  They have generated the configurations covering a large parameter space allowing one to fully control the physical pion mass, infinite volume and continuum limit extrapolations~\cite{Bazavov:2010ru,Bazavov:2012xda}.
The good chiral symmetry properties of the Domain Wall (DW) action~\cite{Kaplan:1992bt,Shamir:1993zy,Furman:1994ky} significantly suppress sources of chiral symmetry breaking from any sea-quark action, motivating the use of this mixed-action setup.
While this action is not unitary at finite lattice spacing, we have tuned the valence quark masses such that the valence pion mass matches the taste-5 HISQ pion mass within a few percent, so as the continuum limit is taken, we recover a unitary theory.

EFT can be used to understand the salient features of such mixed-action lattice QCD (MALQCD) calculations. Chiral perturbation theory ($\chi$PT)~\cite{Langacker:1973hh,Gasser:1983yg,Leutwyler:1993iq} can be extended to incorporate discretization effects into the analytic formula describing the quark-mass dependence of various hadronic quantities~\cite{Sharpe:1998xm}.
The MA EFT~\cite{Bar:2003mh} for DW valence fermions on dynamical rooted staggered fermions is well developed~\cite{Bar:2005tu,Tiburzi:2005is,Chen:2005ab,Chen:2006wf,Orginos:2007tw,Jiang:2007sn,Chen:2007ug,Chen:2009su}.
The use of valence fermions which respect chiral symmetry leads to a universal form of the MA EFT extrapolation formulas at next-to-leading order (NLO) in the joint quark mass and lattice spacing expansions~\cite{Chen:2006wf,Chen:2007ug}, which follows from the suppression of chiral symmetry breaking discretization effects.

\subsection{Correlation function construction and analysis \label{sec:two_points}}
The correlation function construction and analysis follows closely the strategy of Ref.~\cite{Berkowitz:2017opd} and \cite{Berkowitz:2017gql,Chang:2018uxx}. Here we summarize the relevant details for this work.

The pseudoscalar decay constants $F$ can be obtained from standard two-point correlation functions by making use of the 5D Ward-Takahashi identity~\cite{Blum:2000kn, Aoki:2002vt}
\begin{equation}
F^{q_1 q_2} = z_{0p}^{q_1 q_2} \frac{m^{q_1} +m_{\textrm{res}}^{q_1} + m^{q_2} + m_{\textrm{res}}^{q_2}}{\left(E_0^{q_1 q_2}\right)^{3/2}},
\end{equation}
where $q_1$ and $q_2$ denote the quark content of the meson with lattice input masses $m_{q_1}$ and $m_{q_2}$ respectively.
The point-sink ground-state overlap-factor $z_{0p}$ and ground-state energy $E_0$ are extracted from a two-point correlation function analysis with the model
\begin{equation}
C^{q_1 q_2}_{(ss)ps}(t) = \sum_n z_{n(s)p}^{q_1 q_2} z_{ns}^{q_1 q_2 \dagger} \left(e^{-E_n^{q_1 q_2}t} + e^{-E_n^{q_1 q_2}(T-t)}\right),
\end{equation}
where $n$ encompasses in general an infinite tower of states, $t$ is the source-sink time separation, $T$ is the temporal box size and we have both smeared ($s$) and point ($p$) correlation functions which both come from smeared sources.
From Ref.~\cite{Berkowitz:2017opd}, we show that gradient-flow smearing leads to the suppression of the domain-wall fermion oscillating mode (which also decouples as $M_5\rightarrow1$, at least in free-field~\cite{Syritsyn:2007mp}), and therefore this mode is not included in the correlator fit model.
Finally, the residual chiral symmetry breaking $m_{\textrm{res}}$ is calculated by the ratio of two-point correlation functions evaluated at the midpoint of the fifth dimension $L_5/2$ and bounded on the domain wall~\cite{Brower:2012vk}
\begin{equation}\label{Eq:mres_correlator}
m_{\textrm{res}}(t) = \frac{\sum_{\mathbf{x}}\langle \pi(t, \mathbf{x}, L_5/2) \pi(0, \mathbf{0}, 0)\rangle}{\sum_{\mathbf{x}}\langle \pi(t, \mathbf{x}, 0) \pi(0, \mathbf{0}, 0)\rangle},
\end{equation}
where $\pi(t, \mathbf{x}, s) \equiv \bar{q}(t, \mathbf{x}, w) \gamma_5 q(t, \mathbf{x}, w)$ is the pseudoscalar interpolating operator at time $t$, space $\mathbf{x}$ and fifth dimension $s$.
We extract $m_{\textrm{res}}$ by fitting Eq.~(\ref{Eq:mres_correlator}) to a constant.

\subsubsection{Analysis strategy}
For all two-point correlation function parameters (MDWF and mixed MDWF-HISQ), we infer posterior parameter distributions in a Bayesian framework using a 4-state model which simultaneously describes the smeared- and point-sink two-point correlation functions (the source is always smeared).
The joint posterior distribution is approximated by a multivariate normal distribution (we later refer to this procedure as \textit{fitting}).
The two-point correlation functions are folded in time to double the statistics.
The analysis of the pion, kaon, $\bar{s}\gamma_{5} s$, and mixed MDWF-HISQ mesons are performed independently, with correlations accounted for under bootstrap resampling.

\begin{figure}
\includegraphics[width=\columnwidth,valign=t]{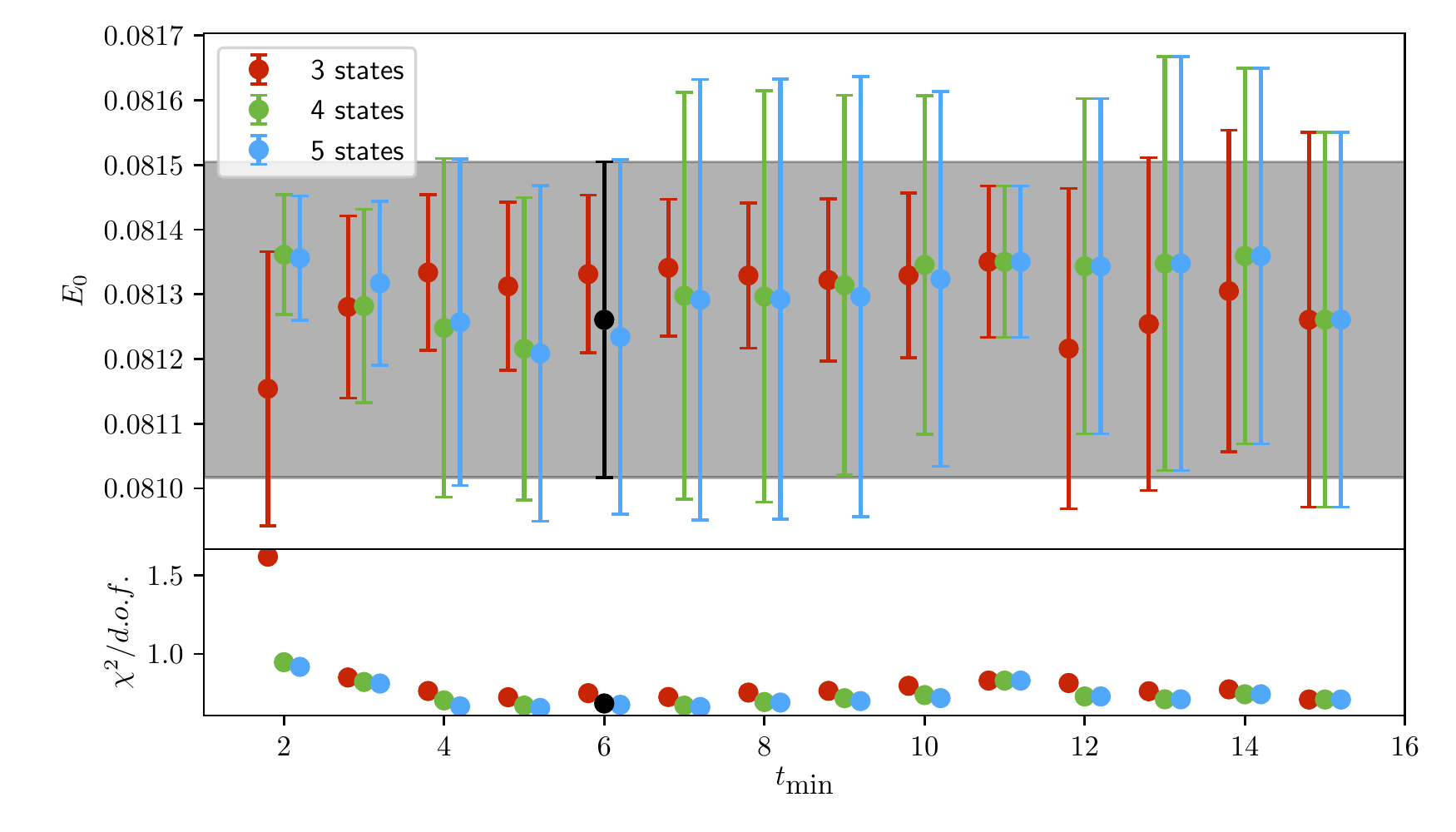}
\includegraphics[width=\columnwidth,valign=t]{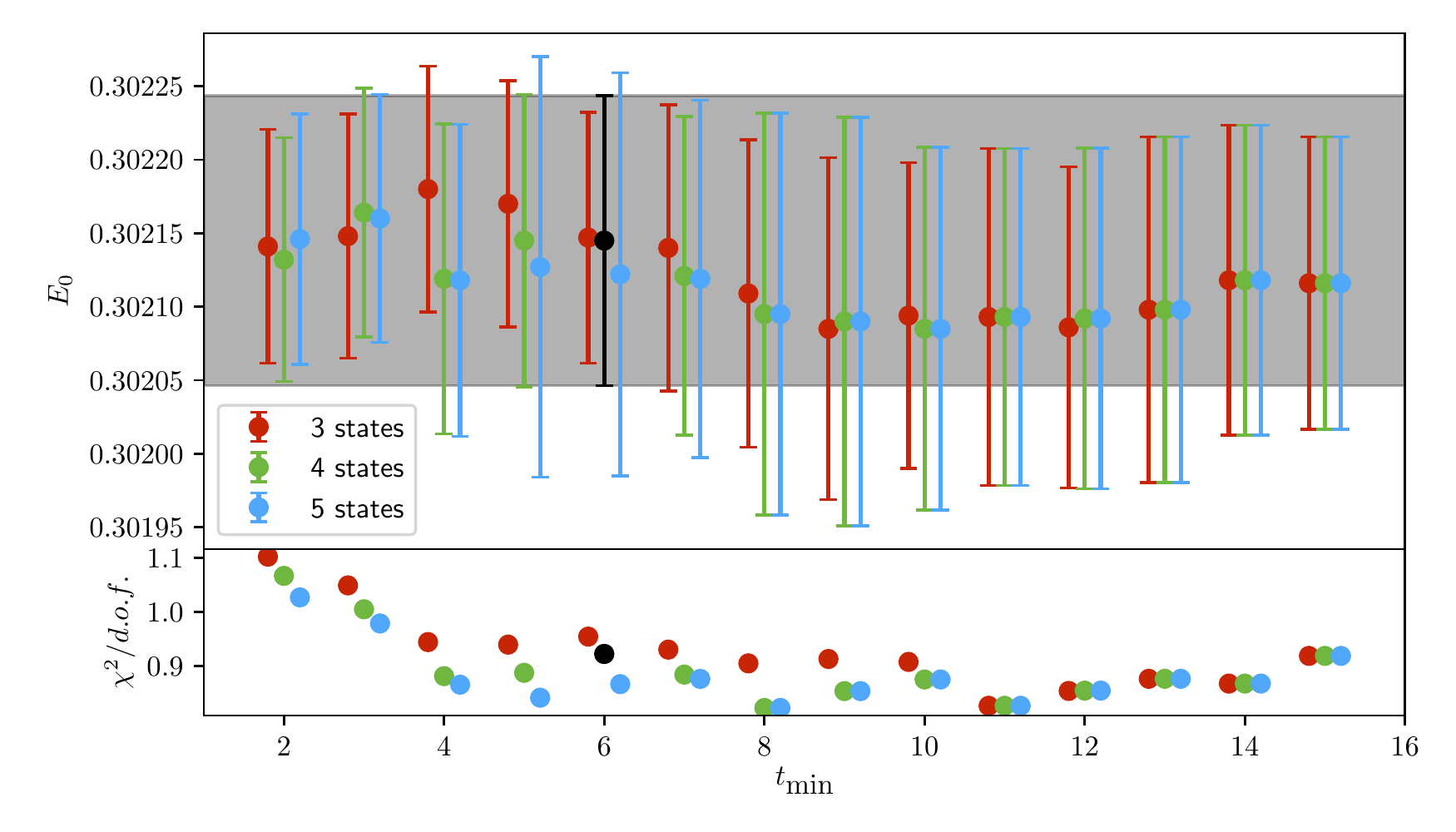}
\caption{\label{fig:corr_stability}
Stability of the ground-state mass determination of the pion (top) and kaon (bottom) on the a12m130 ensemble.
The x-axis is the value of $t_{\rm min}$ used in the analysis and the resulting $E_0$ for a given $t_{\rm min}$ and number of states in the analysis is plotted.
The 68\% confidence interval of the chosen fit (black) is plotted as a horizontal band to guide the eye.
}
\end{figure}

We analyze data of source-sink time separations between 0.72 and 3.6~fm for all 0.09~fm and 0.12~fm lattice spacing two-point correlation functions, and separations between 0.75 and 3.6~fm for all 0.15~fm lattice spacing two-point correlation functions.

We choose normally distributed priors for the ground-state energy and all overlap factors, and log-normal distributions for excited-state energy priors. The ground-state energy and overlap factors are motivated by the plateau values of the effective masses and scaled correlation function, and a prior width of 10\% of the central value. The excited-state energy splittings are set to the value of two pion masses with a width allowing for fluctuations down to one pion mass within one standard deviation. The excited-state overlap factors are set to zero, with a width set to the mean value of the ground-state overlap factor.

Additionally, we fit a constant to the correlation functions in Eq.~(\ref{Eq:mres_correlator}). For the 0.09 and 0.12~fm ensembles, we analyze source-sink separations that are greater than 0.72~fm. For the 0.12~fm ensemble, the minimum source-sink separation is 0.75~fm. The prior distribution for the residual chiral symmetry breaking parameter is set to the observed value per ensemble, with a width that is 100\% of the central value. The uncertainty is propagated with bootstrap resampling.

We emphasize that all input fit parameters (i.e. number of states, fit region, priors) are chosen to have the same values in physical units for all observables, to the extent that a discretized lattice allows. Additionally, we note that the extracted ground-state observables from these correlation functions are insensitive to variations around the chosen set of input fit parameters.
\figref{fig:corr_stability} shows the stability of the determination of $E_0$ for the pion and kaon on the a12m130 ensemble versus $t_{\rm min}$ and the number of states.

\begin{table*}
\caption{\label{tab:lattice_fits}
Extracted masses and decay constants from correlation functions.
An HDF5 file is provided with this publication which includes the resulting bootstrap samples which can be used to construct the correlated uncertainties.  The small parameters in this table are defined as $\e_{\pi,K} = m_{\pi,K} / (4\pi F_\pi)$, $\e_a = a / (2w_0)$.  The normalization of $\e_a$ is chosen such that as a small parameter, it spans the range of $\e_\pi^2 \lesssim \e_a^2 \lesssim \e_K^2$.
}
\begin{ruledtabular}
\begin{tabular}{lcccccccccc}
Ensemble& $am_\pi$& $am_K$& $\e_\pi^2$& $\e_K^2$& $m_\pi L$& $\e_a^2$& $\a_S$& $aF_\pi$& $aF_K$&  $F_K / F_\pi$\\
\hline
  a15m400& 0.30281(31)& 0.42723(27)& 0.09216(33)& 0.18344(62)& 4.85& 0.19378(13)& 0.58801& 0.07938(12)& 0.08504(09)& 1.0713(09)\\
  a15m350& 0.26473(30)& 0.41369(28)& 0.07505(28)& 0.18326(60)& 4.24& 0.19378(13)& 0.58801& 0.07690(11)& 0.08370(09)& 1.0884(09)\\
  a15m310& 0.23601(29)& 0.40457(25)& 0.06223(17)& 0.18285(48)& 3.78& 0.19378(13)& 0.58801& 0.07529(09)& 0.08293(09)& 1.1015(13)\\
  a15m220& 0.16533(19)& 0.38690(21)& 0.03269(11)& 0.17901(48)& 3.97& 0.19378(13)& 0.58801& 0.07277(08)& 0.08196(10)& 1.1263(15)\\
a15m135XL& 0.10293(07)& 0.38755(14)& 0.01319(05)& 0.18704(59)& 4.94& 0.19378(13)& 0.58801& 0.07131(11)& 0.08276(10)& 1.1606(18)\\
\hline
  a12m400& 0.24347(16)& 0.34341(14)& 0.08889(30)& 0.17685(63)& 5.84& 0.12376(18)& 0.53796& 0.06498(11)& 0.06979(07)& 1.0739(17)\\
  a12m350& 0.21397(20)& 0.33306(16)& 0.07307(37)& 0.17704(83)& 5.14& 0.12376(18)& 0.53796& 0.06299(14)& 0.06851(07)& 1.0876(27)\\
  a12m310& 0.18870(17)& 0.32414(21)& 0.05984(25)& 0.17657(69)& 4.53& 0.12376(18)& 0.53796& 0.06138(11)& 0.06773(10)& 1.1033(21)\\
 a12m220S& 0.13557(32)& 0.31043(22)& 0.03384(19)& 0.1774(10)\phantom{0} & 3.25& 0.12376(18)& 0.53796& 0.05865(16)& 0.06673(11)& 1.1378(27)\\
 a12m220L& 0.13402(15)& 0.31021(19)& 0.03289(15)& 0.17621(79)& 5.36& 0.12376(18)& 0.53796& 0.05881(13)& 0.06631(17)& 1.1276(29)\\
  a12m220& 0.13428(17)& 0.31001(17)& 0.03314(15)& 0.17666(81)& 4.30& 0.12376(18)& 0.53796& 0.05870(13)& 0.06636(11)& 1.1306(22)\\
  a12m130& 0.08126(16)& 0.30215(11)& 0.01287(08)& 0.17788(71)& 3.90& 0.12376(18)& 0.53796& 0.05701(11)& 0.06624(08)& 1.1619(21)\\
\hline
  a09m400& 0.18116(15)& 0.25523(13)& 0.08883(32)& 0.17633(59)& 5.80& 0.06515(08)& 0.43356& 0.04837(08)& 0.05229(07)& 1.0810(09)\\
  a09m350& 0.15785(20)& 0.24696(12)& 0.07256(32)& 0.17761(68)& 5.05& 0.06515(08)& 0.43356& 0.04663(08)& 0.05127(07)& 1.0994(10)\\
  a09m310& 0.14072(12)& 0.24106(14)& 0.06051(22)& 0.17757(59)& 4.50& 0.06515(08)& 0.43356& 0.04552(07)& 0.05053(08)& 1.1101(16)\\
  a09m220& 0.09790(06)& 0.22870(09)& 0.03307(14)& 0.18045(70)& 4.70& 0.06515(08)& 0.43356& 0.04284(08)& 0.04899(07)& 1.1434(18)\\
  a09m135& 0.05946(06)& 0.21850(08)& 0.01346(08)& 0.18175(91)& 3.81& 0.06515(08)& 0.43356& 0.04079(10)& 0.04804(06)& 1.1778(22)\\
\hline
 a06m310L& 0.09456(06)& 0.16205(07)& 0.06141(35)& 0.1803(10)\phantom{0} & 6.81& 0.02726(03)& 0.29985& 0.03037(08)& 0.03403(07)& 1.1205(17)\\
\end{tabular}
\end{ruledtabular}
\end{table*}

\section{Extrapolation Functions}
We now turn to the extrapolation/interpolation to the physical point.
We have three ensembles at the physical pion mass with relatively high statistics and precise determinations of $F_K/F_\pi$ (a15m135XL,  a12m130, and a09m135, see \tabref{tab:lattice_fits}) such that the physical quark mass extrapolation is an interpolation.
Nevertheless, we explore how the ensembles with heavier pion masses impact the physical point prediction and we use our dataset to explore uncertainty arising in the $SU(3)$-flavor chiral expansion.

We begin by assuming a canonical power-counting scheme for our MALQCD action~\cite{Bar:2005tu} in which $\mathrm{O}(m_\pi^2) \sim \mathrm{O}(m_K^2) \sim \mathrm{O}(a^2 \L_{\rm QCD}^4)$ are all treated as small scales.
For the quark mass expansion, the dimensionless small parameters $(m_P / 4\pi F)^2$ naturally emerge from $\chi$PT where $P\in\{\pi, K, \eta\}$.  For the discretization corrections, while $a\L_{\rm QCD}^2$ is often used to estimate the relative size of corrections compared to typical hadronic mass scales, it is a bit unnatural to use this in a low-energy EFT as $\L_{\rm QCD}$ is a QCD scale that does not emerge in $\chi$PT.

We chose to use another hadronic scale to form a dimensionless parameter with $a$, that being the gradient flow scale $w_0\sim0.17$~fm~\cite{Borsanyi:2012zs}.  This quantity is easy to compute, has mild quark mass dependence, and the value is roughly $w_0^{-1} \simeq 4\pi F_\pi$.
We then define the dimensionless small parameters for controlling the expansion to be
\begin{align}\label{eq:eps_ma}
    &\e_{P}^2 = \left(\frac{m_P}{\L_\chi}\right)^2 = \left(\frac{m_P}{4\pi F}\right)^2,&
    &\e_a^2   = \left(\frac{a}{2w_0}\right)^2.&
\end{align}
We leave $F$ ambiguous, as we will explore taking $F=F_\pi$, $F=F_K$ and $F^2=F_\pi F_K$ in our definition of $\L_\chi$.
This particular choice of $\e_a$ is chosen such that the range of values of this small parameter roughly corresponds to
$\e_\pi^2 \lesssim \e_a^2 \lesssim \e_K^2$ as the lattice spacing is varied, similar to the variation of $\e_\pi^2$ itself over the range of pion masses used, see \tabref{tab:lattice_fits}.
As we will discuss in \secref{sec:analysis}, this choice of $\e_a$ seems natural as determined by the size of the discretization low-energy constants (LECs) which are found in the analysis.
Note, this differs from the choice used in our analysis of $g_A$~\cite{Berkowitz:2017gql,Chang:2018uxx}.

With this power-counting scheme, the different orders in the expansion are defined to be
\begin{equation}\label{eq:power_counting}
\begin{array}{rl}
    \text{NLO:}  & \mathrm{O}(\e_P^2) \sim \mathrm{O}(\e_a^2) ,\\
    \text{\nxlo{2}:} & \mathrm{O}(\e_P^4) \sim \mathrm{O}(\e_P^2 \e_a^2) \sim \mathrm{O}(\e_a^4) ,\\
    \text{\nxlo{3}:}& \mathrm{O}(\e_P^6) \sim \mathrm{O}(\e_P^4 \e_a^2)
        \sim \mathrm{O}(\e_P^2 \e_a^4) \sim \mathrm{O}(\e_a^6) .
\end{array}
\end{equation}
Even at finite lattice spacing, $F_K=F_\pi$ in the $SU(3)$ flavor symmetry limit, also known as the $SU(3)$ vector limit $SU(3)_V$, and so there cannot be a pure $\mathrm{O}(\e_a^2)$ correction as it must accompany terms which vanish in the $SU(3)_V$ limit, such as $\e_K^2 - \e_\pi^2$.  Therefore, at NLO, there cannot be any counterterms proportional to $\e_a^2$ and the only discretization effects that can appear at NLO come through modification of the various meson masses that appear in the MA EFT.

We find that the precision of our results requires including terms higher than NLO, and we have to work at a hybrid \nxlo{3} order to obtain a good description of our data.
Therefore, we will begin with a discussion of the full \nxlo{2} $\chi$PT theory expression for $F_K/F_\pi$ in the continuum limit~\cite{Amoros:1999dp,Ananthanarayan:2017yhz,Ananthanarayan:2018irl,Ananthanarayan:2017qmx}.

\subsection{\nxlo{2} $\chi$PT \label{sec:n2lo_xpt}}
The analytic expression for $F_K/F_\pi$ up to \nxlo{2} is~\cite{Ananthanarayan:2017qmx}
\begin{align}\label{eq:FKFpi_nnlo_Fpi}
    \frac{F_K}{F_\pi} &= 1
    +\frac{5}{8}\ell_\pi - \frac{1}{4}\ell_k -\frac{3}{8}\ell_\eta
    +4 \Lbar_5 (\e_K^2 - \e_\pi^2)
\nonumber\\&\phantom{=}
    +\e_K^4 F_F\left(\frac{m_\pi^2}{m_K^2}\right)
    +\hat{K}_1^r \l_\pi^2
    +\hat{K}_2^r \l_\pi \l_K
\nonumber\\&\phantom{=}
    +\hat{K}_3^r \l_\pi \l_\eta
    +\hat{K}_4^r \l_K^2
    +\hat{K}_5^r \l_K \l_\eta
    +\hat{K}_6^r \l_\eta^2
\nonumber\\&\phantom{=}
    +\hat{C}_1^r \l_\pi
    +\hat{C}_2^r \l_K
    +\hat{C}_3^r \l_\eta
    +\hat{C}_4^r\, .
\end{align}
The first line is the LO (1) plus NLO terms, while the next three lines are the \nxlo{2} terms.
Several nonunique choices were made to arrive at this formula.  Prior to discussing these choices, we first define the parameters appearing in \eqnref{eq:FKFpi_nnlo_Fpi}.
First, the small parameters were all defined as
\begin{equation}
    \e_P^2 = \left(\frac{m_P}{4\pi F_\pi(m_P)}\right)^2\, ,
\end{equation}
where $F_\pi(m_P)$ is the ``on-shell'' pion decay constant at the masses $m_P$.
The quantities $\ell_P$ are defined as
\begin{equation}\label{eq:ell_P}
\ell_P = \e_P^2 \ln\left(\frac{m_P^2}{\mu^2}\right)\, ,
\end{equation}
where $\mu$ is a renormalization scale.
The coefficient $\bar{L}_5 = (4\pi)^2 L_5^r(\mu)$ is one of the regulated Gasser-Leutwyler LECs~\cite{Gasser:1984gg} which has a renormalization scale dependence that exactly cancels against the dependence arising from the logarithms appearing at the same order.  In the following, we define all of the Gasser-Leutwyler LECs with the extra $(4\pi)^2$ for convenience:
\begin{equation}
    \bar{L}_i \equiv (4\pi)^2 L_i^r(\mu)\, .
\end{equation}
The $\eta$ mass has been defined through the Gell-Mann--Okubo (GMO) relation
\begin{equation}\label{eq:gmo}
    m_\eta^2 \equiv \frac{4}{3}m_K^2 - \frac{1}{3}m_\pi^2\, ,
\end{equation}
with the corrections to this relation being propagated into \eqnref{eq:FKFpi_nnlo_Fpi} for consistency at \nxlo{2}.
The logs are
\begin{equation}
    \l_P \equiv \ln \left( \frac{m_P^2}{\mu^2} \right)\, .
\end{equation}
The $\ln^2$ terms are encapsulated in the $F_F(x)$ function, defined in Eqs.~(8-17) of Ref.~\cite{Ananthanarayan:2017qmx},%
\footnote{They also provide an approximate formula which is easy to implement, but our numerical results are sufficiently precise to require the exact expression. To implement this function in our analysis, we have modified an interface \texttt{C++} file provided by J.~Bijnens to \texttt{CHIRON}~\cite{Bijnens:2014gsa}, the package for two-loop $\chi$PT functions. We have provided a \texttt{Python} interface as well so that the function can be called from our main analysis code, which is provided with this article.}
and the $\hat{K}_i^r \l_P \l_{P^\prime}$ terms whose coefficients are given by%
\footnote{We correct a typographical error in the $K_6^r$ term presented in Ref.~\cite{Ananthanarayan:2017qmx}: a simple power-counting reveals the $\xi_K^2=\e_K^4$ accompanying this term should not be there.}
\begin{align}
&\hat{K}_1^r = \phantom{-}\frac{11}{24}\e_\pi^2 \e_K^2  -\frac{131}{192}\e_\pi^4 ,&
&\hat{K}_2^r = -\frac{41}{96}\e_\pi^2 \e_K^2 -\frac{3}{32}\e_\pi^4 ,&
\nonumber\\
&\hat{K}_3^r = \phantom{-}\frac{13}{24}\e_\pi^2 \e_K^2  +\frac{59}{96}\e_\pi^4,&
&\hat{K}_4^r = \frac{17}{36}\e_K^4           +\frac{7}{144}\e_\pi^2\e_K^2,&
\nonumber\\
&\hat{K}_5^r = -\frac{163}{144}\e_K^4 -\frac{67}{288}\e_\pi^2\e_K^2 +\frac{3}{32}\e_\pi^4,\hspace{-60pt}&
\nonumber\\
&\hat{K}_6^r = \phantom{-}\frac{241}{288}\e_K^4 -\frac{13}{72}\e_\pi^2 \e_K^2 -\frac{61}{192}\e_\pi^4.\hspace{-60pt}&
\end{align}
The single log coefficients $\hat{C}^r_{1-3}$ are combinations of the NLO Gasser-Leutwyler coefficients
\begin{equation}\label{eq:C_i_terms}
\hat{C}_i^r = c_i^{\pi\pi} \e_\pi^4 + c_i^{K\pi} \e_K^2 \e_\pi^2 + c_i^{KK} \e_K^4\, ,
\end{equation}
where
\begin{align}\label{eq:c_i_coefficients}
c_1^{\pi\pi} &= -\frac{113}{72} -2(2\Lbar_1 +5\Lbar_2) -\frac{13}{2}\Lbar_3 +\frac{21}{2}\Lbar_5\, ,
\nonumber\\
c_1^{K\pi}   &= -\frac{7}{9} -\frac{11}{2}\Lbar_5\, ,
\nonumber\\
c_1^{KK}     &= \phantom{-}c_2^{\pi\pi} = 0\, ,
\nonumber\\
c_2^{K\pi}   &= \phantom{-}\frac{209}{144} +3\Lbar_5\, ,
\nonumber\\
c_2^{KK}     &= \phantom{-}\frac{53}{96} +2(2\Lbar_1 +5\Lbar_2) +5\Lbar_3 - 5\Lbar_5\, ,
\nonumber\\
c_3^{\pi\pi} &= \phantom{-}\frac{19}{288} + \frac{1}{6}\Lbar_3 +\frac{11}{6}\Lbar_5 -8(2\Lbar_7 +\Lbar_8)\, ,
\nonumber\\
c_3^{K\pi}   &= -\frac{4}{9} -\frac{4}{3}\Lbar_3 -\frac{25}{6}\Lbar_5 +16(2\Lbar_7 +\Lbar_8)\, ,
\nonumber\\
c_3^{KK}     &= \phantom{-}\frac{13}{18} +\frac{8}{3}\Lbar_3 -\frac{2}{3}\Lbar_5 -8(2\Lbar_7 +\Lbar_8)\, .
\end{align}
Finally, $\hat{C}_4^r$ is a combination of these $L_i^r$ coefficients as well as counterterms appearing at \nxlo{2}.
At \nxlo{2}, only two counterterm structures can appear due to the $SU(3)_V$ constraints:
\begin{equation}\label{eq:n2lo_xpt_ct}
\hat{C}_4^{r} =
    (\e_K^2 -\e_\pi^2) \left[ (A_K^4+L_K^4) \e_K^2 + (A_\pi^4+L_\pi^4) \e_\pi^2 \right]
\end{equation}
which are linear combinations of the \nxlo{2} counterterms
\begin{align}
A_K^4 &= 16 (4\pi)^4 (C_{14}^r +C_{15}^r)\, ,
\nonumber\\
A_\pi^4 &= \phantom{1}8 (4\pi)^4 (C_{15}^r + 2C_{17}^r)\, ,
\end{align}
and contributions from the Gasser-Leutwyler LECs (Eq.~(7) of Ref.~\cite{Ananthanarayan:2017qmx})
\begin{align}\label{eq:nnlo_GL}
L_K^4 &= 8 \Lbar_5( 8(\Lbar_4-2\Lbar_6) +3\Lbar_5 -8\Lbar_8)
\nonumber\\&\phantom{=}
    -2\Lbar_1 -\Lbar_2 -\frac{1}{18}\Lbar_3 +\frac{4}{3}\Lbar_5 -8(2\Lbar_7 +\Lbar_8)\, ,
\nonumber\\
L_\pi^4 &= 8\Lbar_5 (4(\Lbar_4-2\Lbar_6) +5\Lbar_5 -8\Lbar_8)
\nonumber\\&\phantom{=}
    -2\Lbar_1 -\Lbar_2 -\frac{5}{18}\Lbar_3 -\frac{4}{3}\Lbar_5 +8(2\Lbar_7 +\Lbar_8)\, .
\end{align}

There were several nonunique choices that went into the determination of \eqnref{eq:FKFpi_nnlo_Fpi}.
When working with the full \nxlo{2} $\chi$PT expression, the different choices one can make result in different \nxlo{3} or higher corrections and exploring these different choices in the analysis will expose sensitivity to higher-order contributions that are not explicitly included.
The first choice we discuss is the Taylor expansion of the ratio of $F_K / F_\pi$
\begin{align}\label{eq:fkfpi_fully_expanded}
\frac{F_K}{F_\pi} &= \frac{1 + \d F_K^{\rm NLO} + \d F_K^\text{\nxlo{2}} +\cdots}
    {1 + \d F_\pi^{\rm NLO} + \d F_\pi^\text{\nxlo{2}}+\cdots}
\nonumber\\&=
    1 + \d F_{K-\pi}^{\rm NLO} +\d F_{K-\pi}^\text{\nxlo{2}}
    \nonumber\\&\phantom{=}\,
    + \left(\d F_\pi^{\rm NLO}\right)^2
    -\d F_\pi^{\rm NLO} \d F_K^{\rm NLO}
    +\cdots\, ,
\end{align}
where the $\cdots$ represent higher-order terms in the expansion and $\d F_{K-\pi}^\text{\nxlo{2}} = \d F_{K}^\text{\nxlo{2}} -\d F_{\pi}^\text{\nxlo{2}}$.
\eqnref{eq:FKFpi_nnlo_Fpi} has been derived from this standard Taylor-expanded form with the choices mentioned above: the use of the on-shell renormalized value of $F\rightarrow F_\pi$ and the definition of the $\eta$ mass through the GMO relation.
The NLO expressions are the standard ones~\cite{Gasser:1984gg}
\begin{align}\label{eq:F_nlo}
\d F_K^{\rm NLO} &= -\frac{3}{8}\ell_\pi -\frac{3}{4}\ell_K -\frac{3}{8}\ell_\eta
    +4\Lbar_5 \e_K^2
\nonumber\\&\phantom{=}
    +4\bar{L}_4 (\e_\pi^2 + 2\e_K^2)\, ,
\nonumber\\
\d F_\pi^{\rm NLO} &= -\ell_\pi -\frac{1}{2}\ell_K
    +4\Lbar_5 \e_\pi^2
    +4\bar{L}_4 (\e_\pi^2 + 2\e_K^2)\, ,
\nonumber\\
\d F_{K-\pi}^{\rm NLO} &= \frac{5}{8} \ell_\pi
    -\frac{1}{4} \ell_K -\frac{3}{8} \ell_\eta
    +4 \bar{L}_5 (\e_K^2 - \e_\pi^2)\, .
\end{align}
The $\d F_P^\text{\nxlo{2}}$ terms have been determined in Ref.~\cite{Amoros:1999dp} and cast into analytic forms in Refs.~\cite{Ananthanarayan:2017yhz,Ananthanarayan:2018irl}.
The NLO terms are of $\mathrm{O}(20\%)$ and so Taylor expanding this ratio leads to sizable corrections from the $\left(\d F_\pi^{\rm NLO}\right)^2 -\d F_\pi^{\rm NLO} \d F_K^{\rm NLO}$ contributions.
Utilizing the full ratio expression could in principle lead to a noticeable difference in the analysis (a different determination of the values of the LECs for example).
Rather than implementing the full $\d F_P^\text{\nxlo{2}}$ expressions for kaon and pion, we explore this convergence by instead just resumming the NLO terms which will dominate the potential differences in higher-order corrections.
A consistent expression at \nxlo{2} is
\begin{equation}\label{eq:fkfpi_nlo_ratio}
\frac{F_K}{F_\pi}[\text{\eqref{eq:FKFpi_nnlo_Fpi}}]
    =
    \frac{1 + \d F_K^{\rm NLO}}{1 +\d F_\pi^{\rm NLO}} + \d F^\text{\nxlo{2}}
    + \d_{\rm ratio}^\text{\nxlo{2}}
    +\cdots\, ,
\end{equation}
where $\d F^\text{\nxlo{2}}$ is the full \nxlo{2} expression in \eqnref{eq:fkfpi_fully_expanded}
\begin{equation}
\d F^\text{\nxlo{2}} = \d F_{K-\pi}^\text{\nxlo{2}}
    + \left(\d F_\pi^{\rm NLO}\right)^2
    -\d F_\pi^{\rm NLO} \d F_K^{\rm NLO}\, ,
\end{equation}
and the ratio correction is given by
\begin{equation}
\d_{\rm ratio}^\text{\nxlo{2}}
    =
    \d F_\pi^{\rm NLO} \d F_K^{\rm NLO}
    -\left(\d F_\pi^{\rm NLO}\right)^2\, .
\end{equation}

Another choice we explore is the use of $F\rightarrow F_\pi$ in the definition of the small parameters.
Such a choice is very convenient as it allows one to express the small parameters entirely in terms of observables one can determine in the lattice calculation (unlike the bare parameters, such as $\chi$PT's $F_0$ and $Bm_q$, which must be determined through extrapolation analysis).
Equally valid, one could have chosen $F\rightarrow F_K$ or $F^2 \rightarrow F_\pi F_K$.
Each choice induces explicit corrections one must account for at \nxlo{2} to have a consistent expression at this order.  The NLO corrections in \eqnref{eq:FKFpi_nnlo_Fpi} are proportional to
\begin{align}
\e_P^2 &= \frac{m_P^2}{(4\pi F_\pi)^2}
\nonumber\\&
    = \frac{m_P^2}{(4\pi)^2 F_\pi F_K} \frac{F_K}{F_\pi}
    = \frac{m_P^2}{(4\pi)^2 F_\pi F_K} \left(1 + \d F_{K-\pi}^{\rm NLO}\right)
\nonumber\\&
    = \frac{m_P^2}{(4\pi F_K)^2} \frac{F_K^2}{F_\pi^2}
    = \frac{m_P^2}{(4\pi F_K)^2} \left(1 + 2\d F_{K-\pi}^{\rm NLO}\right)\, ,
\end{align}
plus higher-order corrections.

Related to this choice, \eqnref{eq:FKFpi_nnlo_Fpi} is implicitly defined at the standard renormalization scale~\cite{Ananthanarayan:2017qmx}
\begin{equation}\label{eq:mu_rho}
    \mu_0^\rho = m_\rho = 770 \textrm{ MeV}\, .
\end{equation}
While $F_K/F_\pi$ of course does not depend upon this choice, the numerical values of the LECs do.  Further, a scale setting would be required to utilize this or any fixed value of $\mu$.
Instead, as was first advocated in Ref.~\cite{Beane:2006kx} to the best of our knowledge,
it is more convenient to set the renormalization scale on each ensemble with a lattice quantity.
For example, Ref.~\cite{Beane:2006kx} used $\mu=f_\pi^{\rm latt} = \sqrt{2}F_\pi^{\rm latt}$ where $F_\pi^{\rm latt}$ is the lattice-determined value of the pion decay constant on a given ensemble.
The advantage of this choice is that the entire extrapolation can be expressed in terms of ratios of lattice quantities such that a scale setting is not required to perform the extrapolation to the physical point.

At NLO in the expansion, one is free to make this choice as the corrections appear at \nxlo{2}.
In the present work, we must account for these corrections for a consistent expression at this order, which is still defined at a fixed renormalization scale.
To understand these corrections, we take as our fixed scale
\begin{equation}\label{eq:mu_0}
    \mu_0 = 4\pi F_0\, ,
\end{equation}
where $F_0$ is the decay constant in the $SU(3)$ chiral limit.
Define $\mu_\pi = 4\pi F_\pi$ and consider the NLO expression
\begin{align}
\frac{F_K}{F_\pi} &= 1
    +\frac{5}{8}\ell_\pi^{\mu_0}
    -\frac{1}{4}\ell_K^{\mu_0}
    -\frac{3}{8}\ell_\eta^{\mu_0}
    +4(\e_K^2 - \e_\pi^2) \Lbar_5(\mu_0)
\nonumber\\&=
    +\frac{5}{8}\e_\pi^2 \ln \left(\e_\pi^2 \frac{\mu_\pi^2}{\mu_0^2}\right)
    -\frac{1}{4}\e_K^2 \ln \left( \e_K^2 \frac{\mu_\pi^2}{\mu_0^2}\right)
    \nonumber\\&\phantom{=}\,
    -\frac{3}{8}\e_\eta^2 \ln \left( \e_\eta^2 \frac{\mu_\pi^2}{\mu_0^2}\right)
    +4(\e_K^2 - \e_\pi^2) \Lbar_5(\mu_0)
    \nonumber\\&=
    1
    +\frac{5}{8}\ell_\pi^{\mu_\pi}
    -\frac{1}{4}\ell_K^{\mu_\pi}
    -\frac{3}{8}\ell_\eta^{\mu_\pi}
    +4(\e_K^2 - \e_\pi^2) \Lbar_5(\mu_0)
    \nonumber\\&\phantom{=}\,
    +\ln \left(\frac{\mu_\pi^2}{\mu_0^2} \right)\left[
    \frac{5}{8}\e_\pi^2 -\frac{1}{4}\e_K^2 -\frac{3}{8}\e_\eta^2
    \right]\, ,
\end{align}
where we have introduced the notation
\begin{equation}
    \ell_P^\mu = \e_P^2 \ln\left(\frac{\e_P^2}{\mu^2}\right)\, .
\end{equation}
If we chose the renormalization scale $\mu_\pi$ and add the second term of the last equality, then this expression is equivalent to working with the scale $\mu_0$ through \nxlo{2}.
The convenience of this choice becomes clear as $\mu_\pi/\mu_0$ has a familiar expansion
\begin{equation}
    \frac{\mu_\pi}{\mu_0} = 1 + \d F_\pi^{\rm NLO} +\cdots\, .
\end{equation}
Using the GMO relation \eqnref{eq:gmo} and expanding $\ln(1+x)$ for small $x$, this expression becomes
\begin{align}
\frac{F_K}{F_\pi} &= 1
    +\frac{5}{8}\ell_\pi^{\mu_\pi}
    -\frac{1}{4}\ell_K^{\mu_\pi}
    -\frac{3}{8}\ell_\eta^{\mu_\pi}
    +4(\e_K^2 - \e_\pi^2) \Lbar_5(\mu_0)
\nonumber\\&\phantom{=}
    -\frac{3}{2} (\e_K^2 -\e_\pi^2) \d F_\pi^{\rm NLO}\, .
\end{align}
Similar expressions can be derived for the choices $\mu_{\pi K}=4\pi F_{\pi K}$ (where $F_{\pi K}=\sqrt{F_\pi F_K}$) and $\mu_K = 4\pi F_K$ which are made more convenient if one also makes the replacements $F^2_\pi \rightarrow \{F_\pi F_K, F_K^2\}$ in the definition of the small parameters plus the corresponding \nxlo{2} corrections that accompany these choices.

If we temporarily expose the implicit dependence of the expression for $F_K/F_\pi$ on the choices of $F$ and $\mu$, such that \eqnref{eq:FKFpi_nnlo_Fpi} is defined as
\begin{equation}
    \frac{F_K}{F_\pi}[\text{\eqref{eq:FKFpi_nnlo_Fpi}}]
    = \frac{F_K}{F_\pi}(F_\pi,\mu_0^\rho)\, ,
\end{equation}
then the following expressions are all equivalent at \nxlo{2}
\begin{align}\label{eq:FKFpi_F_variation}
    \frac{F_K}{F_\pi}(F_\pi, \mu_0) &=
    \frac{F_K}{F_\pi}(F_K, \mu_{K})
    + \d_{F_{K}}^\text{\nxlo{2}}
    \nonumber\\&=
    \frac{F_K}{F_\pi}(F_{\pi K}, \mu_{\pi K})
    + \d_{F_{\pi K}}^\text{\nxlo{2}}
    \nonumber\\&=
    \frac{F_K}{F_\pi}(F_{\pi}, \mu_{\pi})
    + \d_{F_{\pi}}^\text{\nxlo{2}}\, ,
\end{align}
where
\begin{align}\label{eq:n2lo_xpt_scale}
\d_{F_{K}}^\text{\nxlo{2}} &=
    -\frac{3}{2}(\e_K^2 - \e_\pi^2) \d F_K^{\rm NLO}
    +2(\d F_{K-\pi}^{\rm NLO})^2
\nonumber\\
\d_{F_{\pi K}}^\text{\nxlo{2}} &=
    -\frac{3}{4}(\e_K^2 - \e_\pi^2) (\d F_K^{\rm NLO} + \d F_\pi^{\rm NLO})
    +(\d F_{K-\pi}^{\rm NLO})^2
\nonumber\\
\d_{F_{\pi}}^\text{\nxlo{2}} &=
    -\frac{3}{2}(\e_K^2 - \e_\pi^2) \d F_\pi^{\rm NLO}
\end{align}
and the LECs in these expressions are related to those at the standard scale by evolving them from $\mu_0^\rho \rightarrow \mu_0$ with their known scale dependence~\cite{Gasser:1984gg}.
Implicit in these expressions is the normalization of the small parameters
\begin{align}
\e_P^2 = \left\{\begin{array}{cl}
    \frac{m_P^2}{(4\pi F_\pi)^2}, & \textrm{for } F \rightarrow F_\pi\\
    \frac{m_P^2}{(4\pi)^2 F_\pi F_K}, & \textrm{for } F \rightarrow \sqrt{F_\pi F_K}\\
    \frac{m_P^2}{(4\pi F_K)^2}, & \textrm{for } F \rightarrow F_K\\
\end{array}\right. \, .
\end{align}

We have described several choices one can make in parametrizing the $\chi$PT formula for $F_K/F_\pi$.
The key point is that if the underlying chiral expansion is well behaved, the formulas resulting from each choice are all equivalent through \nxlo{2} in the $SU(3)$ chiral expansion, with differences only appearing at \nxlo{3} and beyond.  Therefore, by studying the variance in the extrapolated answer upon these choices, one is assessing some of the uncertainty arising from the truncation of the chiral extrapolation formula.

\subsection{Discretization corrections \label{sec:discretization}}

\begin{table*}
\caption{\label{tab:mixed_mesons}
Extracted masses of the mixed MDWF-HISQ mesons.
We use the notation from Ref.~\cite{Chen:2001yi} in which $m_\pi$ and $m_K$ denote the masses of the valence pion and kaon and $j$ and $r$ denote the light and strange flavors of the sea quarks while $u$ and $s$ denote the light and strange flavors of the valence quarks.  Since we have tuned the valence MDWF pion and $\bar{s}s$ mesons to have the same mass as the HISQ sea pion and $\bar{s}s$ mesons within a few percent, the quantities $m_{ju}^2 - m_\pi^2$ and other splittings provide an estimate of the additive mixed-meson mass splitting due to discretization effects, $a^2 \D_{\rm Mix}$~\cite{Bar:2005tu} and additional additive corrections~\cite{Chen:2009su}.
At LO in MA EFT, these splittings are predicted to be quark mass independent, which we find to be approximately true, with a notable decrease in the splitting as the valence quark mass is increased as first observed in Ref.~\cite{Orginos:2007tw} as well as a milder decrease as the seq-quark mass is increased.
}
\begin{ruledtabular}
\begin{tabular}{lllllllllll}
    Ensemble& $am_{ju}$& $am_{js}$& $am_{ru}$& $am_{rs}$& $am_{ss}$& $w_0^2 \D_{\rm Mix, ju}^2$& $w_0^2\D_{\rm Mix, js}^2$& $w_0^2\D_{\rm Mix, ru}^2$& $w_0^2\D_{\rm Mix, rs}^2$& $w_0^2 a^2 \D_{\rm I}$\\
    \hline
      a15m400& 0.3597(17)& 0.4586(24)& 0.4717(19)& 0.5537(11)& 0.5219(02)& 0.0486(15)& 0.0359(28)& 0.0516(23)& 0.0440(16)& 0.112(14)\\
      a15m350& 0.3308(23)& 0.4463(14)& 0.4598(16)& 0.5526(10)& 0.5201(02)& 0.0508(20)& 0.0362(17)& 0.0519(19)& 0.0451(15)& 0.112(14)\\
      a15m310& 0.3060(17)& 0.4345(16)& 0.4508(14)& 0.5490(12)& 0.5188(02)& 0.0489(13)& 0.0324(18)& 0.0511(17)& 0.0416(16)& 0.112(14)\\
      a15m220& 0.2564(27)& 0.4115(17)& 0.4320(29)& 0.5420(08)& 0.5150(01)& 0.0495(18)& 0.0253(19)& 0.0476(33)& 0.0368(11)& 0.112(14)\\
    a15m135XL& 0.232(13) & 0.4058(56)& 0.4337(84)& 0.5560(31)& 0.5257(02)& 0.0559(75)& 0.0187(59)& 0.0489(94)& 0.0423(45)& 0.112(14)\\
    \hline
      a12m400& 0.2678(06)& 0.3560(08)& 0.3624(07)& 0.4333(06)& 0.4207(01)& 0.0251(07)& 0.0177(12)& 0.0271(10)& 0.0217(11)& 0.063(05)\\
      a12m350& 0.2303(08)& 0.3446(07)& 0.3454(10)& 0.4322(05)& 0.4197(01)& 0.0147(07)& 0.0158(10)& 0.0168(15)& 0.0214(09)& 0.063(05)\\
      a12m310& 0.2189(09)& 0.3344(10)& 0.3439(09)& 0.4305(05)& 0.4180(02)& 0.0248(08)& 0.0136(14)& 0.0266(13)& 0.0213(09)& 0.063(05)\\
     a12m220S& 0.1774(14)& 0.3187(12)& 0.3323(17)& 0.4286(10)& 0.4158(02)& 0.0264(10)& 0.0105(16)& 0.0283(24)& 0.0219(18)& 0.063(05)\\
     a12m220L& 0.1774(14)& 0.3187(12)& 0.3323(17)& 0.4286(10)& 0.4156(02)& 0.0273(10)& 0.0107(16)& 0.0286(23)& 0.0222(18)& 0.063(05)\\
      a12m220& 0.1774(14)& 0.3187(12)& 0.3323(17)& 0.4286(10)& 0.4154(01)& 0.0272(10)& 0.0110(16)& 0.0289(23)& 0.0225(18)& 0.063(05)\\
      a12m130& 0.1491(20)& 0.3080(15)& 0.3240(26)& 0.4271(08)& 0.4141(01)& 0.0316(12)& 0.0073(19)& 0.0276(34)& 0.0220(14)& 0.063(05)\\
    \hline
      a09m400& 0.1878(05)& 0.2581(06)& 0.2607(06)& 0.3162(05)& 0.3133(01)& 0.0094(07)& 0.0056(12)& 0.0109(11)& 0.0071(12)& 0.020(02)\\
      a09m350& 0.1654(06)& 0.2498(05)& 0.2526(06)& 0.3159(04)& 0.3124(01)& 0.0093(07)& 0.0054(10)& 0.0108(12)& 0.0083(11)& 0.020(02)\\
      a09m310& 0.1485(06)& 0.2428(05)& 0.2472(10)& 0.3150(04)& 0.3117(01)& 0.0086(07)& 0.0032(10)& 0.0114(20)& 0.0080(09)& 0.020(02)\\
      a09m220& 0.1090(09)& 0.2303(06)& 0.2334(07)& 0.3115(03)& 0.3094(01)& 0.0088(07)& 0.0028(10)& 0.0083(12)& 0.0051(08)& 0.020(02)\\
      a09m135& 0.0786(15)& 0.2187(11)& 0.2270(15)& 0.3079(05)& 0.3027(07)& 0.0102(09)& 0.0004(19)& 0.0146(26)& 0.0123(19)& 0.020(02)\\
    \hline
     a06m310L& 0.0957(08)& 0.1619(11)& 0.1619(12)& 0.2103(10)& 0.2098(01)& 0.0020(14)& \hspace{-0.25em}-0.0004(34)& \hspace{-0.25em}-0.0004(34)& 0.0020(40)& 0.004(00)\\
\end{tabular}
\end{ruledtabular}
\end{table*}

We now turn to the discretization corrections.
We explore two parametrizations for incorporating the corrections arising at finite lattice spacing.
The simplest approach is to use the continuum extrapolation formula and enhance it by adding contributions from all allowed powers of $\e_P^2$ and $\e_a^2$ to a given order in the expansion.  This is very similar to including only the contributions from local counterterms that appear at the given order.
At \nxlo{2}, the set of discretization corrections is given by%
\footnote{\label{fn:symanzik}
One can use the renormalization-group to resum corrections from radiative gluons that modify the leading asymptotic scaling behavior~\cite{Balog:2009yj,Balog:2009np}.
For actions without dimension-5 operators in the Symanzik EFT, these resummed scaling violations are known to be proportional to
\begin{equation}\label{eq:improved_symanzik}
\d_{a}^{\rm Symanzik} = c_2^{\mathrm{O}} a^2 \a_S^{n+\hat{\g}_1}\, ,
\end{equation}
where $c_2^{\mathrm{O}}$ is an LEC for operator $\mathrm{O}$ and whose value depends upon the lattice action.  The power $n=0$ for unimproved actions (such as our MDWF valence action), $n=1$ for tree-level improved actions (such as the HISQ action) and $n=2$ for one-loop improved actions.  The anomalous dimension $\hat{\g}_1$ can be determined in the asymptotic scaling regime which has been recently done for Yang-Mills and Wilson fermion actions with $\hat{\g}_1^{\rm YM} = 7/11$~\cite{Husung:2019ytz}.
This anomalous dimension is not known for our action.  In principle, one could perform a fit where instead of treating the $a^2$ and $\a_S a^2$ terms with different LECs, one could combine them as in \eqnref{eq:improved_symanzik} and try and fit both $c_2^{\mathrm{O}}$ and $\hat{\g}_1$.  We leave this to future studies and in this work, we use \eqnref{eq:n2lo_a}.}
\begin{equation}\label{eq:n2lo_a}
\d_a^\text{\nxlo{2}} =
    A_s^4 \e_a^2 (\e_K^2 - \e_\pi^2)
    +A_{\a_S}^4 \a_S \e_a^2 (\e_K^2 - \e_\pi^2)\, ,
\end{equation}
where $A_s^4$ and $A_{\a_S}^4$ are the LECs and $\a_S$ is the running QCD coupling that emerges in the Symanzik expansion of the lattice expansion through loop corrections.
Each contribution at this order must vanish in the $SU(3)_V$ limit because the discretization corrections are flavor blind and so we have the limiting constraint
\begin{equation}
    \lim_{m_{l}\rightarrow m_s} \frac{F_K}{F_\pi} = 1\, ,
\end{equation}
at any lattice spacing.

From a purist EFT perspective, we should instead utilize the MA EFT expression.
Unfortunately, the MA EFT expression is only known at NLO~\cite{Bar:2005tu} and our results require higher orders to provide good fits.
Nevertheless, we can explore the utility of the MA EFT by replacing the NLO $\chi$PT expression with the NLO MA EFT expression and using the continuum expression enhanced with the local discretization corrections at higher orders, \eqnref{eq:n2lo_a}.

Using the parametrization of the hairpin contributions from Ref.~\cite{Chen:2006wf}, the NLO MA EFT expressions are
\begin{align}
\d F_\pi^{\rm MA} &= -\ell_{ju} -\frac{\ell_{ru}}{2}
    +4\bar{L}_5 \e_\pi^2 +4\bar{L}_4 (\e_\pi^2+2\e_K^2)
    +\e_a^2 \bar{L}_a\, ,
\nonumber\\
\d F_K^{\rm MA} &=
    -\frac{\ell_{ju}}{2}
    +\frac{\ell_\pi}{8}
    -\frac{\ell_{ru}}{4}
    -\frac{\ell_{js}}{2}
    -\frac{\ell_{rs}}{4}
    +\frac{\ell_{ss}}{4}
    -\frac{3\ell_{X}}{8}
\nonumber\\&\phantom{=}
    +4\bar{L}_5 \e_K^2 +4\bar{L}_4 (\e_\pi^2+2\e_K^2)
    +\e_a^2 \bar{L}_a
\nonumber\\&\phantom{=}
    -\frac{\d_{ju}^2}{8} \left(
        d\ell_\pi
        -2\mc{K}_{\pi X}\right)
    -\frac{\d_{ju}^4}{24} \mc{K}_{\pi X}^{(2,1)}
\nonumber\\&\phantom{=}
    +\frac{\d_{rs}^2}{4} \left( \mc{K}_{ss X}
        -\frac{2}{3}(\e_K^2-\e_\pi^2) \mc{K}_{ss X}^{(2,1)} \right)
\nonumber\\&\phantom{=}
    +\frac{\d_{ju}^2 \d_{rs}^2}{12} \left(
        \mc{K}_{ss X}^{(2,1)}
        -2\mc{K}_{\pi ss X}
    \right)\, .
\end{align}
In these expressions, we use the partially quenched flavor notation~\cite{Chen:2001yi} in which
\begin{align}
\begin{array}{rl}
\pi:& \textrm{valence-valence pion}\\
K:  & \textrm{valence-valence kaon}\\
u:  & \textrm{valence light flavor quark}\\
j:  & \textrm{sea light flavor quark}\\
s:  & \textrm{valence strange flavor quark}\\
r:  & \textrm{sea strange flavor quark}\\
X:  & \textrm{sea-sea eta meson}
\end{array}\, ,
\end{align}
and so, for example
\begin{equation}
    \ell_{ju} = \frac{m_{ju}^2}{(4\pi F_\pi)^2} \ln \left( \frac{m_{ju}^2}{(4\pi F_\pi)^2} \right)\, ,
\end{equation}
where $m_{ju}$ is the mass of a mixed valence-sea pion.
The partial quenching parameters $\d_{ju^2}$ and $\d_{rs}^2$ provide a measure of the unitarity violation in the theory.  For our MDWF on HISQ action, at LO in MA EFT, they are given by the splitting in the quark masses plus a discretization correction arising from the taste-identity splitting
\begin{align}
\d_{ju}^2 &= \frac{2B_0 (m_j - m_u) + a^2 \D_{\rm I}}{(4\pi F_\pi)^2}\, ,
\nonumber\\
\d_{rs}^2 &= \frac{2B_0 (m_r - m_s) + a^2 \D_{\rm I}}{(4\pi F_\pi)^2}\, .
\end{align}
For the tuning we have done, setting the valence-valence pion mass equal to the taste-5 sea-sea pion mass, these parameters are given just by the discretization terms as $m_u=m_j$ and $m_s=m_r$ within 1\%-2\%.
The sea-sea eta mass in this tuning is given at LO in MA EFT as
\begin{equation}
    m_X^2 = \frac{4}{3}m_K^2 -\frac{1}{3}m_\pi^2 + a^2 \D_{\rm I}\, .
\end{equation}
These parameters, and the corresponding meson masses are provided in \tabref{tab:mixed_mesons}.
The expressions for $d\ell_\pi$, $\mc{K}_{\phi_1 \phi_2}$, $\mc{K}_{\phi_1 \phi_2}^{(2,1)}$ and $\mc{K}_{\phi_1 \phi_2 \phi_3}$ are provided in Appendix \ref{sec:ma_expressions}.

At NLO in the MA EFT, the LECs which contribute to $\d F_{K}$ and $\d F_\pi$ are the same as in the continuum, $L_4$ and $L_5$, plus a discretization LEC which we have denoted $\bar{L}_a$.  Just like the $L_4$ contribution, the contribution from $\bar{L}_a$ exactly cancels in $\d F_K - \d F_\pi$.
At \nxlo{2}, beyond the continuum counterterm contributions, \eqnref{eq:n2lo_xpt_ct}, there are the two additional LECs contributions, \eqnref{eq:n2lo_a}.

\subsection{Finite volume corrections \label{sec:FV}}

\begin{table}
\caption{\label{tab:cn_weigths}
Multiplicity factors for the finite volume corrections of the first 10 vector lengths, $|\mathbf{n}|$.
}
\begin{ruledtabular}
\begin{tabular}{c|cccccccccc}
$|\mathbf{n}|$& 1 & $\sqrt{2}$& $\sqrt{3}$& $\sqrt{4}$& $\sqrt{5}$& $\sqrt{6}$& $\sqrt{7}$& $\sqrt{8}$& $\sqrt{9}$& $\sqrt{10}$\\
\hline
$c_n$& 6&12& 8& 6& 24& 24& 0& 12& 30& 24
\end{tabular}
\end{ruledtabular}
\end{table}

We now discuss the corrections arising from the finite spatial volume.
The leading finite volume (FV) corrections arise from the tadpole integrals which arise at NLO in both the $\chi$PT and MA expressions.
The well-known modification to the integral can be expressed as~\cite{Gasser:1986vb,Colangelo:2004xr,Colangelo:2005gd}
\begin{equation}\label{eq:ell_P_FV}
    \ell_P^{\mu_\pi, {\rm FV}} = \ell_P^{\mu_\pi}
    + 4 \e_P^2 \sum_{|\mathbf{n}|\neq0} \frac{c_n}{m_P L|\mathbf{n}|} K_1(m_P L|\mathbf{n}|)\, ,
\end{equation}
where the sum runs over all nonzero integer three-vectors.  Each value of $|\mathbf{n}|$ can be thought of as a winding of the meson $P$ around the finite universe.  The $c_n$ are multiplicity factors counting all the ways to form a vector of length $|\mathbf{n}|$ from triplets of integers, see \tabref{tab:cn_weigths} for the first few.
$K_1(x)$ is a modified Bessel function of the second kind.
In the asymptotically large volume limit, the finite volume correction to these integrals is
\begin{align}
\d^{\rm FV} \ell_P &\equiv \ell_P^{\rm FV} - \ell_P
\nonumber\\&=
    \e_P^2\, 2\sqrt{2\pi} \frac{e^{-m_P L}}{(m_P L)^{3/2}}
\nonumber\\&\quad
    +\e_P^2 \times \mathrm{O}\left(
        \frac{e^{-m_P L\sqrt{2}}}{(m_P L\sqrt{2})^{3/2}},
        \frac{e^{-m_P L}}{(m_P L)^{5/2}}
        \right) .
\end{align}

\begin{figure}
\includegraphics[width=\columnwidth,valign=t]{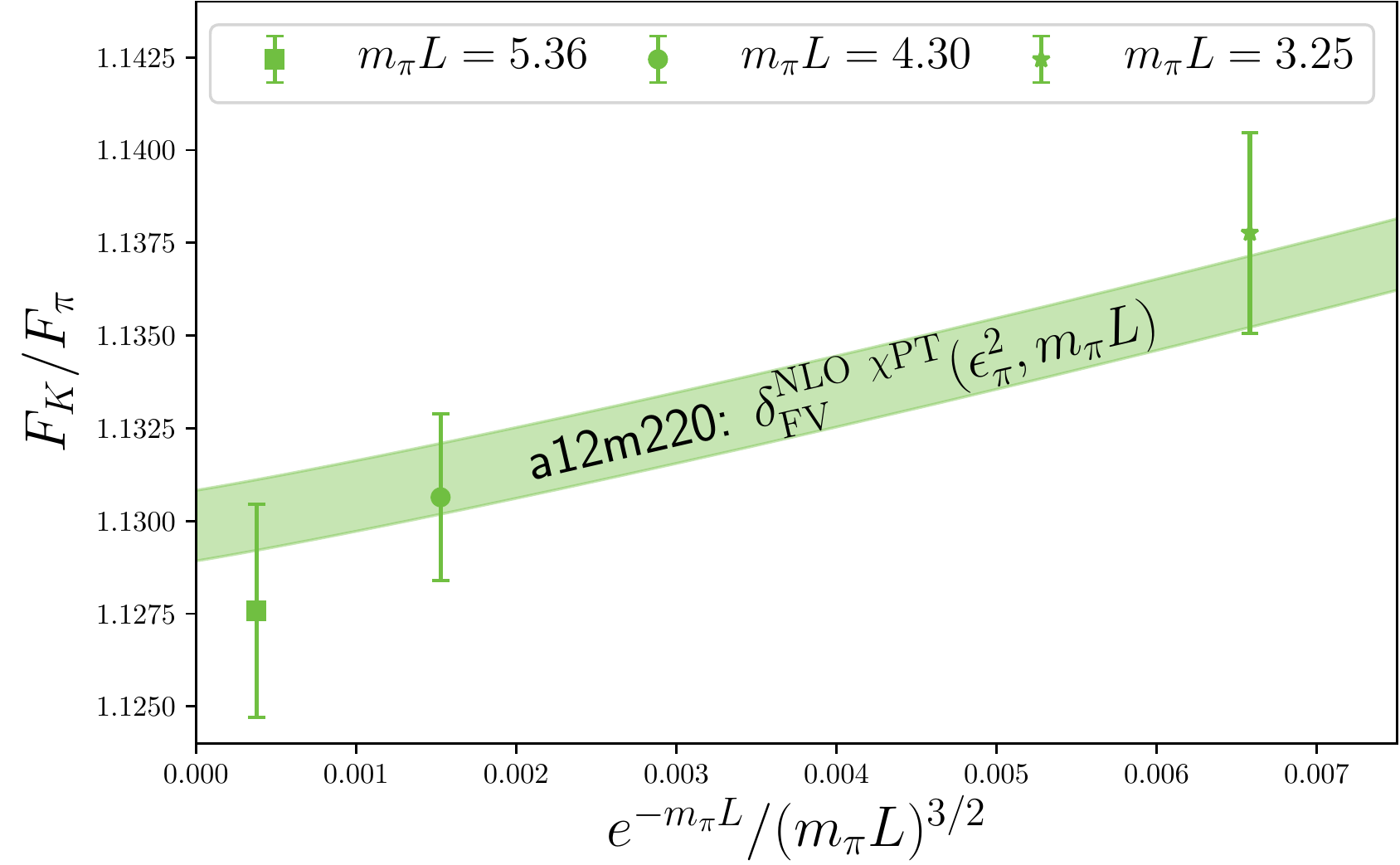}
\caption{\label{fig:fv_extrapolation}
We compare the finite volume results on a12m220L, a12m220 and a12m220S to the predicted finite volume corrections from NLO $\chi$PT.  The uncertainty band is from the full \nxlo{3} $\chi$PT extrapolation, plotted with fixed mesons masses ($\e_P^2$) and fixed lattice spacing ($\e_a^2$), determined from the a12m220L ensemble.  At the one-sigma level, our data are consistent with the leading FV corrections.
}
\end{figure}

The full finite volume corrections to the continuum formula are also known at \nxlo{2}~\cite{Bijnens:2014dea} as well as in the partially quenched $\chi$PT~\cite{Bijnens:2015dra}.
In this work, we restrict the corrections to those arising from the NLO corrections as our results are not sensitive to higher-order FV corrections.
This is because, with the ensembles used in this work, all ensembles except a12m220S satisfy $m_\pi L \gtrsim 4$ (see \tabref{tab:lattice_fits}).  MILC generated three volumes for this a12m220 ensemble series to study FV corrections.
\figref{fig:fv_extrapolation} shows a comparison of the results from the a12m220L, a12m220, and a12m220S along with the predicted volume corrections arising from NLO in $\chi$PT.  The uncertainty band arises from an \nxlo{3} fit using the full \nxlo{2} continuum $\chi$PT formula enhanced with discretization LECs and \nxlo{3} corrections arising from continuum and finite lattice spacing corrections.  Even with one of the most precise fits, we see that the numerical results are consistent with the predicted NLO FV corrections.

\subsection{\nxlo{3} corrections \label{sec:n3lo}}
The numerical dataset in this work requires us to add \nxlo{3} corrections to obtain a good fit quality.
At this order, we only consider local counterterm contributions, of which there are three new continuumlike corrections and three discretization corrections.  A nonunique, but complete parametrization is
\begin{multline}\label{eq:n3lo_ct}
\d^\text{\nxlo{3}} = (\e_K^2 -\e_\pi^2) \bigg\{
    \e_a^4 A_s^6
    + \e_a^2 (A_{s,K}^6 \e_K^2 + A_{s,\pi}^6 \e_\pi^2)
\\
    +A_{K\pi}^6\, \e_K^2 \e_\pi^2
    +(\e_K^2 -\e_\pi^2) (A_K^6 \e_K^2
        +A_\pi^6 \e_\pi^2)
    \bigg\}\, .
\end{multline}
In principle, we could also add counterterms proportional to higher powers of $\a_S$ but with four lattice spacings, we would not be able to resolve the difference between the complete set of operators including all possible additional $\a_S$ corrections.  The set of operators we do include is sufficient to parametrize the approach to the continuum limit.

\section{Extrapolation Details and Uncertainty Analysis \label{sec:analysis}}
We now carry out the extrapolation/interpolation to the physical point, which we perform in a Bayesian framework.
To obtain a good fit, we must work to \nxlo{3} in the mixed chiral and continuum expansion.
The results from the a06m310L ensemble drive this need, in particular, for higher-order discretization corrections to parameterize the results from all the ensembles.
We will explore the impact of the a06m310L ensemble in more detail in this section.
First, we discuss the values of the priors we set and the definition of the physical point.

\subsection{Prior widths for LECs \label{sec:prior_widths}}
The number of additional LECs we need to determine at each order in the expansion is
\begin{center}
\begin{tabular}{rccc}
order& $N_{L_i}$& $N_{\chi}$& $N_{a}$\\
\hline
NLO     & 1     & 0         & 0\\
\nxlo{2}& 7     & 2         & 2\\
\nxlo{3}& 0     & 3         & 3\\
\hline
Total   & 8     & 5         & 5
\end{tabular}.
\end{center}
$N_{L_i}$ is the number of Gasser-Leutwyler coefficients, $N_{\chi}$ the number of counterterms associated with the continuum $\chi$PT expansion and $N_a$ is the number of counterterms associated with the discretization corrections.
In total, there are 18 unknown LECs.  While we utilize 18 ensembles in this analysis, the span of parameter space is not sufficient to constrain all the LECs without prior knowledge.
In particular, the introduction of all 8 $L_i$ coefficients requires prior widths informed from phenomenology.

\begin{table*}
\caption{\label{tab:Li}
$\G_i$ coefficients that appear in the scale dependence of the $L_i(\mu)$.
We evolve the $L_i(\mu)$ from the typical scale $\mu = 770$~MeV, \eqnref{eq:mu_rho} to $\mu_0=4\pi F_0$, beginning with the BE14 estimates from the review~\cite{Bijnens:2014lea} (Table 3), using their known scale dependence~\cite{Gasser:1984gg}, \eqnref{eq:Li_mu}.
We assign the following slightly more conservative uncertainty as a prior width in the minimization: If a value of $L_i$ is less than $0.5\times10^{-3}$, we assign it a 100\% uncertainty at the scale $\mu=770$~MeV; If the value is larger than $0.5\times10^{-3}$, we assign it the larger of 0.5 or 1/3 of the mean value.
}
\begin{ruledtabular}
\begin{tabular}{c|cccccccc}
$L_i$& $L_1$ & $L_2$& $L_3$& $L_4$& $L_5$& $L_6$& $L_7$& $L_8$ \\
\hline
$\G_i$& 3/32& 3/16& 0& 1/8& 3/8& 11/144& 0& 5/48\\
$10^3L_i(m_\rho)$& 0.53(50)& 0.81(50)& -3.1(1.0)& 0.30(30)& 1.01(50)& 0.14(14)& -0.34(34)& 0.47(47)\\
$10^3L_i(\mu_0)$& 0.37(50)& 0.49(50)& -3.1(1.0)& 0.09(30)& 0.38(50)& 0.01(14)& -0.34(34)& 0.29(47)
\end{tabular}
\end{ruledtabular}
\end{table*}

In the literature, the $L_i$ are typically quoted at the renormalization scale $\mu_\rho=770$~MeV while in our work, we use the scale $\mu_{F_0} = 4\pi F_0$.
We can use the BE14 values of the $L_i$ LECs from Ref.~\cite{Bijnens:2014lea} and the known scale dependence~\cite{Gasser:1984gg} to convert them from $\mu_\rho$ to $\mu_{F_0}$:
\begin{equation}\label{eq:Li_mu}
L_i^r(\mu_2) = L_i^r(\mu_1)
    - \frac{\G_i}{(4\pi)^2} \ln \left( \frac{\mu_2}{\mu_1}\right)\, ,
\end{equation}
with the values of $\G_i$ listed in Table~\ref{tab:Li} for convenience.
We use $F_0=80$~MeV, which is the value adopted by FLAG~\cite{Aoki:2019cca}.
We set the central value of all the $L_i$ with this procedure and the widths are set as described in \tabref{tab:Li}.

Next, we must determine priors for the \nxlo{2} and \nxlo{3} local counterterm coefficients, $A^n_{K,\pi,s}$.
We set the central value of all these priors to 0 and then perform a simple grid search varying the widths to find preferred values of the width, as measured by the Bayes factor.  Our goal is not to optimize the width of each prior individually for each model used in the fit, but rather find a set of prior widths that is close to optimal for all models.
To this end, we vary the width of the $\chi$PT LECs together at each order (\nxlo{2}, \nxlo{3}) and the discretization LECs together at each order (\nxlo{2}, \nxlo{3}) for a four-parameter search.  We apply a very crude grid where the values of the widths are taken to be 2, 5, or 10.

We find taking the width of all these $A^n_{K,\pi,s}$ LECs equal to 2 results in good fits with near-optimal values.  This provides evidence the normalization of small parameters we have chosen for $\e_P^2$ and $\e_a^2$, \eqnref{eq:eps_ma}, is ``natural'' and supports the  we have assumed, \eqnref{eq:power_counting}.
The \nxlo{2} LECs mostly favor a width of 2 while the \nxlo{3} discretization LECs prefer 5 and the \nxlo{3} $\chi$PT LECs vary from model to model with 5 a reasonable value for all.
As a result of this search, we pick as our priors
\begin{align}\label{eq:prior_widths}
&\tilde{A}^4_{K,\pi} = 0\pm2,&
&\tilde{A}^4_{s}     = 0\pm2, &
\nonumber\\
&\tilde{A}^6_{K,\pi} = 0\pm5,&
&\tilde{A}^6_{s}     = 0\pm5. &
\end{align}

\subsection{Physical point \label{sec:phys_point}}
As our calculation is performed with isospin symmetric configurations and valence quarks, we must define a physical point to quote our final result.
We adopt the definition of the physical point from FLAG.
FLAG[2017]~\cite{Aoki:2016frl} defines the isospin symmetric pion and kaon masses to be [Eq. (16)]
\begin{align}\label{eq:mKmpi_isospin}
\bar{M}_\pi &= 134.8(3) \textrm{ MeV}\, ,
\nonumber\\
\bar{M}_K   &= 494.2(3) \textrm{ MeV}\, .
\end{align}
The values of $F_{\pi^+}$ and $F_{K^+}$ are taken from the $N_f=2+1$ results from FLAG[2020]~\cite{Aoki:2019cca} (we divide the values by $\sqrt{2}$ to convert to the normalization used in this work)
\begin{align}\label{eq:fkfpi_flag}
F_{\pi^+}^{\rm phys} &= \phantom{1}92.07(57) \textrm{ MeV}\, ,
\nonumber\\
F_{K^+}^{\rm phys}   &= 110.10(49) \textrm{ MeV}\, .
\end{align}
The isospin symmetric physical point is then defined by extrapolating our results to the values (for the choice $F\rightarrow F_\pi$)
\begin{align}\label{eq:iso_Fpi}
(\e_\pi^{\rm phys})^2 &= \left( \frac{\bar{M}_\pi}{4\pi F_{\pi^+}^{\rm phys}} \right)^2\, ,
\nonumber\\
(\e_K^{\rm phys})^2   &= \left( \frac{\bar{M}_K}{4\pi F_{\pi^+}^{\rm phys}} \right)^2\, .
\end{align}

\subsection{Model averaging procedure \label{sec:model_avg}}

Our model average is performed under a Bayesian framework following the procedure described in \cite{BMA, 	Chang:2018uxx}. Suppose we are interested in estimating the posterior distribution of $Y = F_K / F_\pi$, ie. $P(Y|D)$ given our data $D$. To that end, we must marginalize over the different models $M_k$.
\begin{equation}
P(Y|D) = \sum_k P(Y | M_k, D) P(M_k | D)
\end{equation}
Here $P(Y | M_k, D)$ is the distribution of $Y$ for a given model $M_k$ and dataset $D$, while $P(M_k | D)$ is the posterior distribution of $M_k$ given $D$. The latter can be written, per Bayes' theorem, as
\begin{equation}
P(M_k | D) = \frac{P(D | M_k) P(M_k)}{\sum_l P(D | M_l) P(M_l)} \, .
\end{equation}
We can be more explicit with what the latter is in the context of our fits. First, mind that we are \emph{a priori} agnostic in our choice of $M_k$. We thus take the distribution $P(M_k)$ to be uniform over the different models. We calculate $P(D | M_k)$ by marginalizing over the parameters (LECs) in our fits:
\begin{equation}
P(D | M_k) = \int \prod_j \text{d} \theta_j^{(k)} \,  P(D | \theta_j^{(k)}, M_k) P(\theta_j^{(k)} | M_k) \, .
\end{equation}
After marginalization, $P(D | M_k)$ is just a number. Specifically, it is the Bayes factor of $M_k$: $P(D | M_k) = \exp(\texttt{logGBF})_{M_k}$, where \texttt{logGBF} is the log of the Bayes factor as reported by \texttt{lsqfit}~\cite{lsqfit:11.5.1}. Thus
\begin{equation}
P(M_k | D) = \frac{\exp(\texttt{logGBF})_{M_k}}{ \sum_{l=1}^K \exp(\texttt{logGBF})_{M_l}}
\end{equation}
with $K$ the number of models included in our average.
We emphasize that this model selection criterion not only rates the quality of the description of data but also penalizes parameters which do not improve this description.
This helps rule out models which overparametrize data.

Now we can estimate the expectation value and variance of $Y$.
\begin{align}
\text{E}[Y] &= \sum_k \text{E}[Y | M_k] \, P(M_k | D)  \\
\text{Var}[Y] &=
\left[
\sum_k \text{Var}[Y | M_k] P(M_k | D)
\right] \\
&\phantom{=}
+ \left[
\left( \sum_k \text{E}^2[Y | M_k] \, P(M_k | D)\right)
- \text{E}^2[Y | D]
\right] \nonumber
\end{align}
The variance ${\rm Var}[Y]$ results from the total law of variance; the first term in brackets is known as the \emph{expected value of the process variance} (which we refer to as the \emph{model averaged variance}), while the latter is the \emph{variance of the hypothetical means} (the root of which we refer to as the \emph{model uncertainty}). After this work was completed, a similar but more thorough discussion of Bayesian model averaging in the context of lattice QCD was presented~\cite{Jay:2020jkz}.

\subsection{Full analysis and uncertainty breakdown \label{sec:full_analysis}}

In total, we consider 216 different models of extrapolation/interpolation to the physical point.
The various choices for building a $\chi$PT or MA EFT model consist of
\begin{align*}
\begin{array}{rl}
\times2:& \textrm{$\chi$PT or MA EFT at NLO}\\
\times3:& \textrm{use $F^2 = \{F_\pi^2, F_\pi F_K, F_K^2\}$ in defining $\e_P^2$}\\
\times2:& \textrm{fully expanded \eqref{eq:fkfpi_fully_expanded} or ratio \eqref{eq:fkfpi_nlo_ratio} form}\\
\times2:& \textrm{at \nxlo{2}, use full $\chi$PT or just counterterms}\\
\times2:& \textrm{include or not an $\a_S$ term at \nxlo{2}}\\
\times2:& \textrm{include or not the NLO FV corrections}\\
\times2:& \textrm{include \nxlo{3} counterterms or not}\\
\hline
192:& \textrm{total choices}
\end{array}\, .
\end{align*}
We also consider pure Taylor expansion fits with only counterterms and no log corrections.  For these fits, the set of models we explore is
\begin{align*}
\begin{array}{rl}
\times2:& \textrm{work to \nxlo{2} or \nxlo{3}}\\
\times3:& \textrm{use $F^2 = \{F_\pi^2, F_\pi F_K, F_K^2\}$ in defining $\e_P^2$}\\
\times2:& \textrm{include or not an $\a_S$ term at \nxlo{2}}\\
\times2:& \textrm{include or not FV corrections}\\
\hline
24:& \textrm{total choices}
\end{array}\, .
\end{align*}
Based upon the quality of fit (gauged by the Bayesian analog to the $p$-value, $Q$,  or the reduced chi square, $\chi_\nu^2$) and/or the weight determined as discussed in the previous section, we can dramatically reduce the number of models used in the final averaging procedure.
First, any model which does not include the FV correction from NLO is heavily penalized.
This is not surprising given the observed volume dependence on the a12m220 ensembles, \figref{fig:fv_extrapolation}.
However, even if we remove the a12m220S ensemble from the analysis, the Taylor-expanded fits have a relative weight of $e^{-6}$ or less compared to those that have $\chi$PT form at NLO.

If we add FV corrections to the Taylor expansion fits (pure counterterm) and use all ensembles,
\begin{multline}
\frac{F_K}{F_\pi} = 1 + \bar{L}_5 (\e_K^2 - \e_\pi^2) \bigg\{ 1
\\
    +t_{\rm FV} \sum_{|\mathbf{n}|\neq0} \frac{c_n}{m_\pi L|\mathbf{n}|} K_1(m_\pi L|\mathbf{n}|) \bigg\}
    +\cdots
\end{multline}
they still have weights which are $\sim e^{-8}$ over the normalized model distribution and also contribute negligibly to the model average.

\begin{figure}
\includegraphics[width=\columnwidth,valign=t]{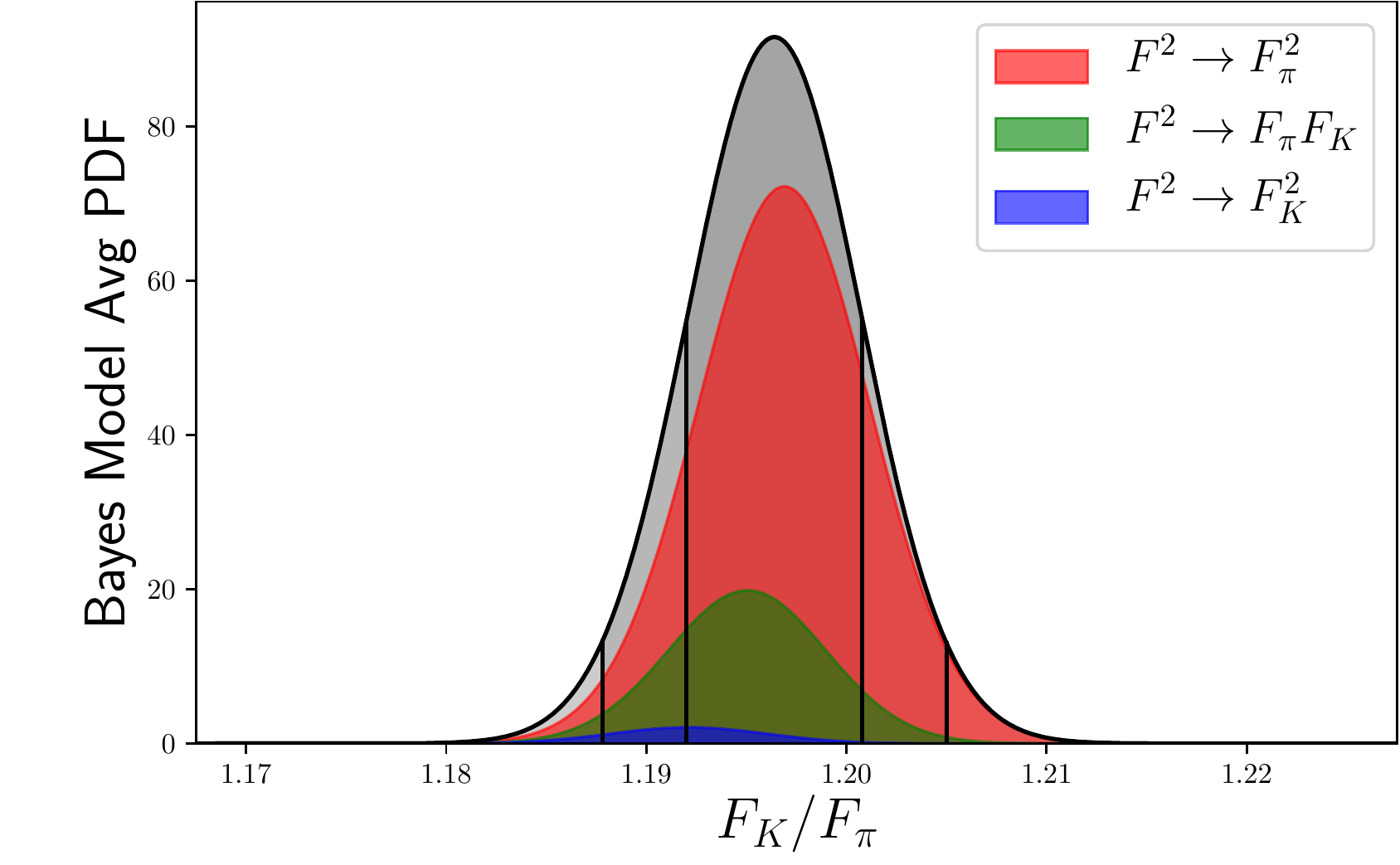}
\caption{\label{fig:bma_histogram}
Final probability distribution giving rise to \eqnref{eq:final_isospin}, separated into the three choices of $F^2=\{F_\pi^2, F_\pi F_K, F_K^2\}$ in the definition of the small parameters, \eqnref{eq:eps_ma}.  The parent ``gray'' distribution is the final PDF normalized to 1 when integrated.}
\end{figure}

We observe that the fits which use the MA EFT at NLO are also penalized with a relative weight of $\sim e^{-8}$, and fits which only work to \nxlo{2} have unfavorable weights by $\sim e^{-5}$ (and are also accompanied by poor $\chi_\nu^2$ values).  Cutting all of these variations reduces our final set of models to be \nxlo{3} $\chi$PT with the following variations
\begin{align*}
\begin{array}{rl}
\times3:& \textrm{use $F^2 = \{F_\pi^2, F_\pi F_K, F_K^2\}$ in defining $\e_P^2$}\\
\times2:& \textrm{fully expanded \eqref{eq:fkfpi_fully_expanded} or ratio \eqref{eq:fkfpi_nlo_ratio} form}\\
\times2:& \textrm{at \nxlo{2}, use full $\chi$PT or just counterterms}\\
\times2:& \textrm{include or not an $\a_S$ term at \nxlo{2}}\\
\hline
24:& \textrm{total choices which enter the model average}
\end{array}\, .
\end{align*}
The final list of models, with their corresponding weights and resulting extrapolated values to the isospin symmetric physical point, is given in \tabref{tab:bma_results} in Appendix~\ref{sec:bma_models}.
Our final result in the isospin symmetric limit, defined as in \eqnref{eq:iso_Fpi} and analogously for other choices of $F^2$, including a breakdown in terms of statistical ($s$), pion mass extrapolation ($\chi$), continuum limit ($a$), infinite volume limit ($V$), physical point (phys) and model selection ($M$) uncertainties, is as reported in \eqnref{eq:final_isospin}
\begin{align*}
\frac{F_K}{F_\pi} &= 1.1964(32)^s(12)^\chi(20)^a(01)^V(15)^{\rm phys}(12)^M
\nonumber\\ &= 1.1964(44)\, .
\end{align*}
The finite volume uncertainty is assessed by removing the a12m220S ensemble from the analysis, repeating the model averaging procedure and taking the difference.
The final probability distribution broken down into the three choices of $F^2$ is shown in \figref{fig:bma_histogram}.

\subsubsection{Impact of a06m310L ensemble \label{sec:a06m310L}}
\begin{figure*}
\includegraphics[width=0.35\textwidth,valign=t]{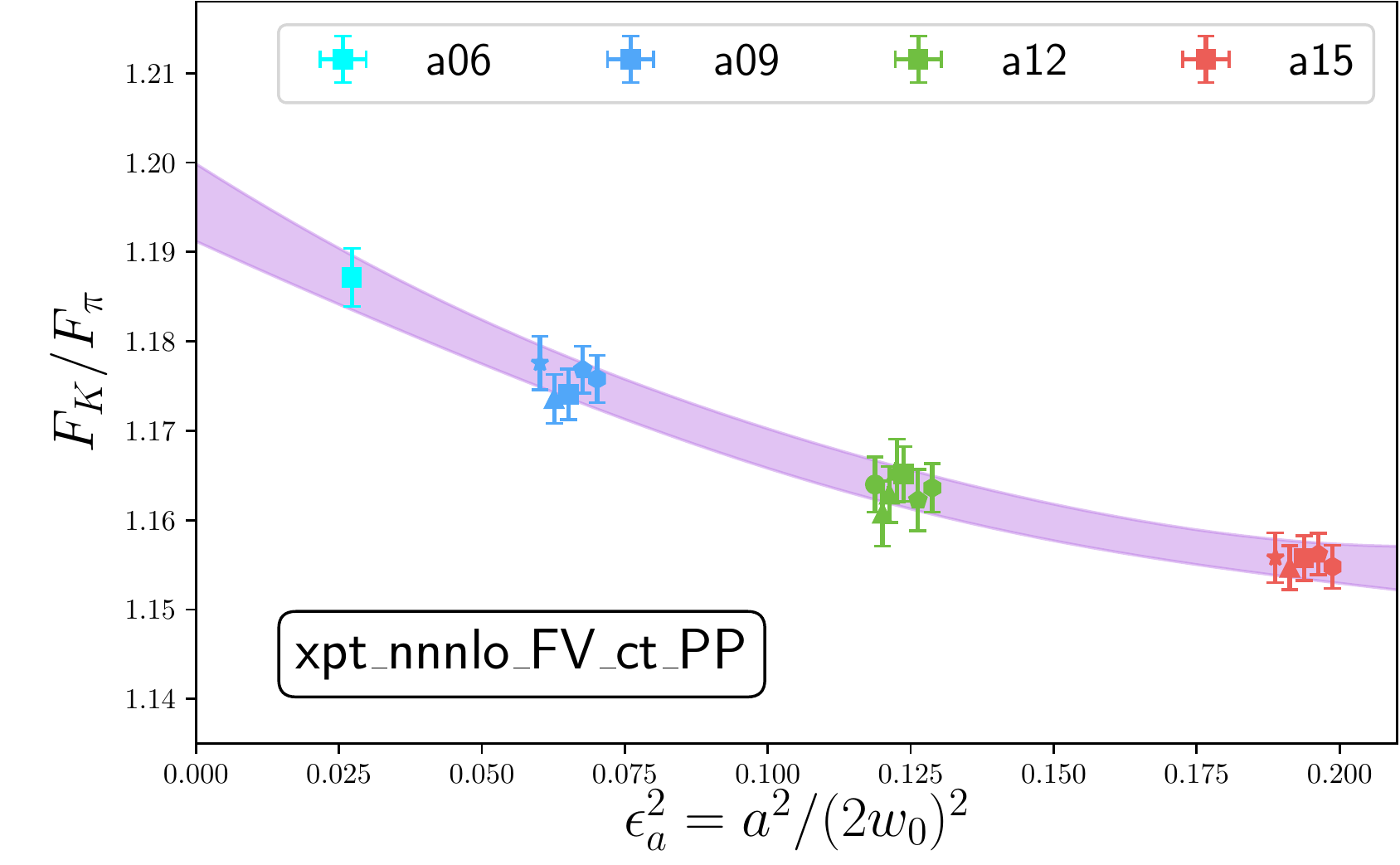}\hspace{-18pt}
\includegraphics[width=0.35\textwidth,valign=t]{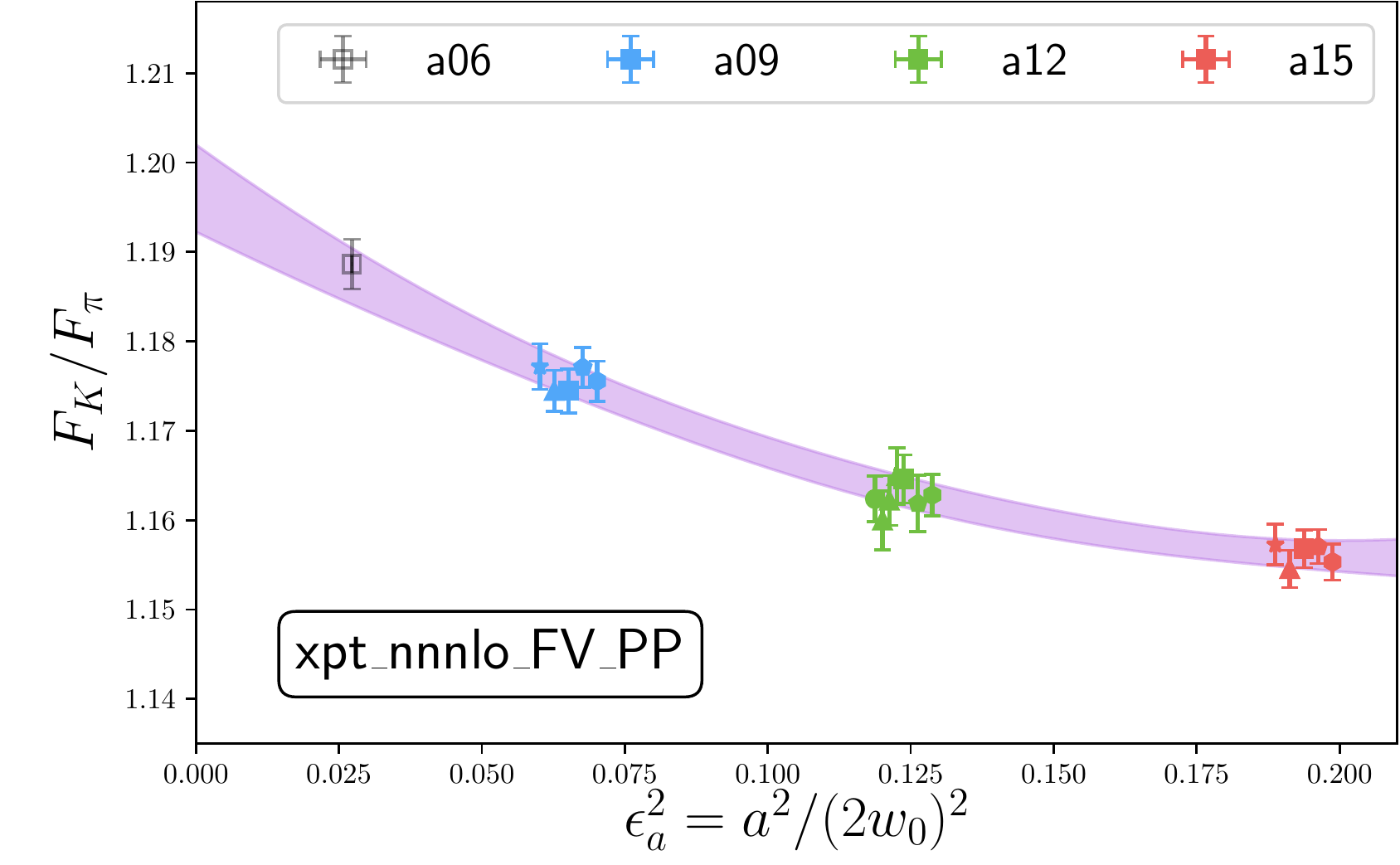}\hspace{-18pt}
\includegraphics[width=0.35\textwidth,valign=t]{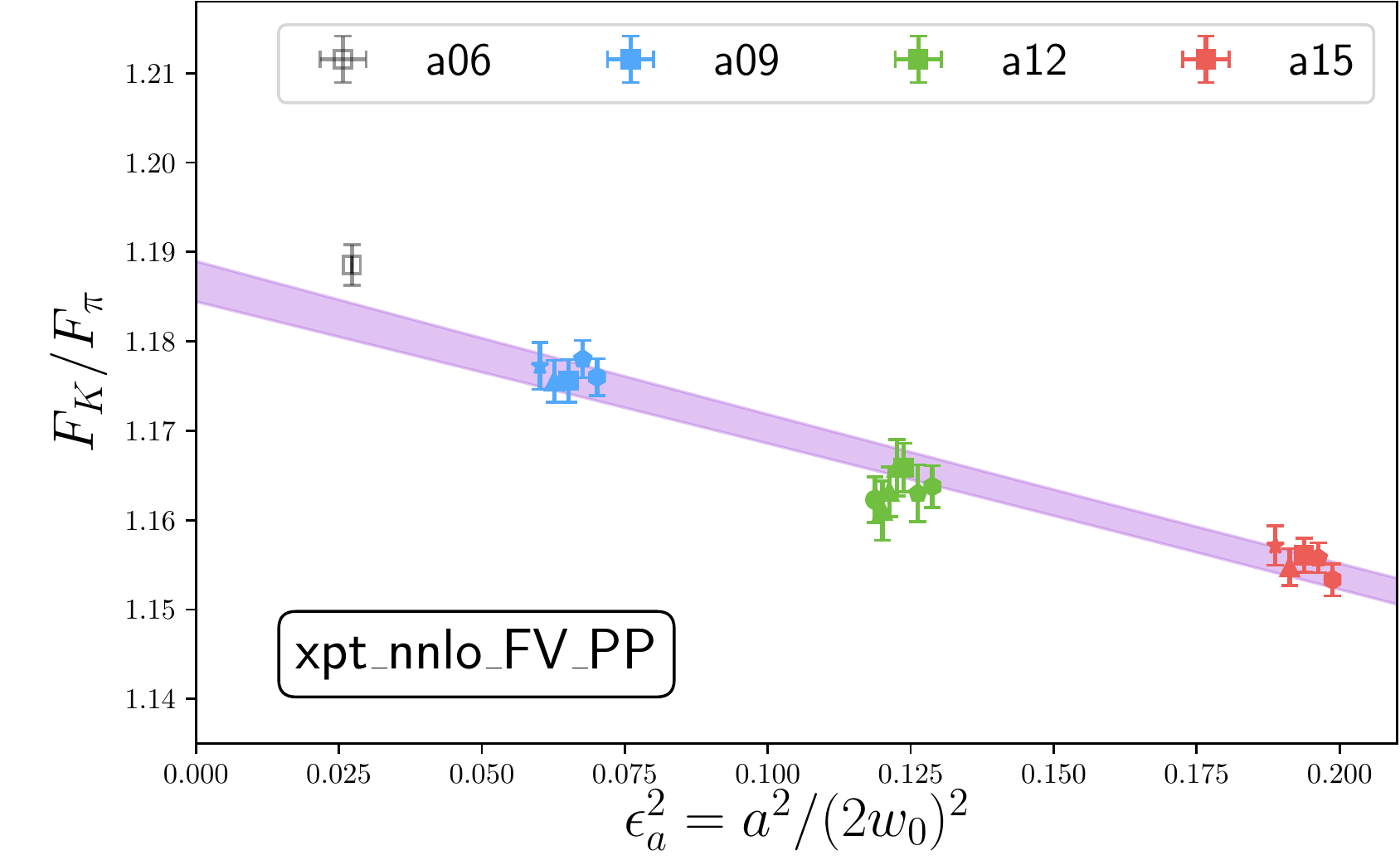}
\caption{\label{fig:impact_a06}
Left: \nxlo{3} fit to all ensembles.
Middle:  same fit to all ensembles excluding  a06m310L.
Right: representative \nxlo{2} fit to all ensembles excluding a06m310L.
In all plots, the results from each ensemble are shifted to the physical values of $\e_\pi^2$ and $\e_K^2$ and the infinite volume limit, with only the $\e_a^2$ dependence remaining.
The labels are are explained in Appendix~\ref{sec:bma_models}; the data points at each spacing are slightly offset horizontally for visual clarity.
}
\end{figure*}

Next, we turn to understanding the impact of the a06m310L ensemble on our analysis.
The biggest difference upon removing the a06m310L ensemble is that the data are not able to constrain the various terms contributing to the continuum extrapolation as well, particularly since there are up to three different types of scaling violations:
\begin{equation*}
    (\e_K^2-\e_\pi^2) \times \{\e_a^2, \alpha_S \e_a^2, \e_a^4\}\, ,
\end{equation*}
and thus, the statistical uncertainty of the results grows as well as the model variance, with a total uncertainty growth from $\sim0.0044$ to $\sim0.0057$, and the mean of the extrapolated answer moves by approximately half a standard deviation.
Furthermore, \nxlo{2} fits become acceptable, though they are still grossly outweighed by the \nxlo{3} fits.
Including both effects, the final model average result shifts from
\begin{equation}
    \frac{F_K}{F_\pi} = 1.1964(44) \rightarrow \frac{F_K}{F_\pi}\bigg|_{\rm no\ a06} = 1.1941(57)\, .
\end{equation}
In \figref{fig:impact_a06}, we show the continuum extrapolation from three fits:
\begin{itemize}[leftmargin=*]
\item Left: all ensembles, \nxlo{3} $\chi$PT with only counterterms at \nxlo{2} and \nxlo{3} and $F=F_\pi$;
\item Middle: no a06m310L, \nxlo{3} $\chi$PT with only counterterms at \nxlo{2} and \nxlo{3} and $F=F_\pi$;
\item Right: no a06m310L, \nxlo{2} $\chi$PT with only counterterms at \nxlo{2} and $F=F_\pi$.
\end{itemize}
As can be seen from the middle plot, the a15, a12 and a09 ensembles prefer contributions from both $\e_a^2$ and $\e_a^4$ contributions and are perfectly consistent with the result on the a06m310L ensemble.
They are also consistent with an \nxlo{2} fit (no $\e_a^4$ contributions) as can be seen in the right figure.
However, the weight of the \nxlo{3} fits is still significantly greater than the \nxlo{2} fits even without the a06m310L data.

We conclude that the a06m310L ensemble is useful, but not necessary to obtain a subpercent determination of $F_K/F_\pi$ with our lattice action.
A more exhaustive comparison can be performed with the analysis notebook provided with this publication.

\begin{figure*}
\includegraphics[width=0.6\textwidth,valign=t]{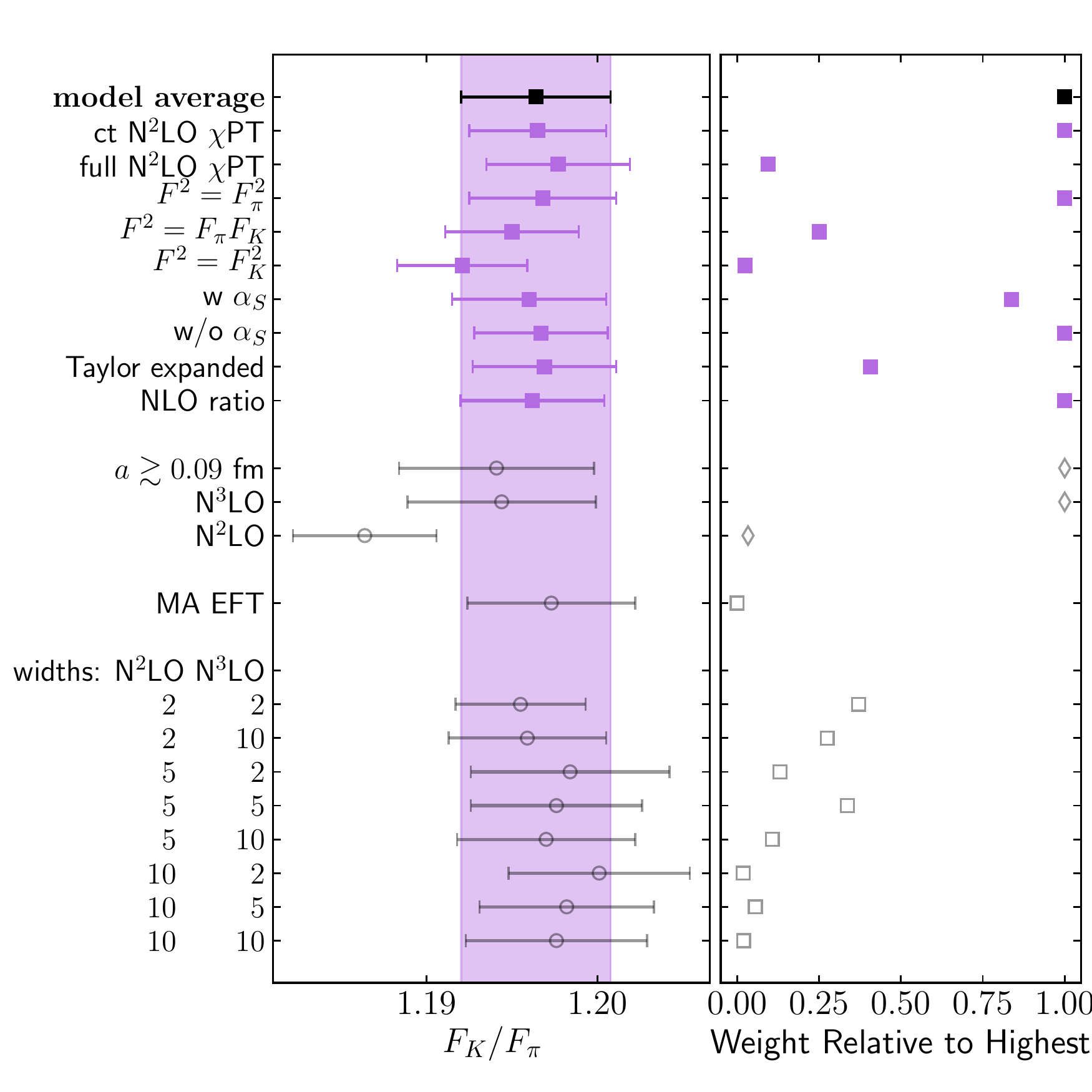}
\caption{\label{fig:stability}
Stability plot of final result compared to various model choices.
The black square at the top is our final answer for the isospin symmetric determination of $F_K/F_\pi$.
The vertical magenta band is the uncertainty of this fit to guide the eye.
The solid magenta squares are various ways of decomposing the model selection that goes into the final average.
The right panel shows the relative weight with respect to the maximum \texttt{logGBF}  value, $\exp\{ \texttt{logGBF}_i - \texttt{logGBF}_{\rm max}\}$.
Below the set of models included in the average, we show sets of analyses that are not included in the average for comparison, which are indicated with open gray symbols.
First, we show the impact of excluding the a06m310L ensemble.  The \texttt{logGBF}  cannot be directly compared between these fits and the main analysis as the number of data points are not the same, so the overall normalization is different; their \texttt{logGBF}  are shown as open diamonds.
The relative \texttt{logGBF}  between the \nxlo{3} and \nxlo{2} analysis can be compared which indicates a large preference for the \nxlo{3} analysis.
We also show the MA EFT analysis, which agrees well with the main analysis.
Finally, we show the results if one were to change the widths of the \nxlo{2} and \nxlo{3} priors from those chosen in \eqnref{eq:prior_widths}.
}
\end{figure*}

In \figref{fig:stability}, we show the stability of our final result for various choices discussed in this section.

\subsubsection{Convergence of the chiral expansion \label{sec:xpt_convergence}}
While the numerical analysis favors a fit function in which only counterterms are used at \nxlo{2} and higher, it is interesting to study the convergence of the chiral expansion by studying the fits which use the full $\chi$PT expression at \nxlo{2}.

\begin{figure}
\includegraphics[width=\columnwidth]{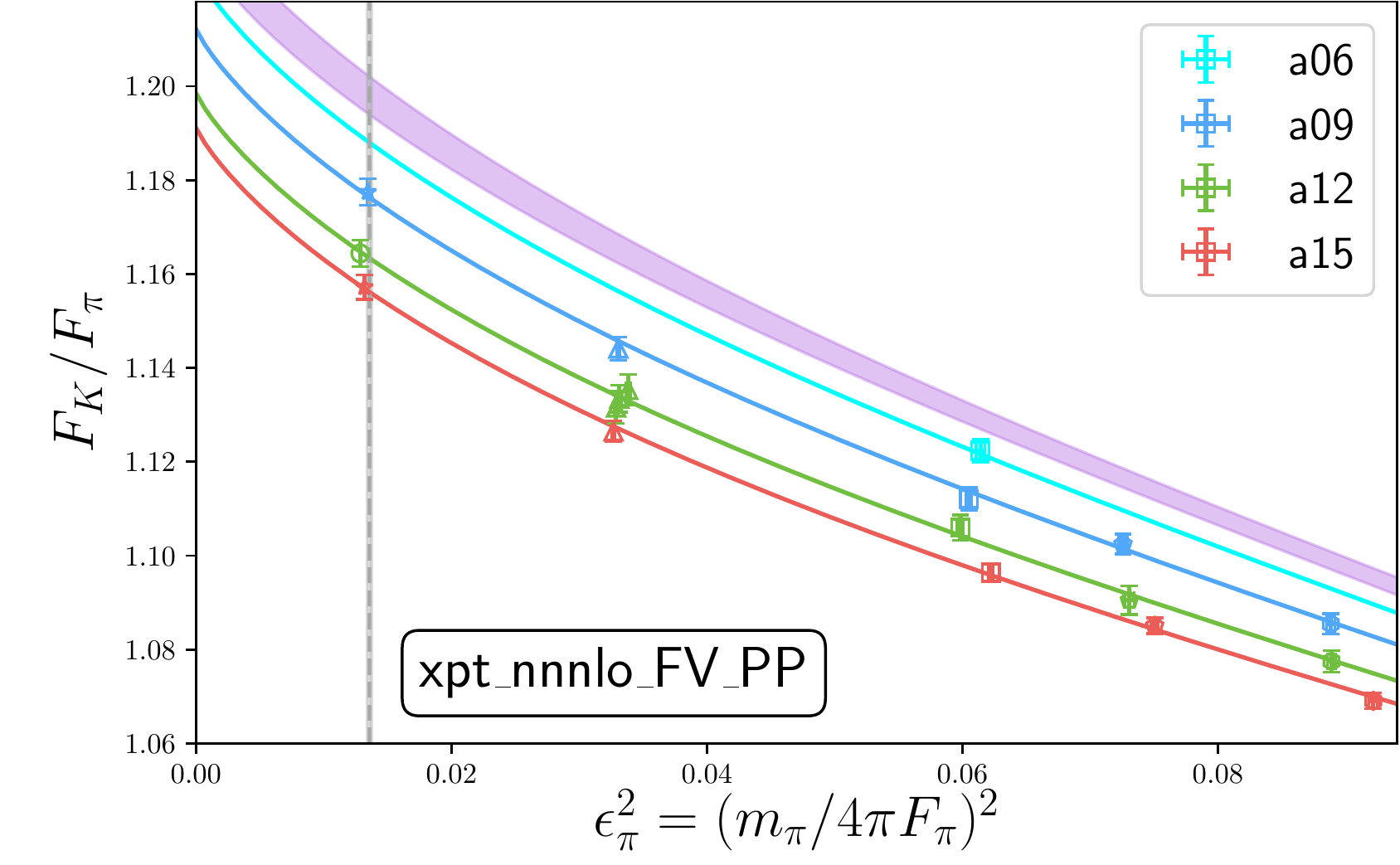}
\includegraphics[width=\columnwidth]{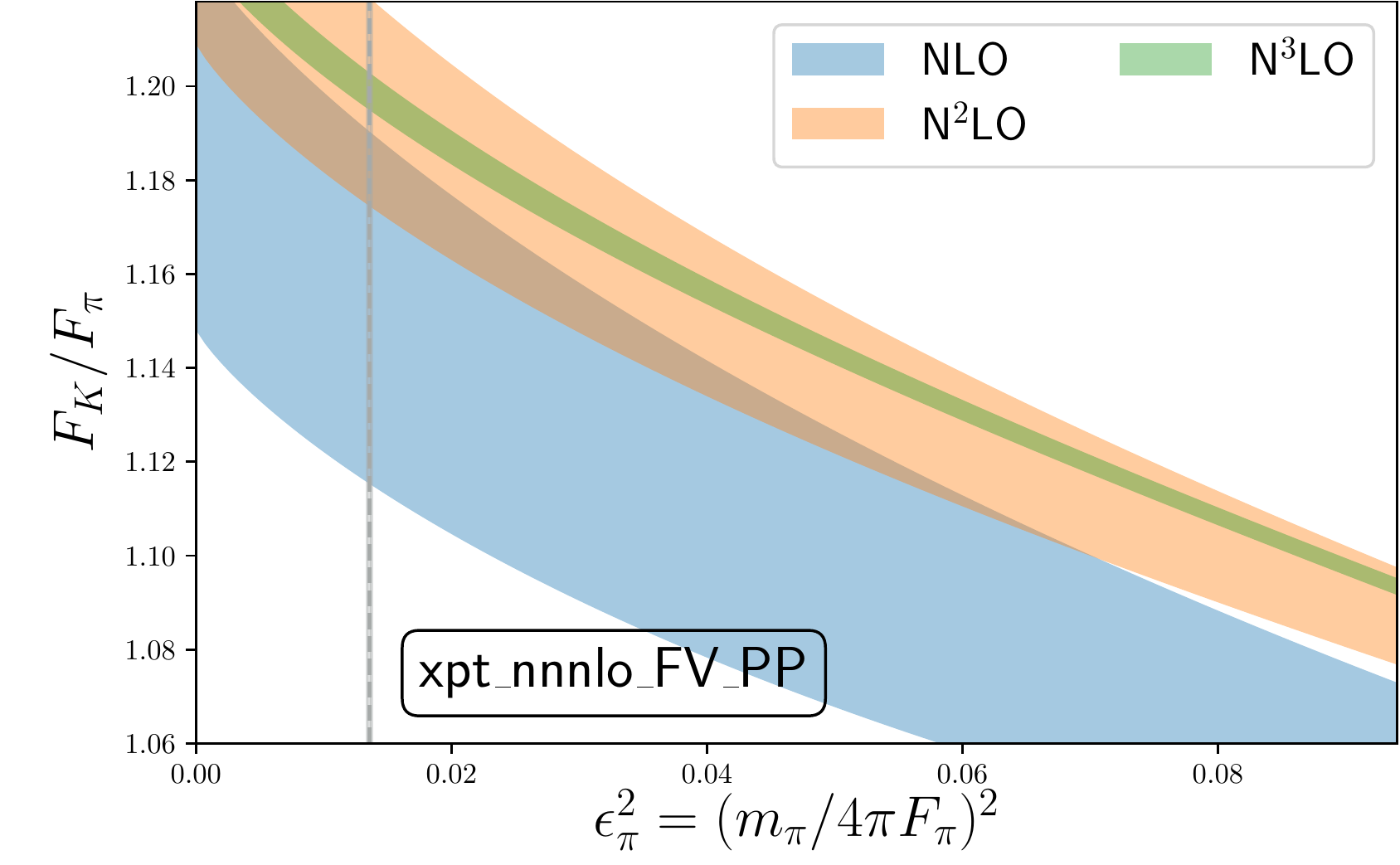}
\caption{\label{fig:mpi_extrapolation}
Sample light quark mass dependence from a $\chi$PT fit with $F=F_\pi$ (\nxlo{2} $\chi$PT + \nxlo{3} counterterms).
Top: curves are plotted with $\e_K^2 = (\e_K^{\rm phys})^2$, at fixed $\e_a^2$ for each lattice spacing, as a function of $\e_\pi^2$.  The magenta band is the full uncertainty in the continuum, infinite volume limit.  The data points have all been shifted from the values of $\e_K^{\rm latt.}$ to $\e_K^{\rm phys}$ and to the infinite volume limit.
Bottom: convergence of the resulting fit as a function of $\e_\pi^2$.  Each band corresponds to all contributions up to that order with the LECs determined from the full fit.
The \nxlo{3} band corresponds to the continuum extrapolated band in the top figure.
}
\end{figure}

In \figref{fig:mpi_extrapolation}, we show the resulting light quark mass dependence using the \nxlo{3} extrapolation with the full \nxlo{2} $\chi$PT formula.
After the analysis is performed, the results from each ensemble are shifted to the physical kaon mass point, leaving only dependence upon $\e_\pi^2$ and $\e_a^2$ as well as dependence upon the $\eta$ mass defined by the GMO relation.
The magenta band represents the full 68\% confidence interval in the continuum, infinite volume limit.  The different colored curves are the mean values as a function of $\e_\pi^2$ at the four different lattice spacings.
We also show the convergence of this fit in the lower panel plot.
From this convergence plot, one sees that roughly that at the physical pion mass (vertical gray line) the NLO contributions add a correction of $\sim0.16$ compared to 1 at LO, the \nxlo{2} contributions add another $\sim0.04$, and the \nxlo{3} corrections are not detectable by eye.  The band at each order represents the sum of all terms up to that order determined from the full fit.
The reduction in the uncertainty as the order is increased is due on large part to the induced correlation between the LECs at different orders through the fitting procedure.

In \figref{fig:bma_histogram}, we observe that the different choices of $F$ are all consistent, indicating higher-order corrections (starting at \nxlo{3} in the noncounterterm contributions) are smaller than the uncertainty in our results.  It is also interesting to note that choosing $F_{\pi K}$ or $F_K$ is penalized by the analysis, indicating the numerical results prefer larger expansion parameters.
In \tabref{tab:LEC_fit}, we show the resulting $\chi$PT LECs determined in this analysis for the two choices $F=\{F_\pi, F_{K\pi}\}$, as well as whether the ratio form of the fit is used, \eqnref{eq:fkfpi_nlo_ratio}.
For the Gasser-Leutwyler LECs, we evolve the values back from $\mu_0\rightarrow\mu_\rho$ for a simpler comparison with the values quoted in literature.
For most of the $L_i$, we observe the numerical results have very little influence on the parameters as they mostly return the prior value (also listed in the table for convenience).
The only LECs influenced by the fit are $L_5$, $L_7$, and $L_8$ with $L_5$ getting pulled about one sigma away from the prior value and $L_7$ and $L_8$ only shifting by a third or half of the prior width.
One interesting observation from our results is that our fit prefers a value of $L_5$ that is noticeably smaller than the value obtained by MILC~\cite{Bazavov:2010hj} and HPQCD~\cite{Dowdall:2013rya} and is also smaller than the BE14 result from Ref.~\cite{Bijnens:2014lea}, although the discrepancy is still less than 2 sigma.
We also note that our value of $L_5$ is very compatible with that determined by RBC/UKQCD with domain-wall fermions and near-physical pion masses~\cite{Blum:2014tka}.  Those interested in exploring this in more detail can utilize our numerical results, and if desired, extrapolation code made available with this publication.

\begingroup \squeezetable
\begin{table}
\caption{\label{tab:LEC_fit}
Resulting LECs from full \nxlo{2} $\chi$PT analysis (also including \nxlo{3} counterterms).
For the Gasser-Leutwyler LECs $L_i$, we evolve them back to the standard scale $\mu=770$~MeV, while for the other LECs, we leave them at the scale $\mu_0=4\pi F_0\simeq1005$~MeV.}
\begin{ruledtabular}
\begin{tabular}{c|rrrrr}
LEC& &\multicolumn{2}{c}{$F^2=F_\pi^2$}& \multicolumn{2}{c}{$F^2=F_\pi F_K$}\\
& &\multicolumn{2}{c}{ratio}& \multicolumn{2}{c}{ratio}\\
\cline{3-4}\cline{5-6}
$\mu=770$& prior& no& yes& no& yes\\
\hline
$10^3L_1$& $0.53(50)$&  $0.47(49)$&  $0.50(49)$&  $0.45(49)$&  $0.48(49)$\\
$10^3L_2$& $0.81(50)$&  $0.77(46)$&  $0.84(46)$&  $0.69(44)$&  $0.77(45)$\\
$10^3L_3$& $-3.1(1.0)$& $-3.02(85)$& $-2.84(86)$& $-3.26(81)$& $-3.05(82)$\\
$10^3L_4$& $0.30(30)$&  $0.24(29)$&  $0.14(29)$&  $0.24(29)$&  $0.16(29)$\\
$10^3L_5$& $1.01(50)$&  $0.48(35)$&  $0.52(34)$&  $0.40(33)$&  $0.47(34)$\\
$10^3L_6$& $0.14(14)$&  $0.14(14)$&  $0.14(14)$&  $0.14(14)$&  $0.14(14)$\\
$10^3L_7$& $-0.34(34)$& $-0.55(32)$& $-0.57(32)$& $-0.52(33)$& $-0.53(33)$\\
$10^3L_8$& $0.47(47)$&  $0.30(46)$&  $0.28(46)$&  $0.35(46)$&  $0.32(46)$\\
\hline
$\mu=\mu_0$\\
\hline
$A_K^4$     & 0(2)& $0.06(1.42)$  & $0.09(1.41)$ &$0.2(1.6)$ &$0.2(1.5)$\\
$A_\pi^4$   & 0(2)& $2.5(1.2)$    & $2.4(1.2)$   &$2.0(1.3)$ &$2.0(1.3)$\\
$A_{K\pi}^6$& 0(5)& $2.8(4.7)$    & $2.8(4.7)$   &$1.9(4.7)$ &$2.0(4.7)$\\
$A_K^6$     & 0(5)& $0.008(4.016)$& $0.3(4.0)$   &$0.1(4.4)$ &$0.2(4.4)$\\
$A_p^6$     & 0(5)& $2.6(4.0)$    & $2.1(4.1)$   &$2.4(4.4)$ &$2.0(4.4)$
\end{tabular}
\end{ruledtabular}
\end{table}
\endgroup

\begin{figure}
\includegraphics[width=0.48\textwidth,valign=t]{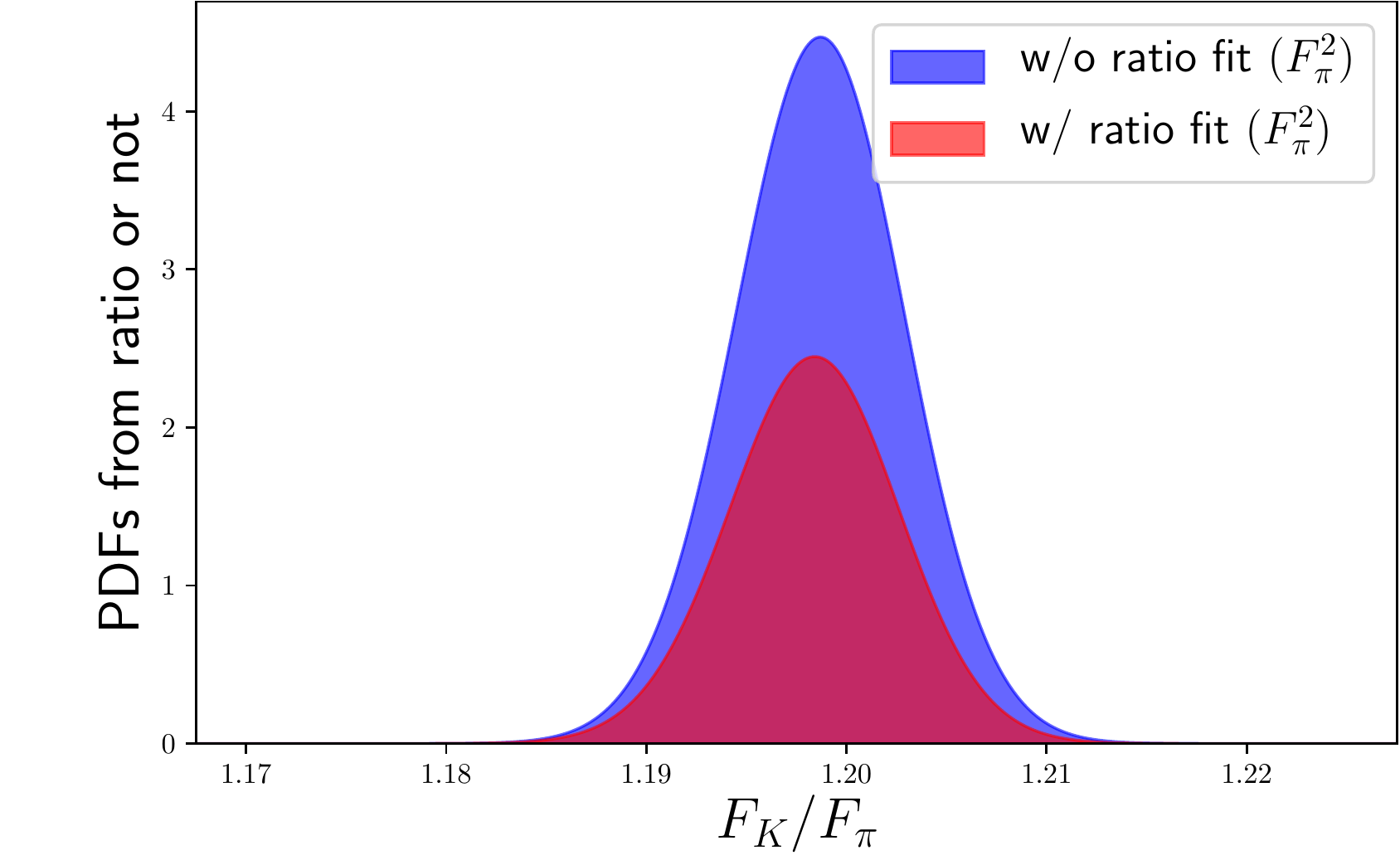}
\caption{\label{fig:xpt_systematics}
Comparison of fits with the fully expanded \eqnref{eq:fkfpi_fully_expanded} and ratio \eqnref{eq:fkfpi_nlo_ratio} expressions, all with the choice $F=F_\pi$.  The PDFs are taken from the parent PDF, \figref{fig:bma_histogram} without renormalizing such that height in this figure reflects the relative weight compared to the total PDF.}
\end{figure}

In \figref{fig:xpt_systematics}, we show the impact of using the fully expanded expression, \eqnref{eq:fkfpi_fully_expanded}, versus the expression in which the NLO terms are kept in a ratio, \eqnref{eq:fkfpi_nlo_ratio}.  To simplify the comparison we restrict it to the choice $F=F_\pi$ and the full \nxlo{2} $\chi$PT expression.
We see that fits without the ratio form are preferred, but the central value of the final result depends minimally upon this choice.

\begin{figure}
\includegraphics[width=\columnwidth,valign=t]{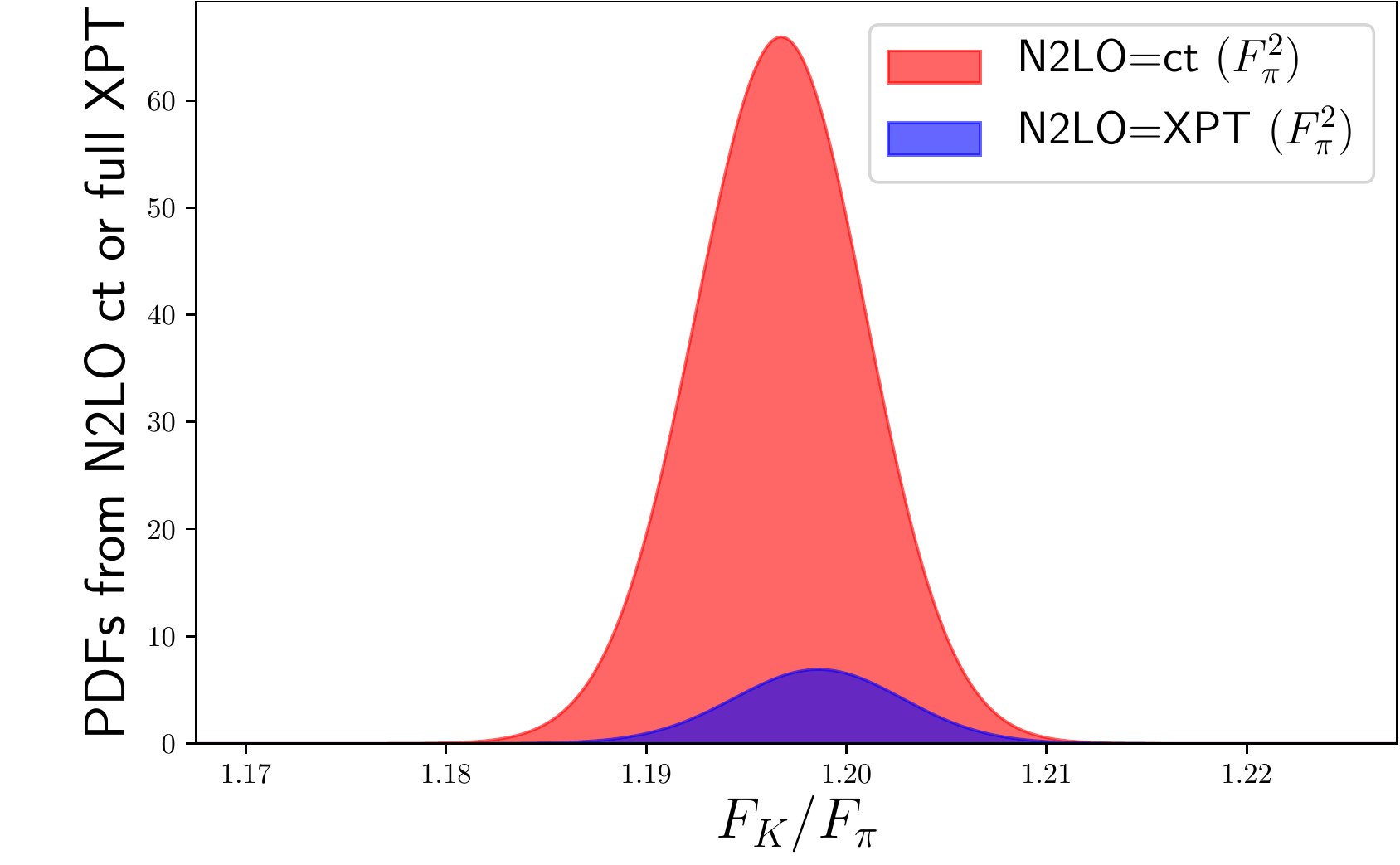}
\caption{\label{fig:xpt_ct}
Comparison of \nxlo{3} $\chi$PT analysis with $F=F_\pi$ using the full \nxlo{2} $\chi$PT expression (smaller histogram) versus only counterterms at \nxlo{2}, Eqs.~\eqref{eq:n2lo_xpt_ct} and \eqref{eq:n2lo_a}.  As in \figref{fig:xpt_systematics}, the PDFs are drawn from the parent PDF.}
\end{figure}

In \figref{fig:xpt_ct}, we show that the results strongly favor the use of only counterterms at \nxlo{2} as opposed to the full $\chi$PT fit function at that order.  We focus on the choice $F=F_\pi$ to simplify the comparison.

Our results are not sufficient to understand why the fit favors only counterterms at \nxlo{2} and higher.
While the linear combination of LECs in \eqnref{eq:n2lo_xpt_ct} are redundant, the $L_i$ LECs also appear in the single-log coefficients, Eqs.~\eqref{eq:C_i_terms} and \eqref{eq:c_i_coefficients} in different linear combinations.
Nevertheless, we double check that the fit is not penalized for the counterterm redundancy, \eqnref{eq:n2lo_xpt_ct}.  Using the priors for $L_i$ from \tabref{tab:Li}, we find the contribution from the Gasser-Leutwyler LECs to these \nxlo{2} counterterms, \eqnref{eq:nnlo_GL}, are given by
\begin{align}
&L_K^4 = 0.3(1.3)\, ,&
&L_\pi^4 = -0.64(94)\, .&
\end{align}
As the $A_P^4$ terms are priored at $0(2)$, it is sufficient to rerun the analysis by simply setting $L_P^4=0$.  We find this result marginally improves the Bayes factors but not statistically significantly, leaving us with the puzzle that the optimal fit is a hybrid NLO $\chi$PT plus counterterms (analytic terms) at higher orders.
We note that it has been known for some time that using $\chi$PT at NLO plus purely analytic terms at NNLO and higher results in good quality extrapolation fits, at least in part because the NNLO chiral logarithms are relatively slowly varying for the range of pion masses for which the NNLO analytic terms are sizable enough to be important~\cite{Aubin:2004fs}.
This is discussed in more detail in the review by Bernard~\cite{Bernard:2015wda}.
The MILC Collaboration no longer reports analysis with just the analytic terms at NNLO~\cite{Bazavov:2010hj} and so it is not clear if other groups observe the same preference for counterterms only at NNLO or not.

If the Taylor expansion fits (pure counterterm) were good and favored over the $\chi$PT fits, this could be a sign that the $SU(3)$ $\chi$PT formula was failing to describe the lattice results.  However, we have to include the NLO $\chi$PT expression, including its predicted (counterterm free) volume dependence to describe the numerical results.
It would be nice to have the full \nxlo{2} MA EFT expression to understand why the hybrid MA EFT fits are so relatively disfavored in the analysis.  There may be compensating discretization effects that cancel against those at NLO to some degree that might allow the full \nxlo{2} MA EFT to better describe the results.
However, at two loops in $\chi$PT, the universality of MA EFT expressions~\cite{Chen:2007ug} breaks down such that the MA EFT expression can no longer be ``derived'' from the corresponding PQ$\chi$PT one (which is known for $F_K/F_\pi$ at two loops~\cite{Bijnens:2004hk,Bijnens:2005ae,Bijnens:2006jv,Bijnens:2015dra}).  It is therefore unlikely that the NNLO MA EFT expression specific to this MALQCD calculation will ever be derived, so this issue will most likely not be resolved with more clarity.

\subsection{QCD isospin breaking corrections \label{sec:isospin}}

Finally, we discuss the correction to our result to obtain a direct determination of $F_{K^+}/F_{\pi^+}$ including strong isospin breaking corrections, but excluding QED corrections.  This is the standard value quoted in the FLAG reviews~\cite{Aoki:2016frl,Aoki:2019cca}.
Our calculations, like most, are performed in the isospin symmetric limit, and therefore, the strong isospin breaking correction must be estimated, rather than having a direct determination.
The optimal approach is to incorporate both QED and QCD isospin breaking corrections into the calculations such that the separation is not necessary, as was done in Ref~\cite{Giusti:2017dwk} by incorporating both types of corrections through the perturbative modification of the path integral and correlation functions~\cite{deDivitiis:2013xla,Giusti:2017dmp}.
In this work, we have not performed these extensive computations and so we rely upon the $SU(3)$ $\chi$PT prediction to estimate the correction due to strong isospin breaking.
As we have observed in \secref{sec:full_analysis}, the $SU(3)$ chiral expansion behaves and converges nicely, so we expect this approximation to be reasonable.

The NLO corrections to $F_K$ and $F_\pi$ including the strong isospin breaking corrections are given by
\begin{align}
\d F_{\pi^\pm}^{\rm NLO} &=
    -\frac{\ell_{\pi^0}}{2} -\frac{\ell_{\pi^\pm}}{2}
    -\frac{\ell_{K^0}}{4} -\frac{\ell_{K^\pm}}{4}
\nonumber\\&\phantom{=}
    +4\bar{L}_4 (\e_{\pi^\pm}^2 + \e_{K^\pm}^2 +\e_{K^0}^2)
    +4\bar{L}_5 \e_{\pi^\pm}^2\, ,
\nonumber\\
\d F_{K^\pm}^{\rm NLO} &=
    -\frac{\ell_{\pi^0}}{8} -\frac{\ell_{\pi^\pm}}{4} -\frac{3\ell_\eta}{8}
    -\frac{\ell_{K^0}}{4} -\frac{\ell_{K^\pm}}{2}
    \nonumber\\&\phantom{=}
    +\frac{1}{4}(\e_{K^0}^2 - \e_{K^\pm}^2) \frac{\ell_\eta - \ell_{\pi^0}}{\e_\eta^2 - \e_{\pi^0}^2}
    \nonumber\\&\phantom{=}
    +4\bar{L}_4 (\e_{\pi^\pm}^2 + \e_{K^\pm}^2 +\e_{K^0}^2)
    +4\bar{L}_5 \e_{K^\pm}^2\, ,
\end{align}
where we have kept explicit the contribution from each flavor of meson propagating in the loop.
There are three points to note in these expressions:
\begin{enumerate}[leftmargin=*]
\item At NLO in the $SU(3)$ chiral expansion, there are no additional LECs that describe the isospin breaking corrections beyond those that contribute to the isospin symmetric limit.  Therefore, one can make a parameter-free prediction of the isospin breaking corrections using lattice results from isospin symmetric calculations, with the only assumption being that $SU(3)$ $\chi$PT converges for this observable;
\item If we expand these corrections about the isospin limit, they agree with the known results~\cite{Cirigliano:2011tm}, and $\d F_{\pi^\pm}$ is free of isospin breaking corrections at this order;
\item We have used the kaon mass splitting in place of the quark mass splitting, which is exact at LO in $\chi$PT $B(m_d-m_u)=(\hat{M}_{K^0}^2-\hat{M}_{K^\pm}^2)$;
\end{enumerate}
The estimated shift of our isospin-symmetric result to incorporate strong isospin breaking is then
\begin{align}\label{eq:iso_1}
\d F^{\rm iso}_{K-\pi} &\equiv \frac{F_{\hat{K}^+}}{F_{\hat{\pi}^+}} - \frac{F_K}{F_\pi}
\nonumber\\&=
    -\frac{1}{4}( \ell_{\hat{K}^+} - \ell_{\bar{K}} )
    +4\bar{L}_5 ( \e_{\hat{K}^+}^2 - \e_{\bar{K}}^2 )
\nonumber\\&\phantom{=}
    +\frac{1}{4}(\e_{K^0}^2 - \e_{K^\pm}^2) \frac{\ell_\eta - \ell_{\pi^0}}{\e_\eta^2 - \e_{\pi^0}^2}
\end{align}
Ref.~\cite{Cirigliano:2011tm} suggested replacing $\bar{L}_5$ with the NLO expression equating it to the isospin symmetric $F_K/F_\pi$ which yields
\begin{multline}\label{eq:iso_2}
\d F^{\rm iso^\prime}_{K-\pi} =
    -\frac{1}{6} \frac{\e_{K^0}^2 - \e_{K^\pm}^2}{\e_\eta^2 - \e_{\pi^0}^2}
        \bigg[ 4\left(\frac{F_K}{F_\pi} -1\right)
\\
            +\e_{\bar{\pi}}^2 \ln \left( \frac{\e_{\bar{K}}^2}{\e_{\bar{\pi}}^2} \right)
            -\e_{\bar{K}}^2 +\e_{\bar{\pi}}^2
        \bigg]\, .
\end{multline}
In this expression, we have utilized the two relations
\begin{align}
\ell_{\hat{K}^+} - \ell_{\bar{K}} &=
    -\frac{2}{3} \frac{\e_{K^0}^2 - \e_{K^\pm}^2}{\e_\eta^2 - \e_{\pi^0}^2}
        (\e_{\bar{K}}^2 -\e_{\bar{\pi}}^2) (\ln\e_{\bar{K}}^2 + 1)
\nonumber\\
    \e_\eta^2 - \e_{\bar{\pi}}^2 &= \frac{4}{3}(\e_{\bar{K}}^2 - \e_{\bar{\pi}}^2)
\end{align}
At this order, both Eqs.~\eqref{eq:iso_1} and \eqref{eq:iso_2} are equivalent.
However, they can result in shifts that differ by more than one standard deviation.  Further, the direct estimate of the strong isospin breaking corrections~\cite{Carrasco:2014poa} is larger in magnitude than either of them.
Therefore, to estimate the strong isospin breaking corrections, we take the larger of the two corrections as the mean and the larger uncertainty of the two, and then add an additional 25\% uncertainty for $SU(3)$ truncation errors.  In \secref{sec:xpt_convergence} we observe the \nxlo{2} correction is $\sim25\%$ of the NLO correction (while NLO is $\sim16\%$ of LO).

In order to evaluate these expressions, we have to define the physical point with strong isospin breaking and without QED isospin breaking.  We employ the values from FLAG[2017]~\cite{Aoki:2016frl} (except $\hat{M}_{\pi^0} = 134.6(3)$~MeV):
\begin{align}
    \hat{M}_{\pi^0} = \hat{M}_{\pi^+} &= 134.8(3) \textrm{ MeV}\, ,
\nonumber\\
    \hat{M}_{K^0}   &= 497.2(4) \textrm{ MeV}\, ,
\nonumber\\
    \hat{M}_{K^+}   &= 491.2(5) \textrm{ MeV}\, .
\end{align}
With this definition of the physical point, we find (under the same model average as \tabref{tab:bma_results})
\begin{align}
\d F^{\rm iso}_{K-\pi} &= -0.00188(51)\, ,
\nonumber\\
\d F^{\rm iso^\prime}_{K-\pi} &= -0.00215(24)\, ,
\end{align}
resulting in our estimated strong isospin breaking correction
\begin{align}
    \frac{F_{\hat{K}^+}}{F_{\hat{\pi}^+}} - \frac{F_K}{F_\pi} &= -0.00215(72)
\end{align}
and our final result as reported in \eqnref{eq:final_FKFpi_plus}
\begin{align*}
\frac{F_{\hat{K}^+}}{F_{\hat{\pi}^+}} &= 1.1942(44)(07)^{\rm iso}
\nonumber\\
    &= 1.1942(45)\, ,
\end{align*}
where the first uncertainty in the first line is the combination of those in \eqnref{eq:final_isospin}.

\section{Summary and Discussion \label{sec:summary}}
\begin{figure}
\includegraphics[width=\columnwidth,valign=t]{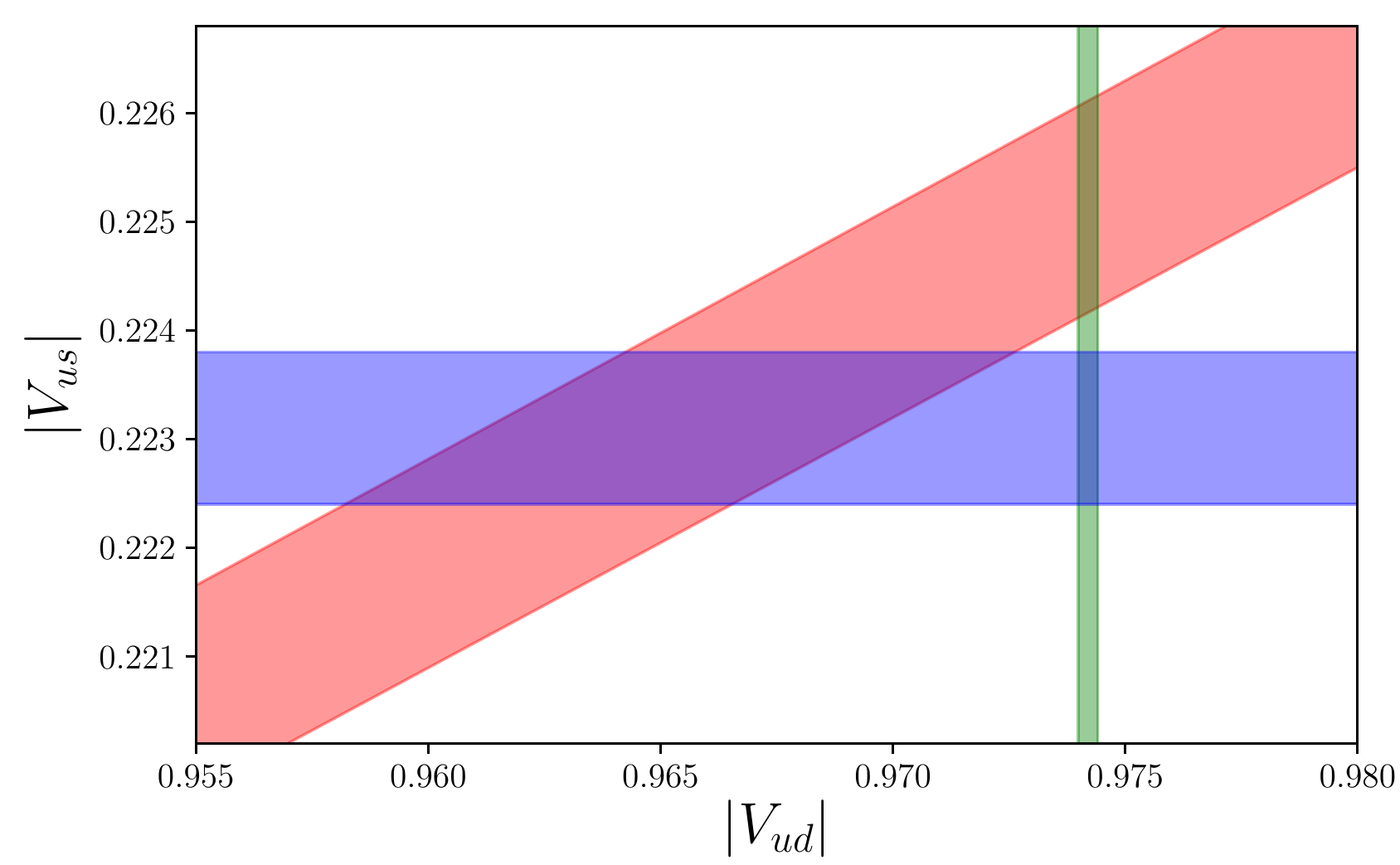}
\caption{\label{fig:vus_vud}
Result for the ratio of CKM matrix elements, $|V_{us}|/|V_{ud}|$, extracted from the ratio $F_K/F_{\pi}$ reported in this work (red band). The global lattice value for $|V_{us}|$ extracted from a semileptonic decay form factor, $f_{+}(0)$~\cite{Aoki:2019cca}, is shown as a horizontal blue band, while the global experimental average for $|V_{ud}|$ from nuclear beta decay~\cite{Tanabashi:2018oca}  is given as a vertical green band. Note that the intersection between the red and green bands agrees well with the unitarity constraint for the CKM matrix, while the intersection between the red and blue bands shows $\sim 2\sigma$ tension.}
\end{figure}

The ratio $F_K/F_\pi$ may be used, in combination with experimental input for leptonic decay widths, to make a prediction for the ratio of CKM matrix elements, $|V_{us}|/|V_{ud}|$. Using the most recent data, \eqnref{eq:vus_vud} becomes~\cite{Tanabashi:2018oca}
\begin{eqnarray}
\frac{|V_{us}|}{|V_{ud}|} \frac{F_{\hat{K}^+}}{F_{\hat{\pi}^+}} = 0.2760(4) \ ,
\end{eqnarray}
where strong isospin breaking effects must be included for direct comparison with experimental data. Combining this expression with our final result, we find
\begin{eqnarray}
\frac{|V_{us}|}{|V_{ud}|} = 0.2311(10) \ .
\end{eqnarray}

Utilizing the current global average, $|V_{ud}| = 0.97420(21)$, extracted from superallowed nuclear beta decays~\cite{Tanabashi:2018oca} results in
\begin{eqnarray}
|V_{us}| = 0.2251(10) \ .
\end{eqnarray}
Finally, we may use our results, combined with the value $|V_{ub}| = \left(3.94(36)\right) \times 10^{-3}$, as a test of unitarity for the CKM matrix, which states that $|V_{ud}|^2 + |V_{us}|^2 + |V_{ub}|^2 = 1$. From our calculation we find
\begin{eqnarray}
|V_{ud}|^2 + |V_{us}|^2 + |V_{ub}|^2 = 0.99977(59) \ .
\end{eqnarray}

Alternatively, rather than using the experimental determination of $|V_{ud}|$ as input for our test of unitarity, we may instead use the global lattice average for $|V_{us}| =0.2231(7)$~\cite{Aoki:2019cca}, extracted via the quantity $f_{+}(0)$, the zero momentum transfer limit of a form factor relevant for the semileptonic decay $K^0 \to \pi^{-} l \nu$. This leads to
\begin{eqnarray}
|V_{ud}|^2 + |V_{us}|^2 + |V_{ub}|^2 = 0.9812(95) \ ,
\end{eqnarray}
leading to a roughly $2\sigma$ tension with unitarity.
Our result, along with with the reported experimental results for $|V_{ud}|$ and lattice results for $|V_{us}|$, are shown in Fig.~\ref{fig:vus_vud}.
One could also combine our results with the more precise average in the FLAG review which would lead to a slight reduction their reported uncertainties, but we will leave that to the FLAG Collaboration in their next update.

\begin{figure*}
\includegraphics[width=0.4\textwidth,valign=t]{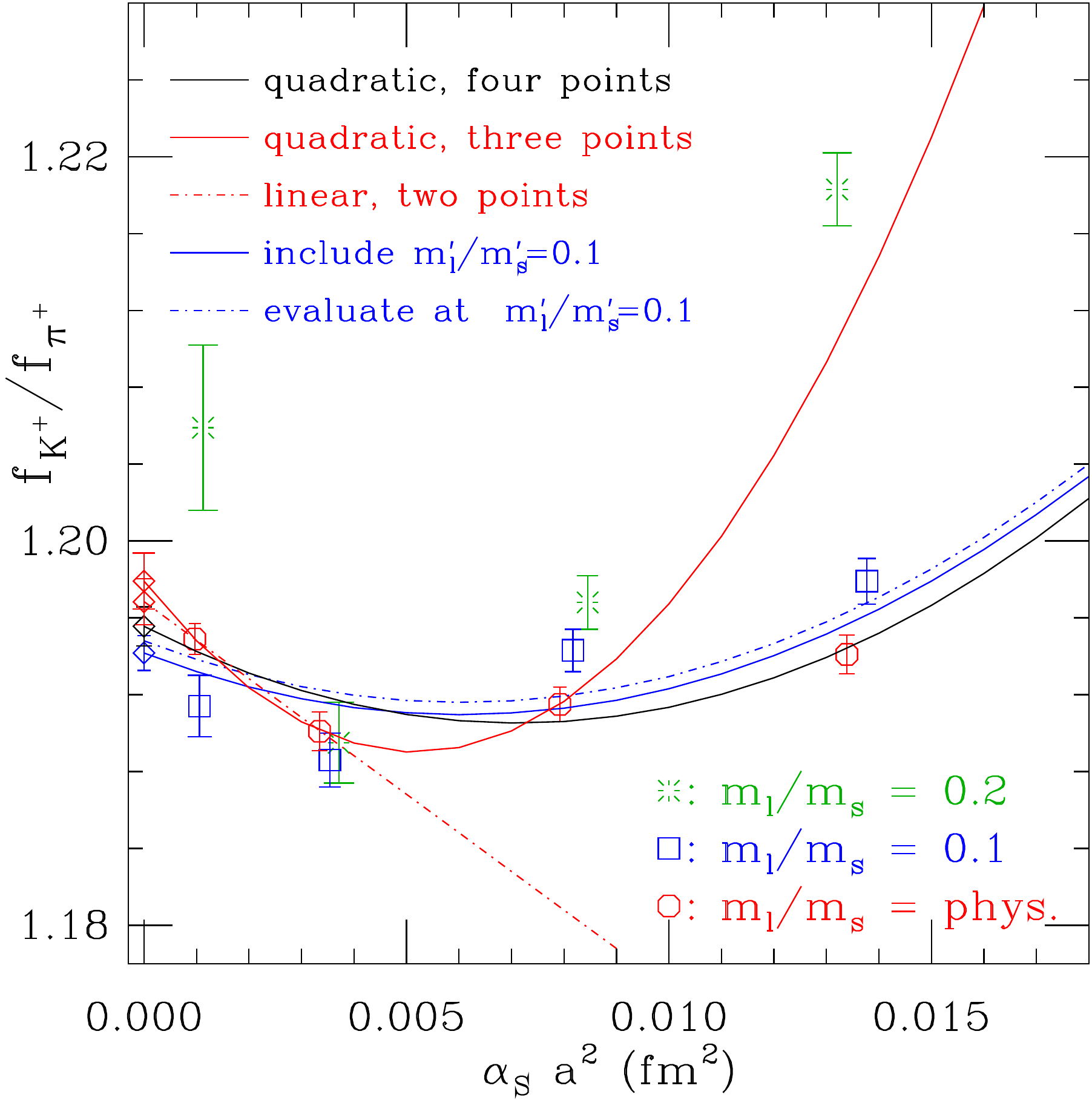}
\includegraphics[width=0.4\textwidth,valign=t]{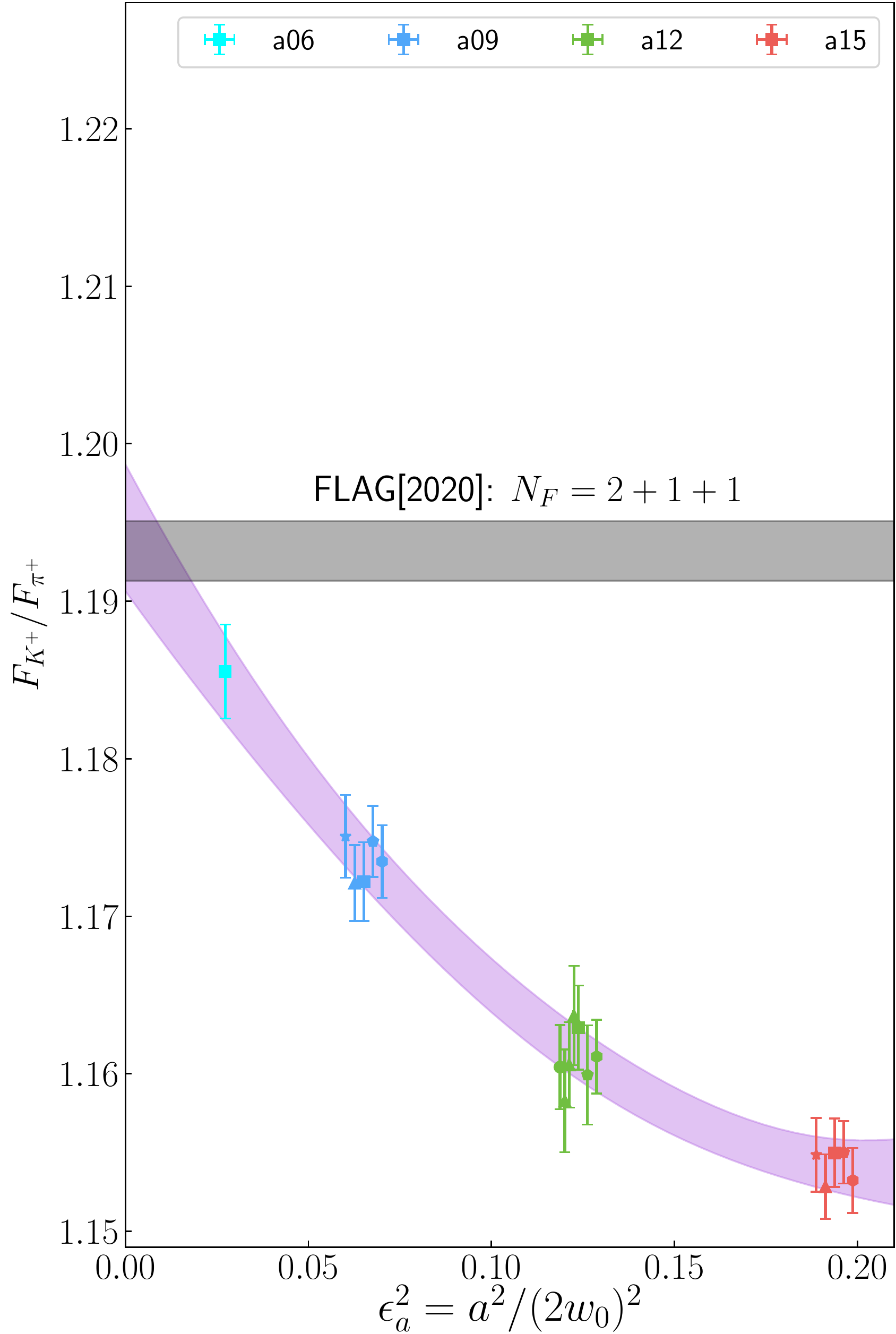}
\caption{\label{fig:continuum_comparison}
Comparison of the continuum extrapolation from the MILC[2014]~\cite{Bazavov:2014wgs} result (LEFT) with the continuum extrapolation in the present work with the MDWF on gradient-flowed HISQ action (right).  In the right plot, we also include the FLAG[2020] average value from the $N_f=2+1+1$ calculations~\cite{Aoki:2019cca}.
While the a06m310L ensemble is not necessary for us to extrapolate to a consistent value as this FLAG average (see \figref{fig:impact_a06}), the overall size of our discretization effects are larger.  This is not necessarily surprising as the HISQ action used by MILC has perturbatively removed all $\mathrm{O}(a^2)$ corrections such that the leading scaling violations begin at $\mathrm{O}(\a_S a^2)$, as implied by the $x$-axis of the left plot.
See Footnote~\ref{fn:symanzik} for a comment on the leading scaling violations~\cite{Balog:2009yj,Balog:2009np,Husung:2019ytz}.
}
\end{figure*}

Another motivation for this work was to precisely test (below 1\%) whether the action we have used for our nucleon structure calculations~\cite{Berkowitz:2017gql,Chang:2018uxx,Berkowitz:2018gqe} can be used to reproduce an accepted value from other lattice calculations that are known at the subpercent level.
Our result provides the first subpercent cross-check of the universality of the continuum limit of this quantity with $N_f=2+1+1$ dynamical flavors, albeit with the same sea-quark action as used by MILC/FNAL and HPQCD~\cite{Dowdall:2013rya,Bazavov:2017lyh}.

Critical in obtaining a subpercent determination of any quantity is control over the continuum extrapolation.
This is relevant to our pursuit of a subpercent determination of $g_A$ as another calculation, utilizing many of the same HISQ ensembles but with a different valence action (clover fermions), obtains a result that is in tension with our own~\cite{Bhattacharya:2016zcn,Gupta:2018qil}.
While there has been speculation that this discrepancy is due to the continuum extrapolations~\cite{Gupta:2018qil}, new work suggests the original work underestimated the systematic uncertainty in the correlation function analysis, and when accounted for, the tension between our results goes away~\cite{Jang:2019vkm}.

In either case, to obtain a subpercent determination of $g_A$, which is relevant for trying to shed light on the neutron lifetime discrepancy~\cite{Czarnecki:2018okw}, it is important to understand the scaling violations of our lattice action.  While a smooth continuum extrapolation in one observable does not guarantee such a smooth extrapolation in another, it at least provides some reassurance of a well-behaved continuum extrapolation.
Furthermore, the determination of $F_K/F_\pi$ involves the same axial current that is relevant for the computation of the nucleon matrix element used to compute $g_A$.

\figref{fig:impact_a06} shows the continuum extrapolation of $F_K/F_\pi$ from our analysis.
The size of the discretization effects are noticeably larger than we observed in our calculation of $g_A$~\cite{Chang:2018uxx}.
In \secref{sec:a06m310L}, we demonstrated that, while helpful, the a06m310L ensemble is not necessary to achieve a subpercent determination of $F_K/F_\pi$.
This is in contrast to the determination by MILC which requires the $a\sim0.06$~fm (or smaller) lattice spacings to control the continuum extrapolation (though we note, the HPQCD calculation~\cite{Dowdall:2013rya}, also performed on the HISQ ensembles, does not utilize the $a\sim0.06$~fm ensembles but agrees with the MILC result).
It should be noted, the MILC result does not rely on the heavier mass ensembles except to adjust for the slight mistuning of the input quark masses on their near-physical point ensembles.
In \figref{fig:continuum_comparison}, we compare our continuum extrapolation to that of MILC~\cite{Bazavov:2014wgs}.

In Ref.~\cite{Bazavov:2014wgs}, they also utilize the same four lattice spacings as in this work (they have subsequently improved their determination with an additional two finer lattice spacings~\cite{Bazavov:2017lyh}.)
A strong competition between the $\mathrm{O}(a^2)$ and $\mathrm{O}(a^4)$ corrections was observed in that work, such that the $a \sim .06$~fm ensemble is much more instrumental for a reliable continuum extrapolation than is the case in our setup.
At the same time, the overall scale of their discretization effects is much smaller than we observe in the MDWF on gradient-flowed HISQ action for this quantity.  This is not entirely surprising as the HISQ action has been tuned to perturbatively remove all $\mathrm{O}(a^2)$ corrections such that the leading corrections formally begin as $\mathrm{O}(\a_S a^2)$.

The analysis and supporting data for this article are openly available~\cite{fkfpi:source}.

\begin{acknowledgments}
We would like to thank V. Cirigliano, S. Simula, J. Simone, and T. Kaneko for helpful correspondence and discussions regarding the strong isospin breaking corrections to $F_K/F_\pi$.
We would like to thank J. Bijnens for helpful correspondence on $\chi$PT and a \texttt{C++} interface to \texttt{CHIRON}~\cite{Bijnens:2014gsa} that we used for the analysis presented in this work.
We thank the MILC Collaboration for providing some of the HISQ configurations used in this work, and A. Bazavov, C. Detar and D. Toussaint for guidance on using their code to generate the new HISQ ensembles also used in this work.
We would like to thank P.~Lepage for enhancements to \texttt{gvar}~\cite{gvar:11.2} and \texttt{lsqfit}~\cite{lsqfit:11.5.1} that enable the pickling of \texttt{lsqfit.nonlinear\_fit} objects.  We also thank C.~Bernard for useful correspondence concerning higher-order extrapolation analysis and R.~Sommer for comments on the leading asymptotic scaling violations.

Computing time for this work was provided through the Innovative and Novel Computational Impact on Theory and Experiment (INCITE) program and the LLNL Multiprogrammatic and Institutional Computing program for Grand Challenge allocations on the LLNL supercomputers.
This research utilized the  NVIDIA GPU-accelerated Titan and Summit supercomputers at Oak Ridge Leadership Computing Facility at the Oak Ridge National Laboratory, which is supported by the Office of Science of the U.S. Department of Energy under Contract No. DE-AC05-00OR22725 as well as the Surface, RZHasGPU, Pascal, Lassen, and Sierra supercomputers at Lawrence Livermore National Laboratory.

The computations were performed utilizing \texttt{LALIBE}~\cite{lalibe} which utilizes the \texttt{Chroma} software suite~\cite{Edwards:2004sx} with \texttt{QUDA} solvers~\cite{Clark:2009wm,Babich:2011np} and HDF5~\cite{hdf5} for I/O~\cite{Kurth:2015mqa}.  They were efficiently managed with \texttt{METAQ}~\cite{Berkowitz:2017vcp,Berkowitz:2017xna} and status of tasks logged with EspressoDB~\cite{Chang:2019khk}.
The hybrid Monte Carlo was performed with the MILC Code~\cite{milc:code}, and for the ensembles new in this work, running on GPUs using \texttt{QUDA}.  The final extrapolation analysis utilized \texttt{gvar} v11.2~\cite{gvar:11.2} and \texttt{lsqfit} v11.5.1~\cite{lsqfit:11.5.1} and \texttt{CHIRON} v0.54~\cite{Bijnens:2014gsa}.

This work was supported by the NVIDIA Corporation (M.A.C.), the Alexander von Humboldt Foundation through a Feodor Lynen Research Fellowship (C.K.), the DFG and the NSFC Sino-German CRC110 (E.B.), the RIKEN Special Postdoctoral Researcher Program (E.R.), the U.S. Department of Energy, Office of Science, Office of Nuclear Physics under Award No.~DE-AC02-05CH11231 (C.C.C., C.K., B.H., A.W.L.), No.~DE-AC52-07NA27344 (D.A.B., D.H, A.S.G., P.V), No.~DE-FG02-93ER-40762 (E.B), No.~DE-AC05-06OR23177 (B.J., C.M., K.O.), No.~DE-FG02-04ER41302 (K.O.); the Office of Advanced Scientific Computing (B.J.); the Nuclear Physics Double Beta Decay Topical Collaboration (D.A.B., H.M.C., A.N., A.W.L.); and the DOE Early Career Award Program (C.C.C., A.W.L.).

\end{acknowledgments}

\appendix
\begin{widetext}

\section{MODELS INCLUDED IN FINAL ANALYSIS \label{sec:bma_models}}
We list the models that have entered the final analysis as described in \secref{sec:full_analysis} and listed in \tabref{tab:bma_results}.
For example, the model
\begin{equation*}
    \text{xpt-ratio\_nnnlo\_FV\_alphaS\_PP}
\end{equation*}
indicates the model uses the continuum $\chi$PT fit function through \nxlo{3} with discretization corrections added as in Eqs.~\eqref{eq:n2lo_a} and \eqref{eq:n3lo_ct}.
The NLO contributions are kept in a ratio form, \eqnref{eq:fkfpi_nlo_ratio}, and we have included the corresponding \nxlo{2} ratio correction $\d_{\rm ratio}^\text{\nxlo{2}}$.
The finite volume corrections have been included at NLO.  The discretization terms at \nxlo{2} include the $\a_S \e_a^2 (\e_K^2-\e_\pi^2)$ counterterm.  The renormalization scale appearing in the logs is $\mu=4\pi F_\pi$ as indicated by \texttt{\_PP}, and we have included the corresponding \nxlo{2} correction $\d_{F_\pi}^\text{\nxlo{2}}$, \eqnref{eq:n2lo_xpt_scale}, to hold the actual renormalization scale fixed at $\mu_0=4\pi F_0$.

When \texttt{\_ct} appears in the model name, the only \nxlo{2} terms that are added are from the local counterterms while all chiral log corrections are set to zero.

\begin{table*}
\caption{\label{tab:bma_results}
List of models used in final result, as described in the text.
}
\begin{ruledtabular}
\begin{tabular}{r|ccccl}
    Model & $\chi^2_\nu$ &   $Q$ &  \texttt{logGBF} & weight& $F_K/F_\pi$\\
\hline
        xpt-ratio\_nnnlo\_FV\_ct\_PP &  0.847   &  0.645&  77.728&  0.273&  1.1968(40)\\
xpt-ratio\_nnnlo\_FV\_alphaS\_ct\_PP &  0.843   &  0.650&  77.551&  0.229&  1.1962(46)\\
              xpt\_nnnlo\_FV\_ct\_PP &  0.908   &  0.569&  76.830&  0.111&  1.1974(40)\\
      xpt\_nnnlo\_FV\_alphaS\_ct\_PP &  0.902   &  0.576&  76.668&  0.095&  1.1966(46)\\
        xpt-ratio\_nnnlo\_FV\_ct\_PK &  1.014   &  0.439&  76.343&  0.068&  1.1952(37)\\
xpt-ratio\_nnnlo\_FV\_alphaS\_ct\_PK &  1.006   &  0.449&  76.234&  0.061&  1.1944(42)\\
                  xpt\_nnnlo\_FV\_PP &  0.949   &  0.517&  75.371&  0.026&  1.1989(40)\\
          xpt\_nnnlo\_FV\_alphaS\_PP &  0.946   &  0.522&  75.196&  0.022&  1.1983(46)\\
              xpt\_nnnlo\_FV\_ct\_PK &  1.135   &  0.309&  75.084&  0.019&  1.1950(36)\\
      xpt\_nnnlo\_FV\_alphaS\_ct\_PK &  1.123   &  0.321&  75.007&  0.018&  1.1941(41)\\
            xpt-ratio\_nnnlo\_FV\_PP &  1.014   &  0.439&  74.765&  0.014&  1.1987(40)\\
    xpt-ratio\_nnnlo\_FV\_alphaS\_PP &  1.009   &  0.445&  74.599&  0.012&  1.1980(46)\\
                  xpt\_nnnlo\_FV\_PK &  1.100   &  0.344&  74.421&  0.010&  1.1969(37)\\
          xpt\_nnnlo\_FV\_alphaS\_PK &  1.093   &  0.352&  74.306&  0.009&  1.1962(42)\\
        xpt-ratio\_nnnlo\_FV\_ct\_KK &  1.262   &  0.202&  74.014&  0.007&  1.1920(36)\\
xpt-ratio\_nnnlo\_FV\_alphaS\_ct\_KK &  1.244   &  0.215&  74.004&  0.007&  1.1912(39)\\
            xpt-ratio\_nnnlo\_FV\_PK &  1.159   &  0.286&  73.880&  0.006&  1.1967(37)\\
    xpt-ratio\_nnnlo\_FV\_alphaS\_PK &  1.150   &  0.295&  73.780&  0.005&  1.1959(41)\\
                  xpt\_nnnlo\_FV\_KK &  1.288   &  0.184&  72.757&  0.002&  1.1938(36)\\
          xpt\_nnnlo\_FV\_alphaS\_KK &  1.273   &  0.194&  72.718&  0.002&  1.1930(40)\\
            xpt-ratio\_nnnlo\_FV\_KK &  1.338   &  0.152&  72.348&  0.001&  1.1938(36)\\
    xpt-ratio\_nnnlo\_FV\_alphaS\_KK &  1.322   &  0.162&  72.323&  0.001&  1.1929(39)\\
      xpt\_nnnlo\_FV\_alphaS\_ct\_KK &  1.536   &  0.068&  71.459&  0.001&  1.1900(38)\\
              xpt\_nnnlo\_FV\_ct\_KK &  1.558   &  0.061&  71.430&  0.001&  1.1909(35)\\\hline
Bayes Model Average         &          &       &        &       &  1.1964(42)(12)
\end{tabular}
\end{ruledtabular}
\end{table*}

\section{NLO MIXED ACTION FORMULAS \label{sec:ma_expressions}}
The expression for $d\ell_\pi$ arises from the integral
\begin{align}
d\ell_\pi &= \int_R \frac{d^d k}{(2\pi)^d} \frac{i}{(k^2 - m_\pi^2)^2}
    = \frac{1 + \ln(m_\pi^2 / \mu^2)}{(4\pi)^2}\, ,
\end{align}
which has been regulated and renormalized with the standard $\chi$PT modified dimensional-regularization scheme~\cite{Gasser:1983yg}.
The finite volume corrections to $\d \ell_\pi$ are given by
\begin{equation}
\d^{\rm FV} d\ell_\pi =
    \sum_{|\mathbf{n}|\neq0} \frac{c_n}{(4\pi)^2} \bigg[
    \frac{2K_1(mL|\mathbf{n}|)}{mL|\mathbf{n}|}
    -K_0(mL|\mathbf{n}|)
    -K_2(mL|\mathbf{n}|)
    \bigg]
\end{equation}
The expression for $\mc{K}_{\phi_1 \phi_2}$ arises from the integral
\begin{align}
\mc{K}_{\phi_1 \phi_2} &= (4\pi)^2\int_R \frac{d^d k}{(2\pi)^d}
    \frac{i}{(k^2 -m_{\phi_1}^2)(k^2 -m_{\phi_2}^2)}
    =\frac{\ell_{\phi_2} - \ell_{\phi_1}}{\e_{\phi_2}^2 - \e_{\phi_1}^2}\, .
\end{align}
Similarly, $\mc{K}_{\phi_1 \phi_2}^{(2,1)}$ is given by
\begin{align}
\mc{K}_{\phi_1 \phi_2}^{(2,1)} &= \int_R \frac{d^d k}{(2\pi)^d}
    \frac{i (4\pi)^2 (4\pi F)^2}{(k^2 -m_{\phi_1}^2)^2 (k^2 - m_{\phi_2}^2)}
    =\frac{\ell_{\phi_2} - \ell_{\phi_1}}{(\e_{\phi_2}^2 - \e_{\phi_1}^2)^2}
    -\frac{d \ell_{\phi_1}}{\e_{\phi_2}^2 - \e_{\phi_1}^2}\, .
\end{align}
Finally, $\mc{K}_{\phi_1 \phi_2 \phi_3}$ is given by
\begin{align}
\mc{K}_{\phi_1 \phi_2 \phi_3} &= \int_R \frac{d^d k}{(2\pi)^d}
    \frac{i (4\pi)^2 (4\pi F)^2}{(k^2 -m_{\phi_1}^2) (k^2 - m_{\phi_2}^2) (k^2 - m_{\phi_3}^2)}
\nonumber\\&
    =\frac{\ell_{\phi_1}}{(\e_{\phi_1}^2 - \e_{\phi_2}^2)(\e_{\phi_1}^2 - \e_{\phi_3}^2)}
    +\frac{\ell_{\phi_2}}{(\e_{\phi_2}^2 - \e_{\phi_1}^2)(\e_{\phi_2}^2 - \e_{\phi_3}^2)}
        +\frac{\ell_{\phi_3}}{(\e_{\phi_3}^2 - \e_{\phi_1}^2)(\e_{\phi_3}^2 - \e_{\phi_2}^2)}\, .
\end{align}
In each of these expressions, the corresponding expression including FV corrections are given by replacing $\ell_\phi \rightarrow \ell^{\rm FV}_\phi$, \eqnref{eq:ell_P_FV}.

\section{HYBRID MONTE CARLO FOR NEW ENSEMBLES \label{sec:hmc}}
\begin{table}
\caption{\label{tab:hmc_parameters}
Input parameters and measured acceptance rate for the new HISQ ensembles.
In addition to the columns standardly reported by MILC (see Table IV of Ref.~\cite{Bazavov:2012xda}), we list the abbreviated ensemble name, the number of streams $N_{\rm stream}$, and the total number of configurations $N_{\rm cfg}$.  For a given ensemble, each stream has an equal number of configurations.
The gauge coupling, light, strange, and charm quark masses on each ensembles are given as well as the tadpole factor $u_0$ and the Naik-term added to the charm quark action $\e_N$.
Here $s$ denotes the total length in molecular time units (MDTU) between each saved configuration, Len. denotes the length between accept/reject steps (in MDTU), and Acc. denotes the fraction of trajectories accepted.  The microstep size $\e$ used in the HMC is provided as Len./$N_{\rm steps}$ which was input with single precision.  The average acceptance rate over all streams is listed as well as the number of streams.
}
\begin{ruledtabular}
\begin{tabular}{lllllllcccccc}
Ensemble& $10/g^2$& $am_l$& $am_s$& $am_c$& $u_0$& $\e_N$& $s$& Len.& $\e$& Acc.& $N_{\rm stream}$& $N_{\rm cfg}$\\
\hline
a15m135XL& 5.80& 0.002426& 0.06730& 0.8447& 0.85535 & $-$0.35892& 5& 0.2& 0.2/150& 0.631& 4& 2000\\
a09m135  & 6.30& 0.001326& 0.03636& 0.4313& 0.874164& $-$0.11586& 6& 1.5& 1.5/130& 0.693& 2& 1010\\
a06m310L & 6.72& 0.0048  & 0.024  & 0.286 & 0.885773& $-$0.05330& 6& 2.0& 2.0/120& 0.765& 2& 1000
\end{tabular}
\end{ruledtabular}
\end{table}

\begin{figure}
\includegraphics[width=\textwidth,valign=t]{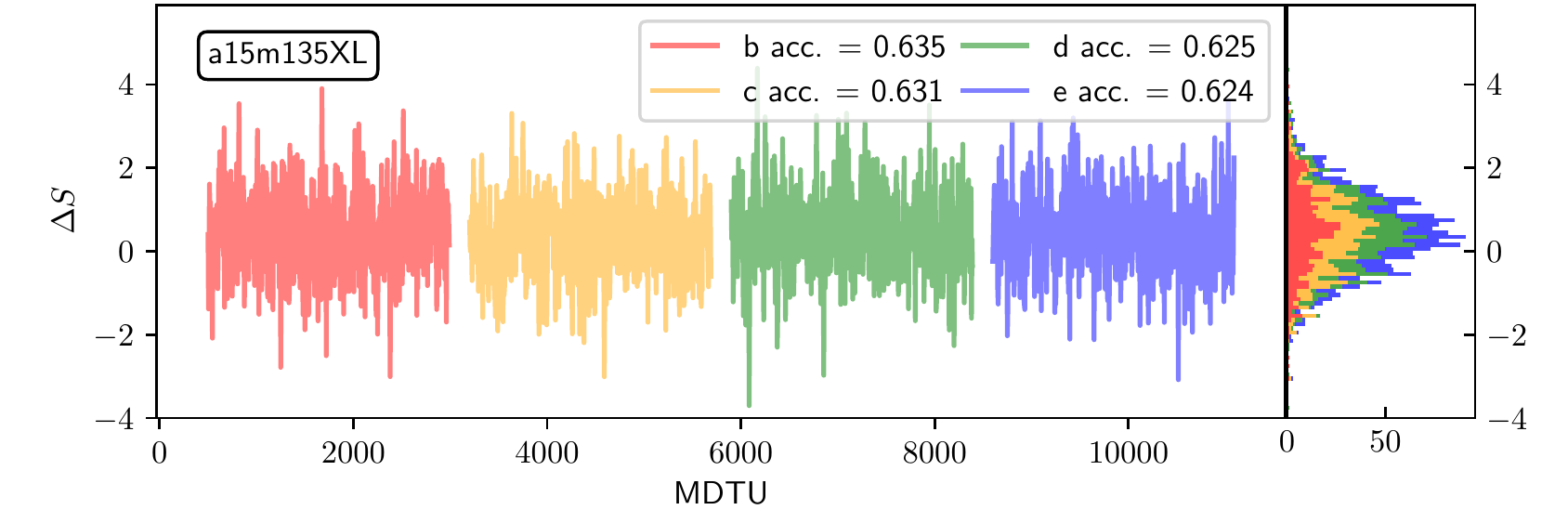}
\includegraphics[width=\textwidth,valign=t]{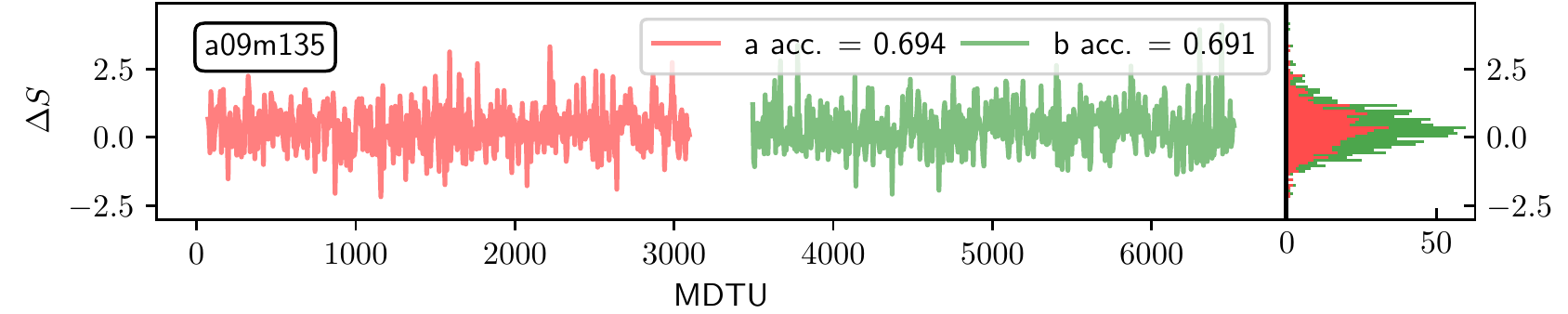}
\includegraphics[width=\textwidth,valign=t]{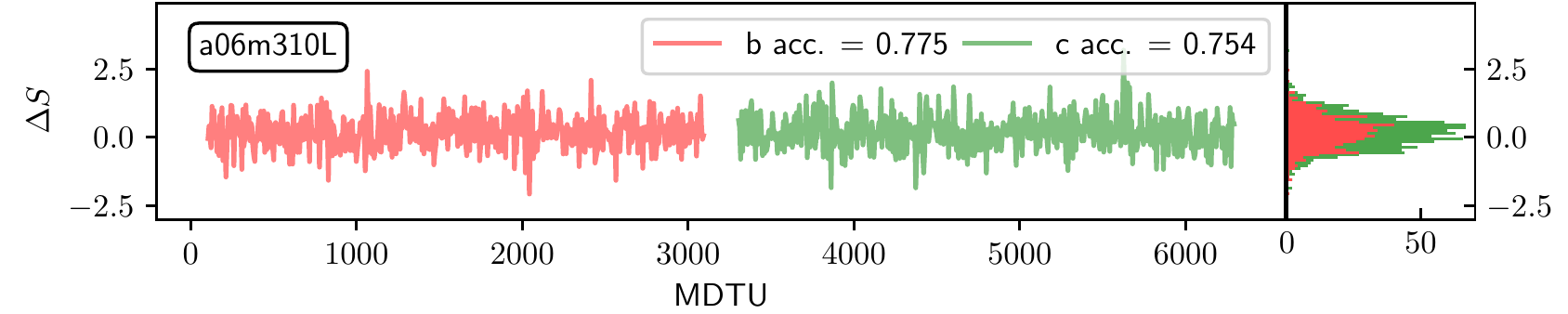}
\caption{\label{fig:deltaS}
The $\Delta S$ values computed in the accept/reject step of the HMC versus MDTU.
The different colors correspond to the different streams which are separated and shifted in MDTU for clarity.
}
\end{figure}

We present various summary information for the three new ensembles used in this work, a06m310L, a15m135XL and a09m135.
In \tabref{tab:hmc_parameters}, we list the parameters of the HISQ ensembles used in the hybrid Monte Carlo (HMC).
In \figref{fig:deltaS}, we show the MDTU history of the $\D S$ for the three ensembles.
For the a15m135XL ensemble, we reduced the trajectory length significantly compared to the a15m130 from MILC to overcome spikes in the HMC force calculations.  To compensate, we lowered the acceptance rate to encourage the HMC to move around parameter space with larger jumps in an attempt to reduce the autocorrelation time.  We ran 25 HMC accept/reject steps before saving a configuration for a total trajectory length of 5.

For each accept/reject step we also measure the quark-antiquark condensate $\bar{\psi}\psi$ using a stochastic estimate with 5 random sources that are averaged together.
We compute it for each of the quark masses $am_l$, $am_s$, and $am_c$.
On the a15m135XL we have measured $\bar{\psi}\psi$ only on every saved configuration for the first half of each stream, while we measured it at each accept/reject step for the second half.
The integrated autocorrelation time, as well as the average and statistical errors of $\bar{\psi}\psi$, are computed using the $\Gamma$-method analysis~\cite{Wolff:2003sm} with the Python package \texttt{unew}~\cite{DePalma:2017lww}.
We report the results in \tabref{tab:pbp_measurements}.
In \figref{fig:pbp_measurements} we report the value of the $\bar{\psi}\psi$ on each saved configuration for the three quark masses on each ensemble.

\begin{table}
\caption{\label{tab:pbp_measurements}
Average values of the quark-antiquark condensate $\bar{\psi}\psi$ with statistical errors and integrated autocorrelation times $\tau$ measured with the $\Gamma$-method analysis.
Each value is averaged over all the available streams (which are all statistically compatible). The integrated autocorrelation time is reported in units of MDTU. The a15m135XL results are obtained from the second half of each stream because we have more measurements.
}
\begin{ruledtabular}
\begin{tabular}{rcrrrrrr}
ensemble& $N_{\rm stream}$& $\bar{\psi}\psi_l$ & $\tau_l$ & $\bar{\psi}\psi_s$ & $\tau_s$ & $\bar{\psi}\psi_c$ & $\tau_c$ \\
\hline
a15m135XL& 4& 0.02390(2)& 11(2) & 0.08928(2)& 71(26) & 0.4800580(5)& $<1$ \\
a09m135  & 2& 0.005761(6) & 9(2) & 0.003935(3) & 32(9) & 0.3205399(5) & 1.5(1) \\
a06m310L & 2& 0.006599(4) & 30(8) & 0.002356(2) & 34(8) & 0.2275664(4) & 2.0(2) \\
\end{tabular}
\end{ruledtabular}
\end{table}

\begin{figure}
\begin{tabular}{ccc}
    $\bar{\psi}\psi_l$ & $\bar{\psi}\psi_s$& $\bar{\psi}\psi_c$\\
\includegraphics[width=0.32\textwidth]{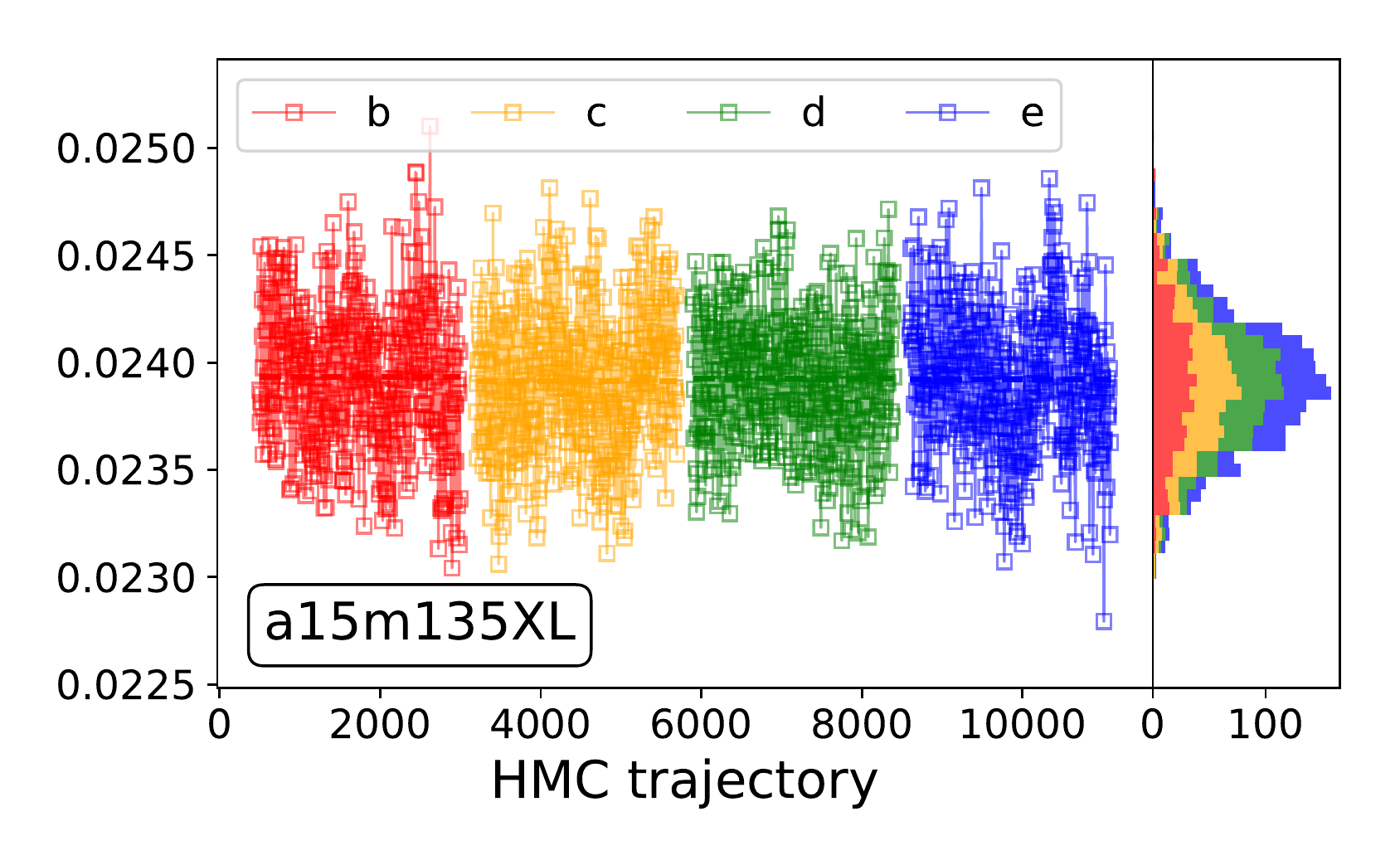}&
\includegraphics[width=0.32\textwidth]{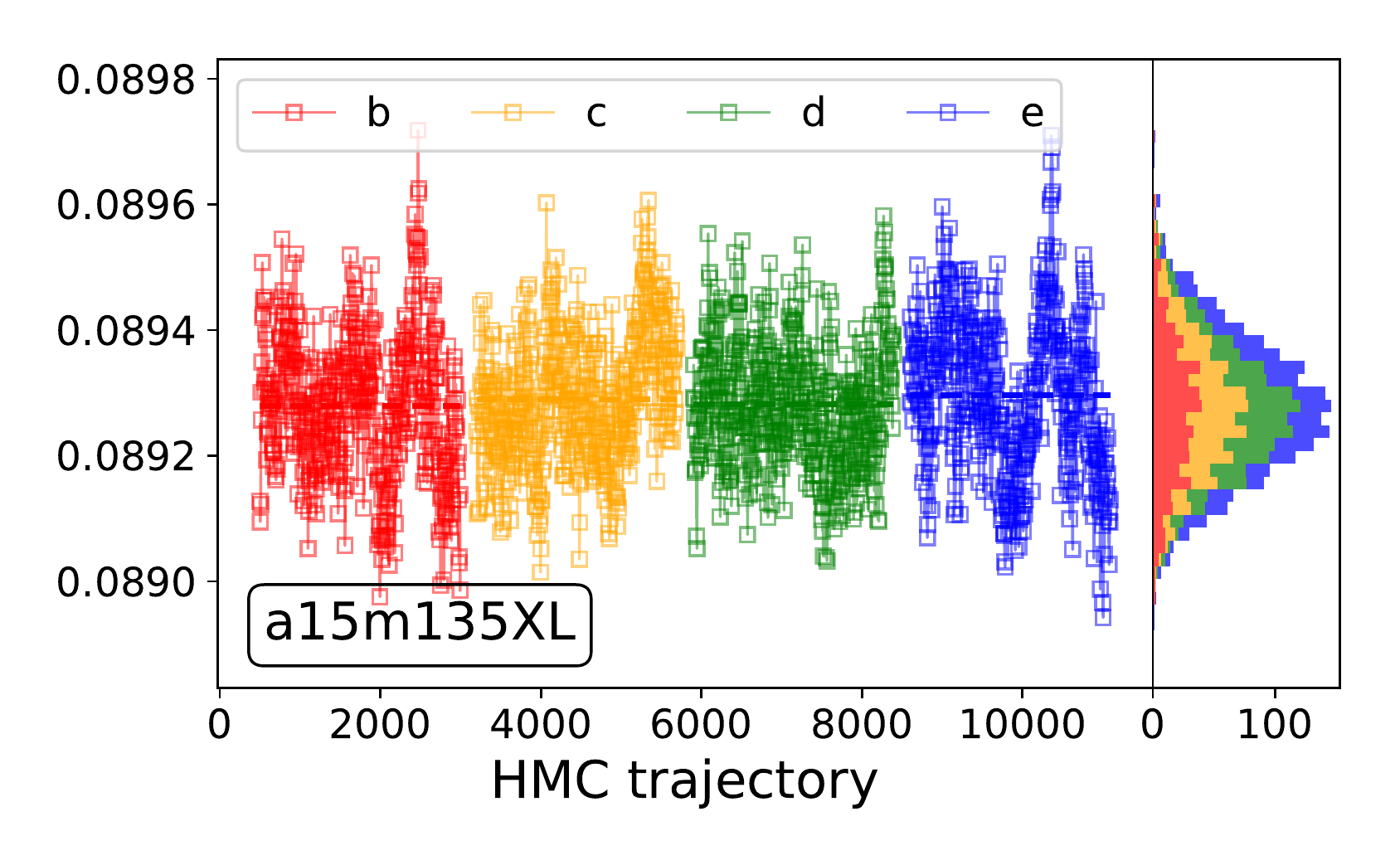}&
\includegraphics[width=0.32\textwidth]{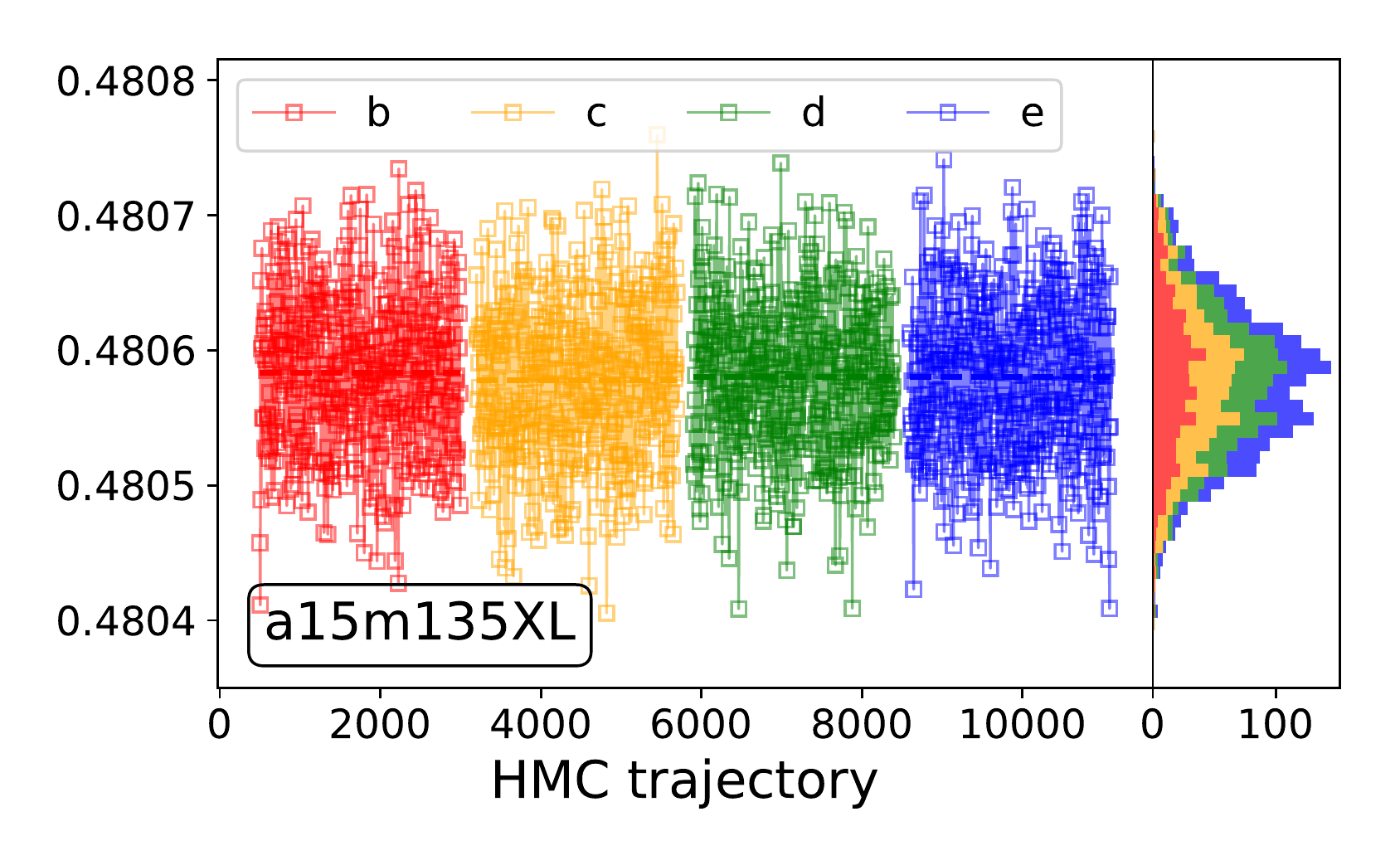}\\
\includegraphics[width=0.32\textwidth]{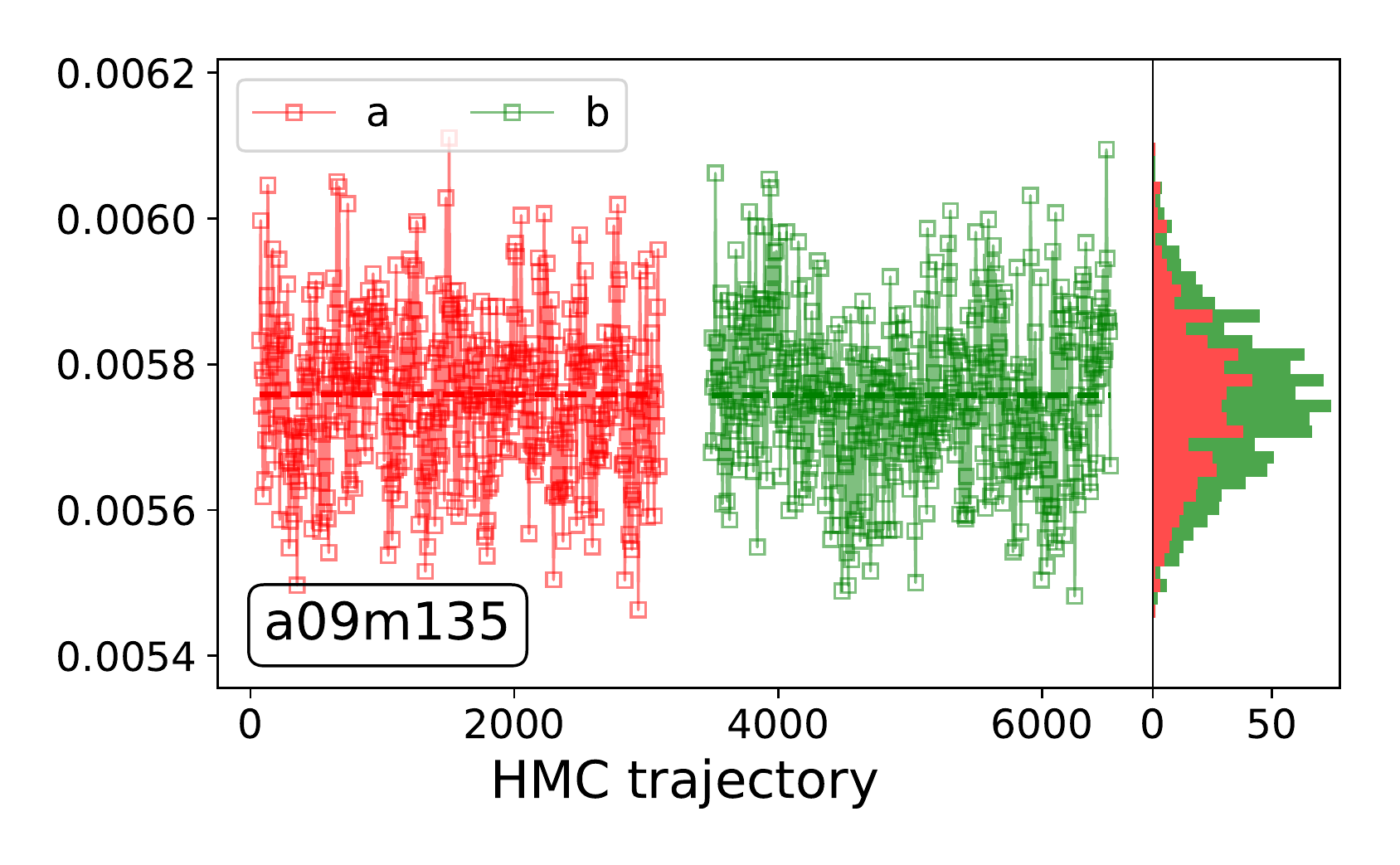}&
\includegraphics[width=0.32\textwidth]{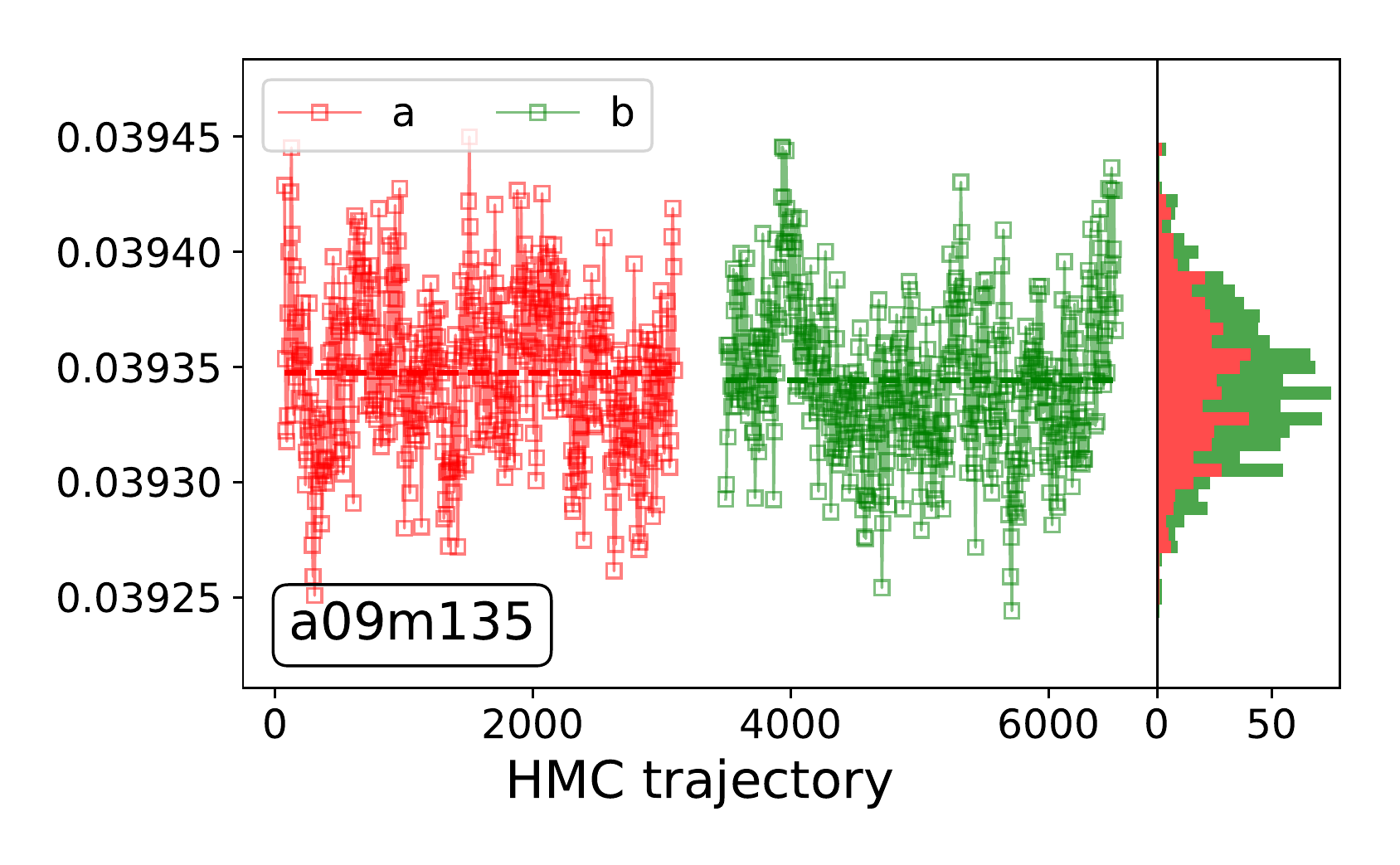}&
\includegraphics[width=0.32\textwidth]{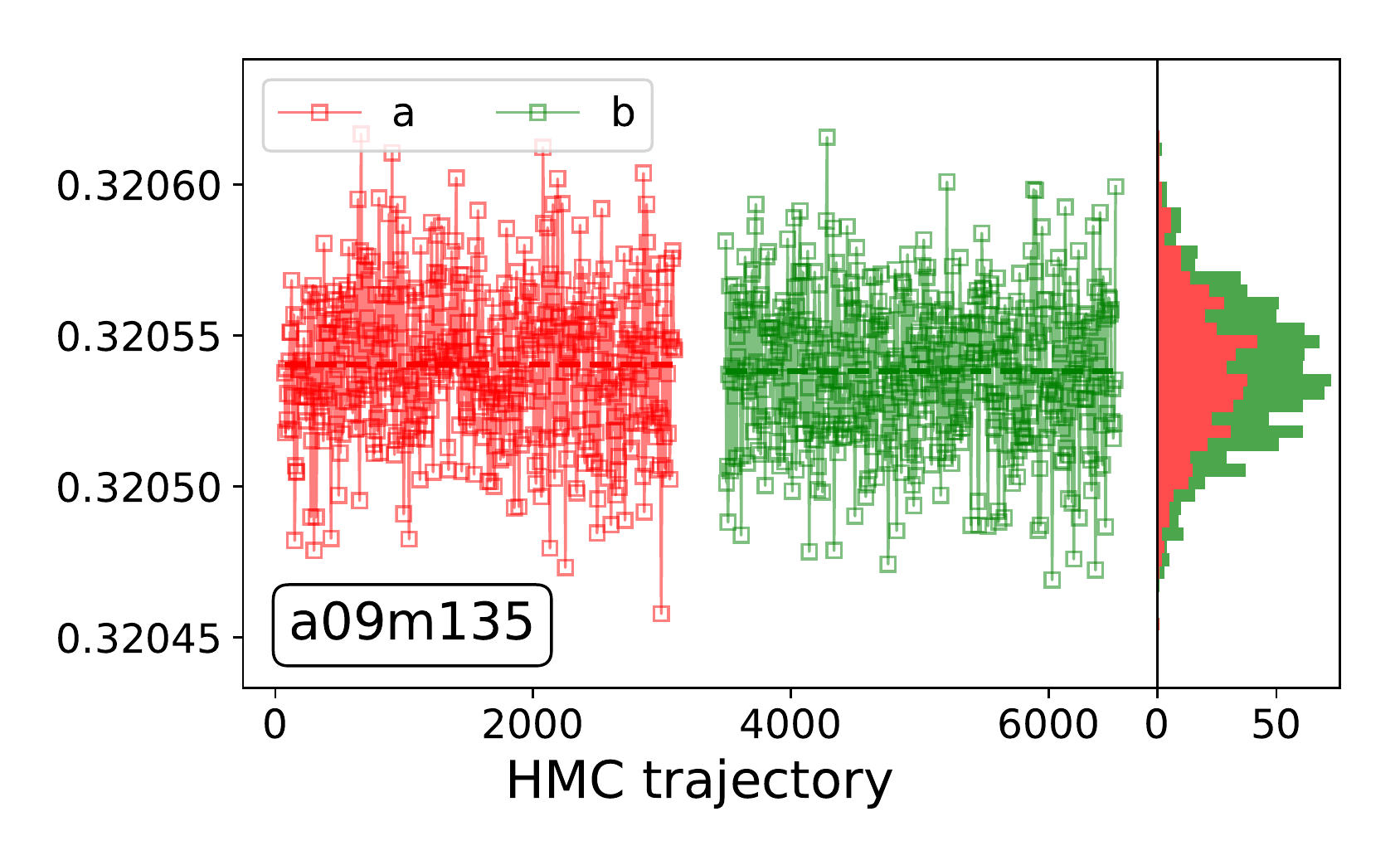}\\
\includegraphics[width=0.32\textwidth]{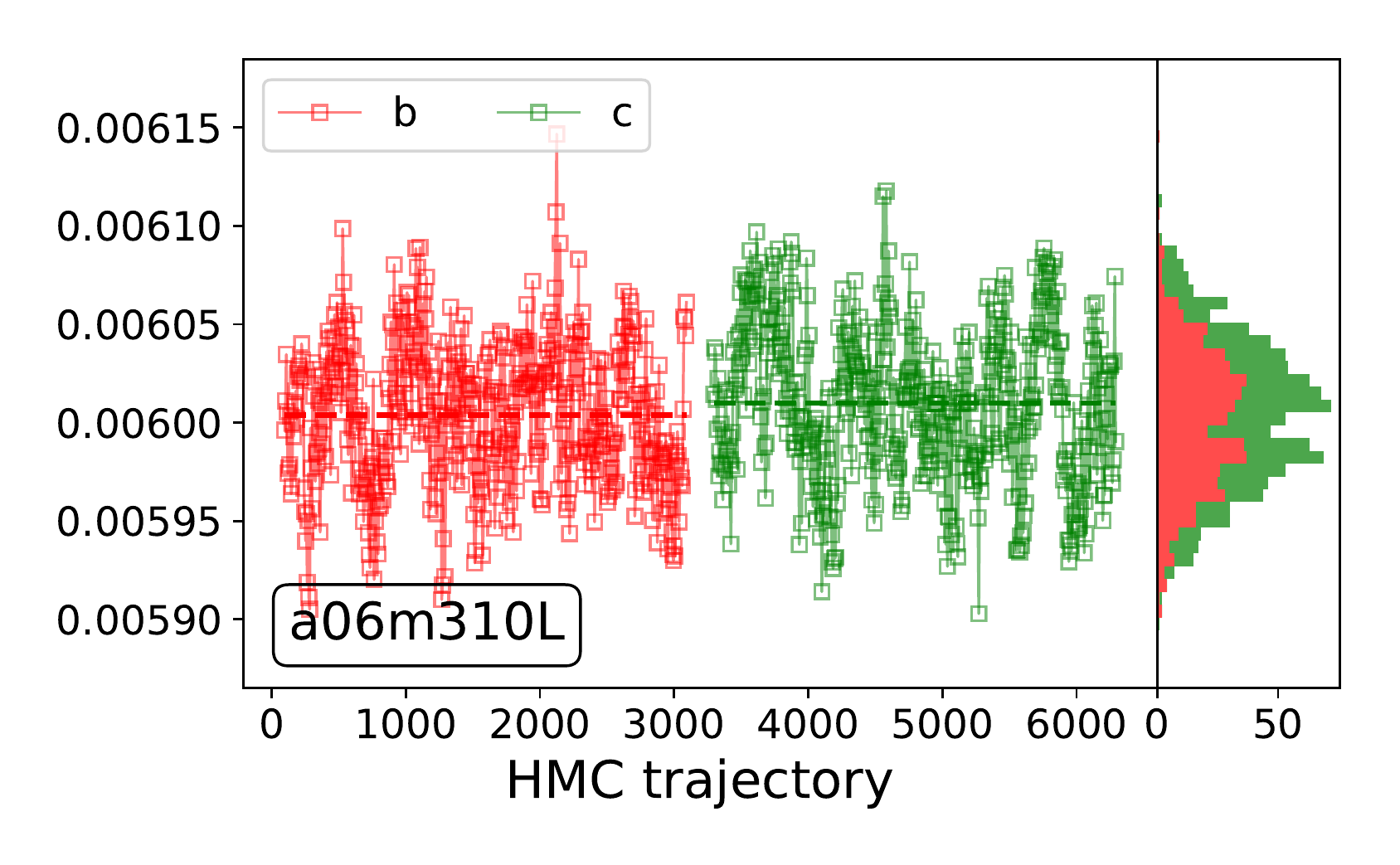}&
\includegraphics[width=0.32\textwidth]{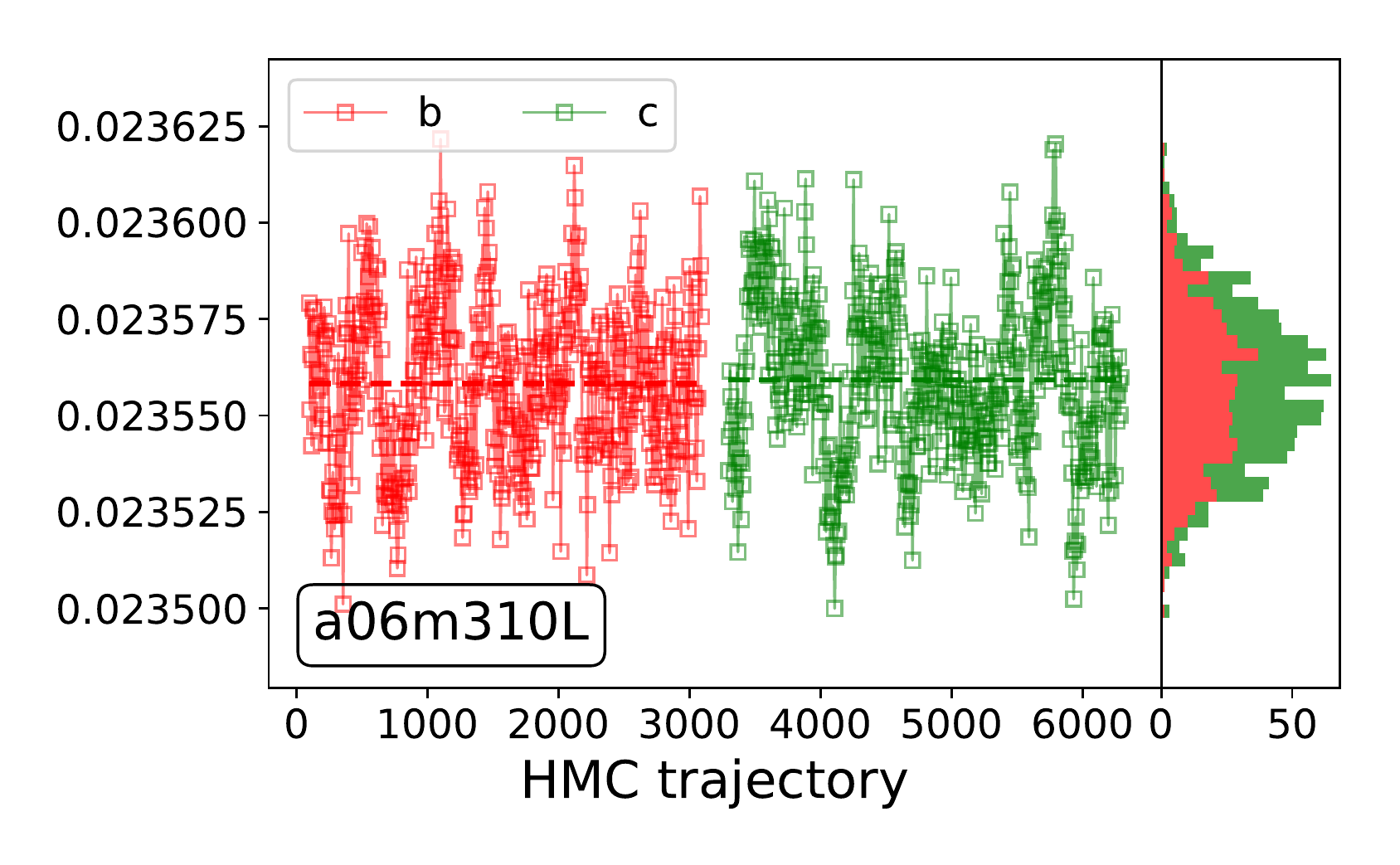}&
\includegraphics[width=0.32\textwidth]{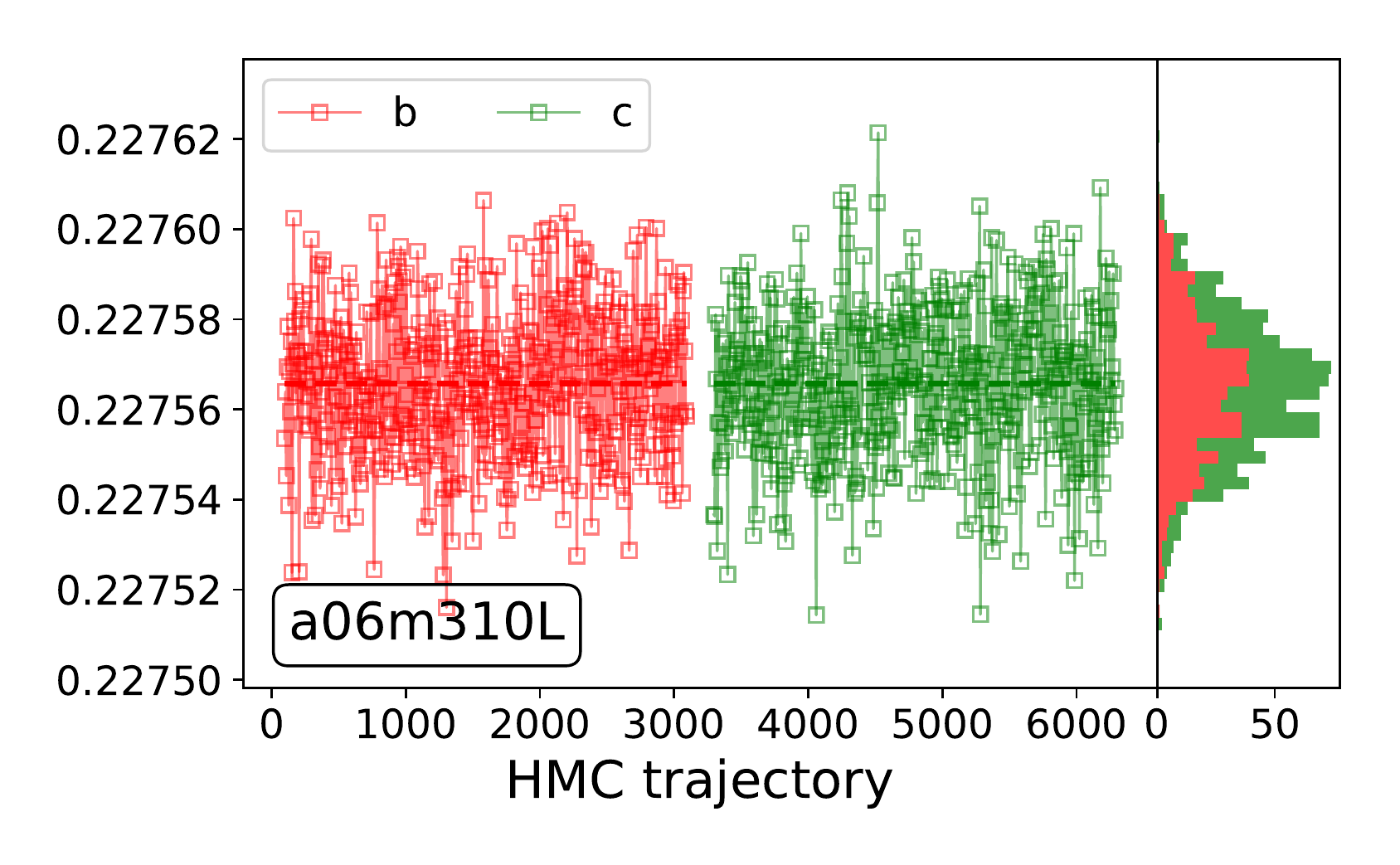}\\
$\bar{\psi}\psi_l$ & $\bar{\psi}\psi_s$& $\bar{\psi}\psi_c$
\end{tabular}
\caption{\label{fig:pbp_measurements}
The quark-antiquark condensate on each configuration of the three ensembles. The different streams are plotted separately for clarity. The plots in each column correspond to the light, strange and charm quark masses, respectively.
}
\end{figure}

Because we observe a long autocorrelation time of the $\langle \bar{\psi}_s\psi_s \rangle$ on the a15m135XL ensemble, we also studied the uncertainty on the extracted pion and kaon effective masses as a function of block size to check for possible longer autocorrelations than usual, with blocking lengths of 10, 25, and 100 MDTU (\figref{fig:a15m135XL_pion_kaon_autocorr}).  We observe that these hadronic quantities have a much shorter autocorrelation time as the uncertainty is independent of $\t_b$ and consistent with the unblocked data.
On this a15m135XL ensemble, while we have generated 2000 configurations, we have only utilized 1000 in this paper (the first half from each of the four streams).

\begin{figure}
\includegraphics[width=\textwidth,valign=t]{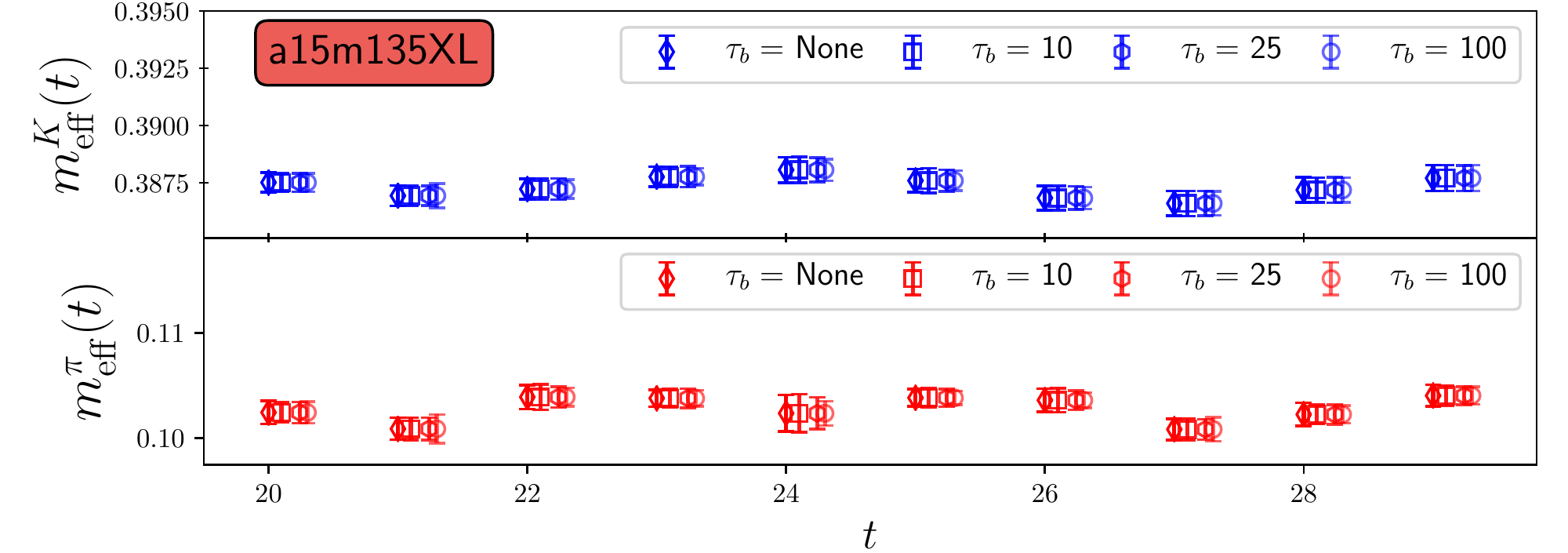}
\caption{\label{fig:a15m135XL_pion_kaon_autocorr}
The effective mass in the mid- to long-time region of the kaon (top) and pion (bottom) on the a15m135XL ensemble are plotted as a function of the blocking time, $\t_b$ in MDTU.  For example, $\t_b=10$ blocks nearest neighbor configurations while $\t_b=100$ is blocking in groups of 20 configurations.  That the uncertainty is independent of $\t_b$ indicates the autocorrelation time for these hadronic quantities is very short.
}
\end{figure}

Finally, in \tabref{tab:Q_measurements}, we list the parameters of the overrelaxed stout smearing used to measure the topological charge $Q$ on each configuration~\cite{Moran:2008ra} and we show the resulting $Q$ distributions in \figref{fig:Q_measurements}.
While the $Q$-distribution on the a06m310L ensemble is less than ideal and the integrated autocorrelation time is long, the volume is sufficiently large ($aL=72a\simeq4.1$~fm) that we do not anticipate any measurable impact from the poorly distributed $Q$-values, which nonetheless average to nearly 0.

\begin{table}
\caption{\label{tab:Q_measurements}
Values of the overrelaxed stout smearing parameters used to measure the topological charge $Q$ and the resulting mean ($\bar{Q}$) and width ($\sigma$) of the the distribution for each stream. The last column reports the integrated autocorrelation time in units of MDTU using the $\Gamma$-method analysis. These measurements were performed with \texttt{QUDA} which is now available via the \texttt{su3\_test} test executable in the \texttt{develop} branch~\cite{Clark:2009wm,Babich:2011np}.
}
\begin{ruledtabular}
\begin{tabular}{rccrrrrrcc}
ensemble& $\rho$& $N_{\rm step}$& \multicolumn{5}{c}{$\bar{Q}_{\rm stream}(\sigma)$}& $\bar{Q}_{\rm all}(\sigma)$ & $\tau_{\rm all}(\sigma)$\\
\cline{4-8}
&&&a& b& c& d& e& & \\
\hline
a15m135XL& $0.068$& 2000& --& -10(34)& -3(35)& -2(33)& 5(32)& -3(33) & 15(3)\\
a09m135  & $0.065$& 2000& 0.5(12.0)& 2(12)& --& --& --& 1(12) & 18(4)\\
a06m310L & $0.066$& 1800& --& 4(12)& -1.2(7.4)& --& --& 1(10) & 420(198)
\end{tabular}
\end{ruledtabular}
\end{table}

\begin{figure}
\includegraphics[width=0.32\textwidth,valign=t]{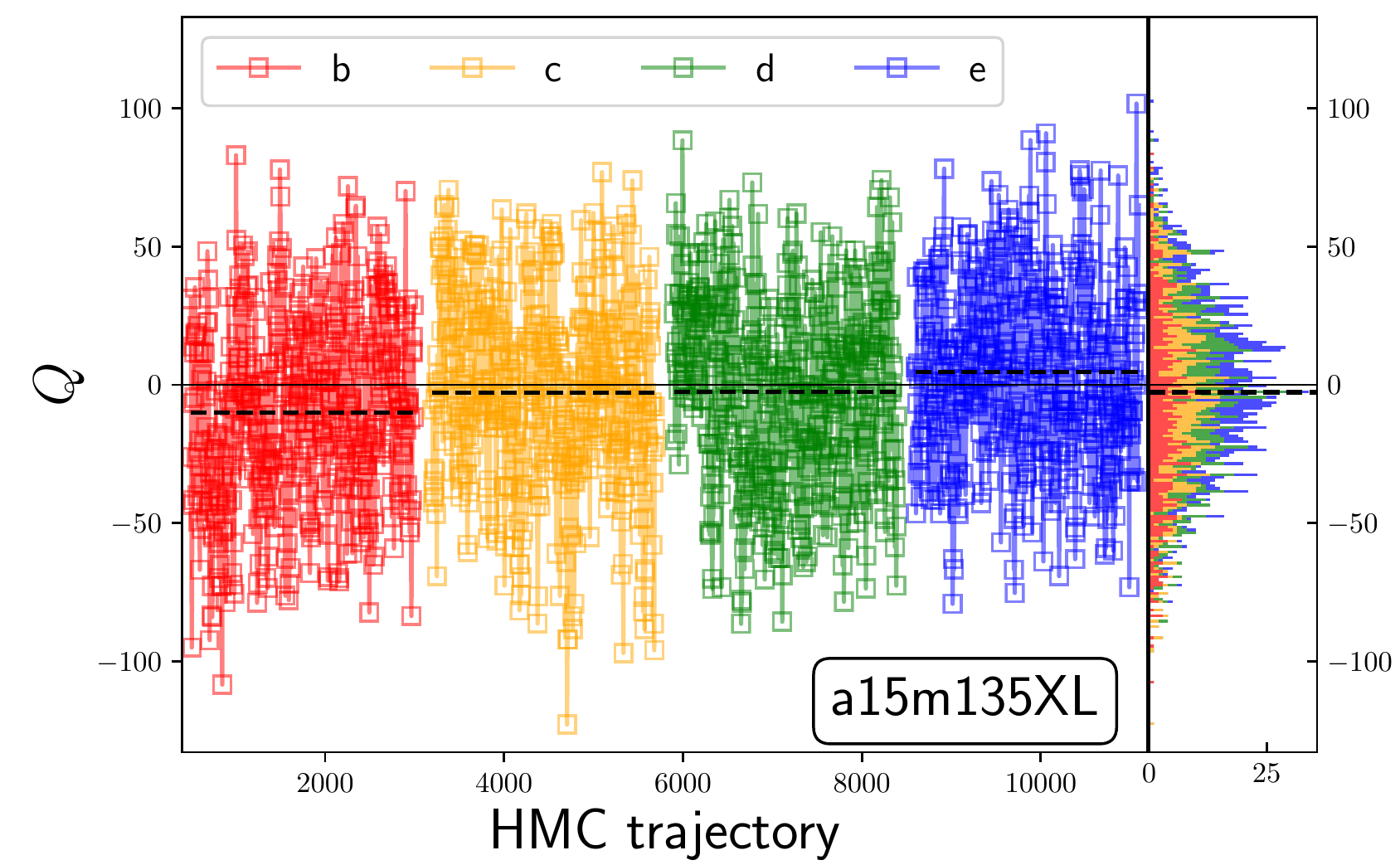}
\includegraphics[width=0.32\textwidth,valign=t]{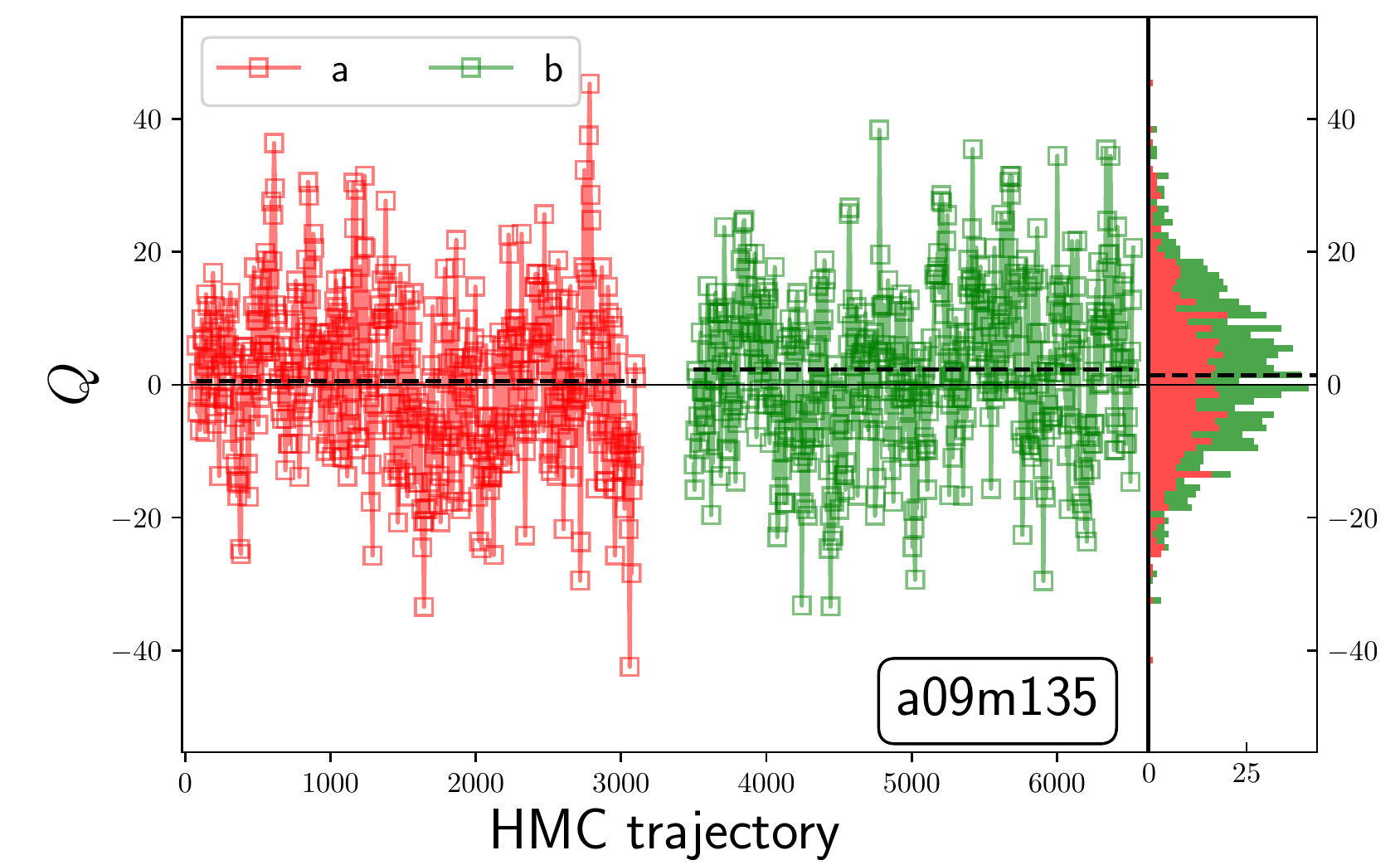}
\includegraphics[width=0.32\textwidth,valign=t]{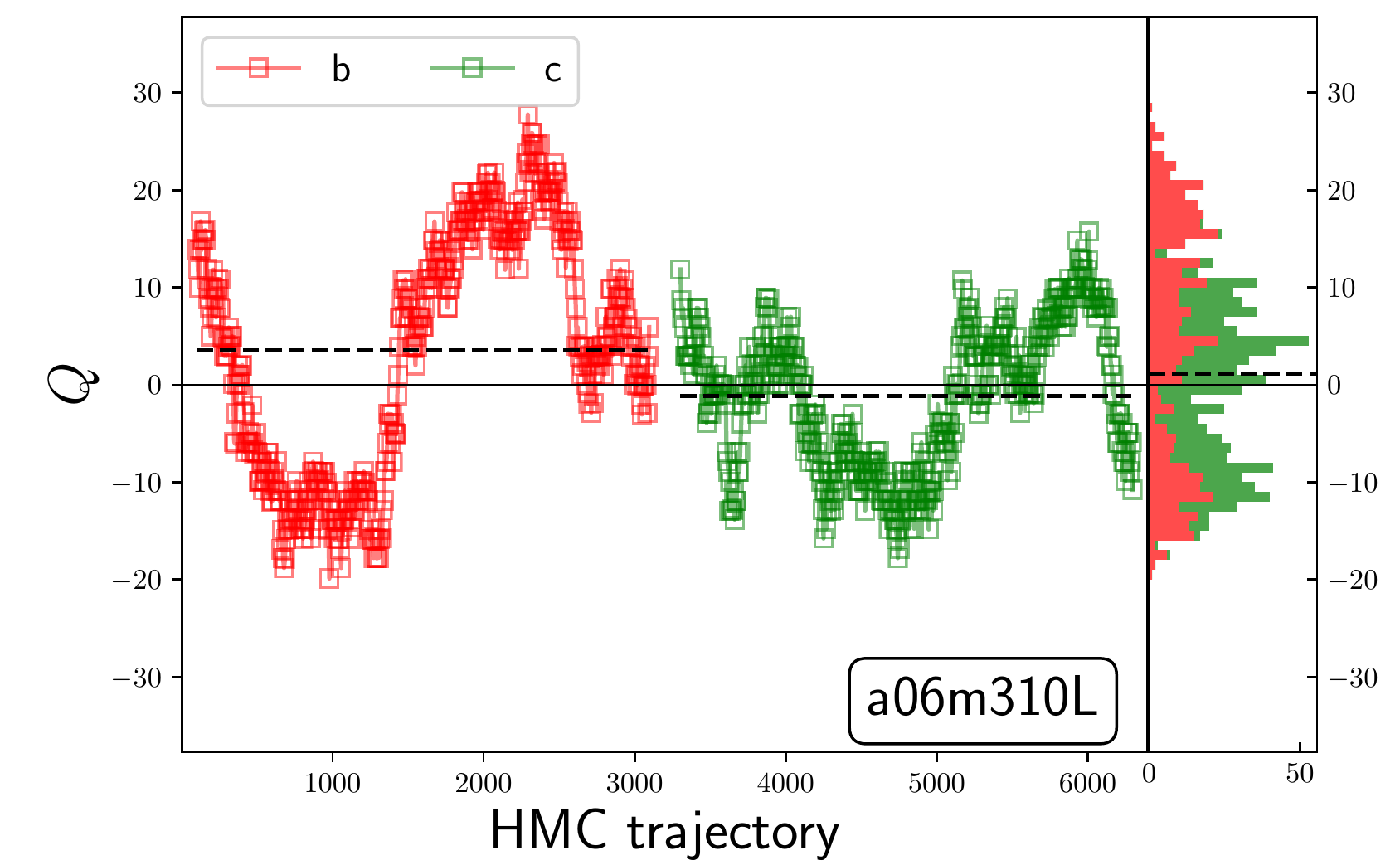}
\caption{\label{fig:Q_measurements}
Distribution of topological charge $Q$ measurements on each configuration.
The $Q$-values were determined by using the overrelaxed stout smearing technique outlined in~\cite{Moran:2008ra} with weight parameters $\rho$ given in \tabref{tab:Q_measurements} and $\varepsilon = -0.25$. We cross-checked a sample of our stout smeared measurements with the more expensive Symanzik flow technique and saw good agreement between the two. We determined the $\rho$ parameter and the number of steps to perform on an ensemble-to-ensemble basis, i.e., for a handful of configurations per ensemble we choose a spread of $\rho$s and step numbers and observe which combination gives the best plateau. These values of $\rho$ and step number ($\varepsilon$ is always $-0.25$) are then applied to the entire ensemble.}
\end{figure}

\end{widetext}

\bibliography{c51_bib}

\end{document}